\documentclass[fleqn,usenatbib]{config/mnras}

\usepackage{newtxtext,newtxmath}

\usepackage[LGR,T1]{fontenc}
\usepackage{anyfontsize}

\DeclareRobustCommand{\VAN}[3]{#2}
\let\VANthebibliography\thebibliography
\def\thebibliography{\DeclareRobustCommand{\VAN}[3]{##3}\VANthebibliography}

\usepackage{hyperref}
\hypersetup{
    colorlinks=true,
    linkcolor=Cerulean,
    citecolor=NavyBlue,
    filecolor=magenta,
    urlcolor=blue,
    pdftitle={R. G. Pascalau \& F. D'Eugenio - GA-NIFS MACS0647-JD1 - JWST/NIRSpec},
    }
\usepackage{graphicx}	
\usepackage{amsmath}	
\usepackage{makecell}	
\usepackage{subcaption}
\usepackage{adjustbox}
\usepackage{tabularx}
\usepackage{placeins}
\usepackage{textgreek}
\usepackage{xargs}
\usepackage{float}
\usepackage[dvipsnames]{xcolor}
\usepackage{xspace}
\usepackage{siunitx}
\DeclareSIUnit\angstrom{\text {Å}}
\usepackage{etoolbox}
\makeatletter
\newcommand\sendemail[4]{
\edef\@tempa{mailto:#1?subject=#2&body=#3 }%
\edef\@tempb{\expandafter\html@spaces\@tempa\@empty}%
\href{\@tempb}{#4}}

\catcode\%=11
\def\html@spaces#1 #2{#1
\catcode\%=14
\makeatother

\newcommand{\todo}[1]{\textcolor{magenta}{[#1]}}
\newcommand{\orcid}[2]{\href{http://orcid.org/#2}{#1}}
\newcommand{\orcidsymb}[2]{\href{http://orcid.org/#2}{#1\adjustbox{trim={-.15\width} {0\height} {-.15\width} {0\height},clip}{\includegraphics[height=10pt]{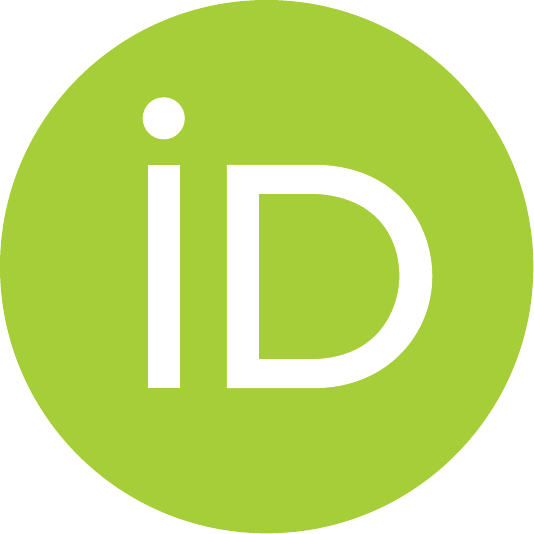}}}}

\newcommand{\citationneeded}{\textcolor{ForestGreen}{$^{\rm citation\;needed}$}}
\let\oldtextsigma\textsigma
\renewcommand{\textsigma}{\oldtextsigma\xspace}
\let\oldAA\AA
\renewcommand{\AA}{\text{\oldAA}\xspace}
\let\oldtextdegree\textdegree
\renewcommand{\textdegree}{\oldtextdegree\xspace}

\newcommand{\kms}{\ensuremath{\mathrm{km\,s^{-1}}}\xspace}
\newcommand{\Msun}{\ensuremath{{\rm M}_\odot}\xspace}
\newcommand{\Zsun}{\ensuremath{{\rm Z}_\odot}\xspace}
\newcommand{\yr}{\ensuremath{{\rm yr}}\xspace}
\newcommand{\Myr}{\ensuremath{{\rm Myr}}\xspace}
\newcommand{\Gyr}{\ensuremath{{\rm Gyr}}\xspace}
\newcommand{\peryr}{\ensuremath{{\rm yr^{-1}}}\xspace}
\newcommand{\Lsun}{\hbox{\,${\rm L}_\odot$}}
\newcommand{\mum}{\text{\textmu m}\xspace}
\newcommand{\kpc}{\text{kpc}\xspace}
\newcommand{\ZH}{\text{[Z/H]}\xspace}
\newcommandx{\pcm}[1][1=3]{\ensuremath{\mathrm{cm}^{-#1}}\xspace}	

\newcommandx{\lambdar}[2][1=R,2=]{\ensuremath{\lambda_{\rm {#1}}{#2}}\xspace}
\newcommand{\eps}{\ensuremath{\epsilon}\xspace}
\newcommand{\mstar}{\ensuremath{M_\star}\xspace}
\newcommand{\mdyn}{\ensuremath{M_\mathrm{dyn}}\xspace}
\newcommand{\re}{\ensuremath{R_\mathrm{e}}\xspace}
\newcommand{\vstar}{\ensuremath{v_\star}\xspace}
\newcommand{\vnai}{\ensuremath{v_{\NaI}}\xspace}
\newcommand{\sigmastar}{\ensuremath{\sigma_\star}\xspace}
\newcommand{\sigmaestar}{\ensuremath{\sigma_{\star,\mathrm{e}}}\xspace}
\newcommand{\vperc}[1]{\ensuremath{v_{#1}}\xspace}

\newcommand{\vesc}{\ensuremath{v_\mathrm{esc}}\xspace}
\newcommand{\nelec}{\ensuremath{n_\mathrm{e}}\xspace}
\newcommand{\Telec}{\ensuremath{T_\mathrm{e}}\xspace}
\newcommand{\Rout}{\ensuremath{R_\mathrm{out}}\xspace}
\newcommand{\vout}{\ensuremath{v_\mathrm{out}}\xspace}
\newcommandx{\Mout}[2][1=,2=]{\ensuremath{M_{\mathrm{out}{#2}}^{#1}}\xspace}
\newcommandx{\Mdotout}[2][1=,2=]{\ensuremath{\dot{M}_{\mathrm{out}{#2}}^{#1}}\xspace}

\newcommandx{\fluxdcgs}[1][1=-20]{$\times 10^{[#1]}$~erg~s$^{-1}$~cm$^{-2}$~\AA$^{-1}$\xspace}
\newcommandx{\fluxcgs}[2][1=-20,2=\ensuremath{\times}]{${#2}10^{#1}$~erg~s$^{-1}$~cm$^{-2}$\xspace}
\newcommandx{\powercgs}[1][1=44]{$\times 10^{#1}$~erg~s$^{-1}$\xspace}
\newcommand{\Av}{\ensuremath{A_V}\xspace}

\newcommand{\jwst}{\textit{JWST}\xspace}
\newcommand{\hst}{\textit{HST}\xspace}
\newcommand{\ppxf}{{\sc ppxf}\xspace}
\newcommand{\prospector}{{\sc prospector}\xspace}
\newcommand{\emcee}{{\sc emcee}\xspace}
\newcommand{\cloudy}{{\sc cloudy}\xspace}
\newcommand{\pyneb}{{\sc pyneb}\xspace}
\newcommandx{\mappings}[1][1=]{{\sc mappings{#1}}\xspace}
\newcommand{\galfit}{{\sc galfit}\xspace}
\newcommand{\qubespec}{{\sc qubespec}\xspace}
\newcommand{\pysersic}{{\sc pysersic}\xspace}

\newcommand{\blackthunder}{BlackTHUNDER\xspace}
\newcommand{\Mdynvalue}{$\Mdyn = 2.0\pm0.5 \times 10^{11}$~\MSun}

\defcitealias{juodzbalis+2024b}{J24}
\defcitealias{gordon+2003}{G03}

\newcommand{\Lyalpha}{\text{Ly\,\textalpha}\xspace}
\newcommand{\Halpha}{\text{H\,\textalpha}\xspace}
\newcommand{\Hbeta}{\text{H\,\textbeta}\xspace}
\newcommand{\Hgamma}{\text{H\,\textgamma}\xspace}
\newcommand{\Hdelta}{\text{H\,\textdelta}\xspace}
\newcommand{\Paalpha}{\text{Pa\,\textalpha}\xspace}
\newcommand{\Pabeta}{\text{Pa\,\textbeta}\xspace}
\newcommand{\Hepsilon}{\text{H\,\textepsilon}\xspace}

\newcommandx{\permittedEL}[6][1=O,2=III,3=,4=,5=,6=]{\text{{#1}\,{\sc {#2}}{#3}{#4}{#5}{#6}}\xspace}
\newcommandx{\semiforbiddenEL}[6][1=O,2=III,3=,4=,5=,6=]{\text{{#1}\,{\sc{#2}}]{#3}{#4}{#5}{#6}}\xspace}
\newcommandx{\forbiddenEL}[6][1=O,2=III,3=,4=,5=,6=]{\text{[{#1}\,{\sc{#2}}]{#3}{#4}{#5}{#6}}\xspace}

\newcommand{\EW}[1]{\text{EW(#1)}\xspace}

\newcommand{\HI}{\permittedEL[H][i]}
\newcommand{\HII}{\permittedEL[H][ii]}

\newcommand{\NV}{\permittedEL[N][v]}
\newcommandx{\NVL}[1][1=1243]{\permittedEL[N][v][\textlambda][#1]}
\newcommandx{\NVall}{\permittedEL[N][v][\textlambda][\textlambda][1239,][1243]}

\newcommandx{\CIIL}[1][1=232x]{\semiforbiddenEL[C][ii][\textlambda][#1]}
\newcommandx{\CIIall}{\semiforbiddenEL[C][ii][\textlambda][\textlambda][2324--][2329]}

\newcommand{\NIV}{\semiforbiddenEL[N][iv]}
\newcommandx{\NIVL}[1][1=1486]{\semiforbiddenEL[N][iv][\textlambda][#1]}

\newcommand{\CIV}{\permittedEL[C][iv]}
\newcommandx{\CIVL}[1][1=1550]{\permittedEL[C][iv][\textlambda][#1]}
\newcommand{\CIVall}{\permittedEL[C][iv][\textlambda][\textlambda][1548,][1551]}

\newcommand{\HeII}{\permittedEL[He][ii]}
\newcommandx{\HeIIL}[1][1=1640]{\permittedEL[He][ii][\textlambda][#1]}

\newcommand{\semiOIII}{\semiforbiddenEL[O][iii]}
\newcommandx{\semiOIIIL}[1][1=1666]{\semiforbiddenEL[O][iii][\textlambda][#1]}
\newcommand{\semiOIIIall}{\semiforbiddenEL[O][iii][\textlambda][\textlambda][1661,][1666]}

\newcommand{\NIII}{\semiforbiddenEL[N][iii]}
\newcommandx{\NIIIL}[1][1=1750]{\semiforbiddenEL[N][iii][\textlambda][#1]}
\newcommand{\NIIIall}{\semiforbiddenEL[N][iii][\textlambda][\textlambda][1747--][1754]}

\newcommandx{\CIII}{\semiforbiddenEL[C][iii]}
\newcommandx{\CIIIL}[1][1=1909]{\semiforbiddenEL[C][iii][\textlambda][#1]}
\newcommand{\CIIIall}{\semiforbiddenEL[C][iii][\textlambda][\textlambda][1907,][1909]}

\newcommand{\NeIV}{\forbiddenEL[Ne][iv]}
\newcommandx{\NeIVL}[1][1=2424]{\forbiddenEL[Ne][iv][\textlambda][#1]}
\newcommand{\NeIVall}{\forbiddenEL[Ne][iv][\textlambda][\textlambda][2422,][2424]}

\newcommand{\MgII}{\permittedEL[Mg][ii]}
\newcommandx{\MgIIL}[1][1=2803]{\permittedEL[Mg][ii][\textlambda][#1]}
\newcommand{\MgIIall}{\permittedEL[Mg][ii][\textlambda][\textlambda][2796,][2803]}

\newcommand{\NeV}{\forbiddenEL[Ne][v]}
\newcommandx{\NeVL}[1][1=3426]{\forbiddenEL[Ne][v][\textlambda][#1]}
\newcommand{\NeVall}{\forbiddenEL[Ne][v][\textlambda][\textlambda][3346,][3426]}

\newcommand{\OII}{\forbiddenEL[O][ii]}
\newcommandx{\OIIL}[1][1=3726]{\forbiddenEL[O][ii][\textlambda][#1]}
\newcommand{\OIIall}{\forbiddenEL[O][ii][\textlambda][\textlambda][3726,][3729]}

\newcommand{\NeIII}{\forbiddenEL[Ne][iii]}
\newcommandx{\NeIIIL}[1][1=3869]{\forbiddenEL[Ne][iii][\textlambda][#1]}
\newcommand{\NeIIIall}{\forbiddenEL[Ne][iii][\textlambda][\textlambda][3869,][3967]}

\newcommand{\OIII}{\forbiddenEL[O][iii]}
\newcommandx{\OIIIL}[1][1=5007]{\forbiddenEL[O][iii][\textlambda][#1]}
\newcommand{\OIIIall}{\forbiddenEL[O][iii][\textlambda][\textlambda][4959,][5007]}

\newcommandx{\NIL}[1][1=5200]{\forbiddenEL[N][i][\textlambda][#1]}
\newcommand{\NIall}{\forbiddenEL[N][i][\textlambda][\textlambda][5198,][5200]}

\newcommand{\OI}{\forbiddenEL[O][i]}
\newcommandx{\OIL}[1][1=6300]{\forbiddenEL[O][i][\textlambda][#1]}
\newcommand{\OIall}{\forbiddenEL[O][i][\textlambda][\textlambda][6300,][6364]}

\newcommand{\HeI}{\permittedEL[He][i]}
\newcommandx{\HeIL}[1][1=5875]{\permittedEL[He][i][\textlambda][#1]}

\newcommand{\OIres}{\permittedEL[O][i]}
\newcommandx{\OIresL}[1][1=8446]{\permittedEL[O][i][\textlambda][#1]}

\newcommand{\NII}{\forbiddenEL[N][ii]}
\newcommandx{\NIIL}[1][1=6583]{\forbiddenEL[N][ii][\textlambda][#1]}
\newcommand{\NIIall}{\forbiddenEL[N][ii][\textlambda][\textlambda][6548,][6583]}

\newcommand{\SII}{\forbiddenEL[S][ii]}
\newcommandx{\SIIL}[1][1=6716]{\forbiddenEL[S][ii][\textlambda][#1]}
\newcommand{\SIIall}{\forbiddenEL[S][ii][\textlambda][\textlambda][6716,][6731]}

\newcommandx{\OIIAuL}[1][1=7325]{\forbiddenEL[O][ii][\textlambda][#1]}
\newcommand{\OIIAuall}{\forbiddenEL[O][ii][\textlambda][\textlambda][7319--][7331]}

\newcommandx{\CIIFIRL}{\forbiddenEL[C][ii][\textlambda][158\,\mum]}

\newcommand{\hda}{\ensuremath{\mathrm{H\text{\textdelta}_A}}\xspace}
\newcommand{\hga}{\ensuremath{\mathrm{H\text{\textgamma}_A}}\xspace}

\newcommand{\target}{MACS-0647-JD1\xspace}
\newcommand{\HST}{\textit{HST}\xspace}
\newcommand{\fde}[1]{\textcolor{DarkOrchid}{FDE: #1}\xspace}
\renewcommand*{\figureautorefname}{Fig.}

\title[MACS0647-JD1: resolved ISM properties]{GA-NIFS: Dissecting The Alchemised: JWST reveals turbulent metal-poor gas fuelling a co-spatial starburst in a complex system at $\bf z=10.17$}

\author[Robert G. Pascalau et al.]{
\parbox{\textwidth}{
\orcidsymb{Robert G. Pascalau,}{0000-0001-9820-5773}\thanks{E-mail: rgp34@cam.ac.uk}$^{1,2}$
\orcidsymb{Francesco D'Eugenio,}{0000-0003-2388-8172}$^{1,2}$
\orcidsymb{Roberto Maiolino,}{0000-0002-4985-3819}$^{1,2,3}$
\orcidsymb{Qiao Duan,}{0009-0009-8105-4564}$^{1,2}$
\orcidsymb{Yuki Isobe,}{0000-0001-7730-8634}$^{1,2,4}$
\orcidsymb{Santiago Arribas,}{0000-0001-7997-1640}$^{5}$
\orcidsymb{Andrew J. Bunker,}{0000-0002-8651-9879}$^{6}$
\orcidsymb{St\'ephane Charlot,}{0000-0003-3458-2275}$^{7}$
\orcidsymb{Michele Perna,}{0000-0002-0362-5941}$^{5}$
\orcidsymb{Bruno Rodr{\'i}guez Del Pino,}{0000-0001-5171-3930}$^{5}$
\orcidsymb{Hannah {\"U}bler,}{0000-0003-4891-0794}$^{8}$
\orcidsymb{Elena Bertola,}{0000-0001-5487-2830}$^{9}$
\orcidsymb{Torsten B{\"o}ker,}{0000-0002-5666-7782}$^{10}$
\orcidsymb{Stefano Carniani,}{0000-0002-6719-380X}$^{11}$
\orcidsymb{Dan Coe,}{0000-0001-7410-7669}$^{10,12,13}$
\orcidsymb{Giovanni Cresci,}{0000-0002-5281-1417}$^{9}$
\orcidsymb{Mirko Curti,}{0000-0002-2678-2560}$^{14}$
\orcidsymb{Tiger Y.Y. Hsiao,}{0000-0003-4512-8705}$^{15}$
\orcidsymb{Lucy R. Ivey,}{0009-0002-5105-1222}$^{1,2}$
\orcidsymb{Gareth C. Jones,}{0000-0002-0267-9024}$^{1,2}$
\orcidsymb{Isabella Lamperti,}{0000-0003-3336-5498}$^{9,16}$
\orcidsymb{Eleonora Parlanti,}{0000-0002-7392-7814}$^{11}$
\orcidsymb{Jan Scholtz,}{0000-0001-6010-6809}$^{1,2}$
\orcidsymb{Sandro Tacchella,}{0000-0002-8224-4505}$^{1,2}$
\orcidsymb{Lorenzo Ulivi,}{0009-0001-3291-5382}$^{5}$
\orcidsymb{Giacomo Venturi,}{0000-0001-8349-3055}$^{11}$
\orcidsymb{Joris Witstok}{0000-0002-7595-121X}$^{17,18}$ and 
\orcidsymb{Sandra Zamora}{0000-0003-4546-897X}$^{11}$
\vspace{0.25cm}
\\
Affiliations are listed at the end of the paper.
}
}
\date{\vspace{-11ex}}

\pubyear{2026}

\begin{document}
\label{firstpage}
\pagerange{\pageref{firstpage}--\pageref{lastpage}}
\maketitle

\begin{abstract}

Recent observations revealed that distant galaxies have bursty star formation histories, regulated by stellar or active galactic nuclei (AGN) feedback and gas inflows. According to theoretical models, feedback preferentially removes metal-rich gas, while subsequent starbursts are triggered by mergers and newly accreted gas that is generally less enriched than the galaxy's interstellar medium (ISM). Therefore, gas-phase metallicity provides key insights into the baryonic processes shaping early galaxies. 
We present the first NIRSpec/IFU study of spatially resolved ISM properties in the MACS0647-JD system ($z=10.17$). The system consists of two stellar components detected in NIRSpec/IFU and NIRCam photometry. The main component ($\log \left(M_{\ast}/M_{\odot}\right)=7.77 \pm0.09$; $12+\log\left(\rm O/H\right)=7.89 \pm 0.16$) is more massive and significantly more metal-rich compared to its companion ($\log \left(M_{\ast}/M_{\odot}\right)=7.42\pm0.07$; $12+\log\left(\rm O/H\right)=7.47 \pm 0.20$), 
suggesting an older stellar population and a prolonged chemical enrichment history. We find that the H$\gamma$ line emission centroid is spatially offset by $\sim 0.1\arcsec$ (150 pc in the source plane) from the stellar continuum centroid; the latter coincides with the location of the main stellar component. This offset provides possible evidence of a merger-driven starburst in this system. By comparing the spatial distributions of the metallicity, 
velocity dispersion, and the burstiness of star formation history, we infer the presence of turbulent, metal-poor gas outside the stellar components. 
This metal-poor, dynamically unstable gas is likely responsible for the enhanced recent star formation in the north-east region of the system. 

\end{abstract}

\begin{keywords}
galaxies: high redshift -- galaxies: evolution -- galaxies: abundances -- galaxies: ISM -- ISM: kinematics and dynamics
\end{keywords}



\section{Introduction}

Theoretical models posit that early galaxies had variable star formation histories (SFHs), characterised by multiple episodes of intense star formation, each followed by periods of lower activity \citep{ma_fire,faucher_giguere+2018,iyer+2020,tacchella+2020,dome+2024,mcclymont2025_SFMS_u}. \textit{JWST} observations are confirming this picture, with the discovery of a plethora of galaxies in each of these phases: starbursts \citep{boyett+24,dressler+24,langeroodi+2024,endsley+25,looser+25,witten+2025} and quenched \citep{strait+23,looser+24,baker+25a,covelo-paz+26}. One of the key tracers of star formation histories is gas-phase metallicity, which 
carries the imprint of a number of mechanisms shaping the evolution of galaxies before Cosmic Noon \citep{dave+2011,lilly+2013,somerville+15,peroux+2020}. SFHs are regulated by a combination of baryonic processes, including mergers, cosmological gas accretion, and gas outflows driven by stellar \citep{baker+23,zamora+25b} or active galactic nuclei (AGN) feedback \citep{fabian+2012,weinberger+2017,bourne+2023}. Such outflows can efficiently remove metal-enriched gas and suppress star formation, leading to quiescent phases \citep{fabian+2012,hopkins+14,hopkins+23}. Conversely, mergers and metal-poor inflows of dense gas can rejuvenate galaxies by triggering intense episodes of star formation \citep{keres+2005,dekel2009,schaye+2010,sparre+2017,sun_g+23,langeroodi_compact_inflows,mason_c+23,yajima+23}. 


It has been suggested that gas accretion and bursty star formation lead to increased scatter of the mass--metallicity relation (MZR; \citealt{Lequeux1979,tremonti+2004,kewley+08,maiolino+2008,sanchez+2017}), as reported by simulations \citep{derossi+2017,agertz+2020,garcia+2024} and semi-analytical models (e.g., \citealt{delucia+2020}). Furthermore, simulations reveal a link between stellar feedback and the spatial distribution of metals: models with enhanced stellar feedback lead to efficient mixing of metals in the interstellar medium (ISM), resulting in flatter radial metallicity gradients on average in these cases \citep{gibson+13,mott+2013,sun_x+2025,garcia+2025b_u}. However, such metallicity gradients are physically stable only if these systems had evolved long enough to reach a nearly steady state. Observational evidence for correlations between very recent star formation activity and metallicity on spatially resolved scales in the early Universe remains scarce (e.g., \citealt{marconcini+2024}).



Following the launch of the James Webb Space Telescope \citep[\textit{JWST};][]{JWST_new,rigby+23}, it became possible to study rest-frame optical and near-infrared spectra and photometry of distant galaxies less than 1 Gyr after the Big Bang \citep{bunker_gnz11,curtis-lake+23,fujimoto+23,kashino+23,oesch+2023,roberts-borsani+23,robertson+23,wang+23,bunker+24,deugenio+24_gsz12,fujimoto+24,hainline+24b,harikane+24,deugenio_jades_dr3,donnan+25u,marconcini_cr7,napolitano+2025,price+25}, even up to $z>14$ \citep[$<300 \ \rm Myr$ after the Big Bang;][]{carniani+24_zg14,naidu_momz14_u}. In order to understand how galaxies evolve across cosmic time, it is necessary to constrain the properties of their interstellar media (ISM). 

Owing to its unprecedented near-infrared sensitivity, and its ability to provide high spatial resolution (sub-kpc scales) and $R\sim 1000-3000$ spectra unaffected by telluric lines and atmospheric effects, {\it JWST}/NIRSpec \citep{nirspec} has completely revolutionised our understanding of galaxy formation and evolution before Cosmic Noon. These early galaxies have truly extreme ISM conditions, including hard radiation fields, high ionisation parameters \citep{cameron+23,mascia+2023,sanders+23,tang+23,boyett+24,calabro+24,topping+24,topping+25_apj}, high electron number densities \citep{isobe+23,reddy+23,li_junyao+2025,topping+25}, and higher electron temperatures combined with lower metallicities compared to lower-$z$ galaxies of similar $M_{\ast}$ and star formation rates \citep{langeroodi+23,trump+23,curti+24,laseter+24,pollock+2025u}. All these observational results are in agreement with the findings of theoretical hydrodynamical simulations such as \textsc{MEGATRON} \citep{choustikov+2025u}. Moreover, \textit{JWST} observations have substantially increased the number of detections of auroral lines at $2<z<10$ \citep{nakajima+23,laseter+24,morishita+24,sanders+24,scholte+25,shapley+25,stiavelli+25}, allowing for accurate calculations of gas-phase metallicities based on the direct-$T_{\rm e}$ method \citep{Peimbert1967,osterbrock+89}. Capitalising on the capabilities of the NIRSpec integral-field unit (IFU) mode \citep{boeker22_IFS} to reveal spatial and spectral information, a number of recent studies have focused on spatially resolved gas-phase metallicity and ISM properties in galaxies at $z=2-8$. One key strength of IFU studies is their ability to provide in-depth analyses of systems with complex morphologies and intricate kinematics \citep{bruno+2024,gareth2024_HFLS3,uebler+24,gareth_z57_SFG,marshall_eiger_u,parlanti+25_z55,jekyll_u,zamora_z7,gareth2024_z715_u}, including the discovery of dual/multiple AGN \citep{perna_dual_AGN,ubler+25sep}. A robust characterisation of the true nature of such systems is challenging without IFU spectroscopy \citep{deugenio_mahler}. 

At higher redshifts, galaxies are, on average, smaller than their low-$z$ counterparts, as demonstrated by observations (e.g., \citealt{shibuya+15,huertas-company+24,matharu+24,miller+25,stephenson+25,yang+2025_u,lola_gal_sizes_u}) and recent simulations (e.g., \citealt{roper+22,costantin+23,shen+2024,will_gal_sizes_u}). Resolving the ISM in such compact early galaxies is therefore difficult, except in cases where gravitational lensing provides sufficiently high magnification to resolve otherwise inaccessible physical scales \citep{mowla+22,claeyssens+23,vanzella+23,hsiao_nirspec,mowla+24,fujimoto_grapes_u}. Furthermore, additional caution is needed when interpreting metallicity gradients and ISM properties in high-$z$ targets because a substantial fraction of them either consist of multiple clumps \citep{le_fevre+20,herrera_camus+25} or are observed during a merger event (as expected given the increasing merger rates at higher redshifts; \citealt{qiao45to115,puskas+25}) during which gas and metals are redistributed across the system. 

MACS0647-JD is a star-forming galaxy (SFG) at $z_{\rm spec}=10.17$ \citep{hsiao+2023a,hsiao_nirspec}, with a stellar mass of $\log\left(M_{\ast}/M_{\odot}\right)=8.1 \pm 0.3$. Its instantaneous star formation rate (SFR) $\rm SFR_{H\alpha}=5.0 \pm 0.6 \ \rm M_{\odot} \ yr^{-1}$ \citep[][probed by $H\alpha$ narrow-line emission]{hsiao+2023b}
places this galaxy 0.6--0.8 dex above the star-forming main sequence at its redshift \citep[e.g.,][]{mcclymont2025_SFMS_u}, implying that the galaxy is undergoing a starburst episode. The system is triply lensed by the foreground galaxy cluster MACSJ0647.7+7015 ($z=0.597$; \citealt{ebeling+2007}). In this study, we analyse the image with the highest magnification, MACS0647-JD1 \citep[$\mu =8 \pm 1$;][]{chan+2017}. This system consists of two closely separated main components (MACS0647-JD A and B) with a third component (C) located farther away from this pair \citep{hsiao+2023a}. It was furthermore determined (based on integrated aperture spectroscopy) that component A of the system has a gas-phase metallicity of $12+\log\left(O/H\right)=7.79 \pm 0.09$ \citep{hsiao+2023b} and an ionisation parameter of $\log U=-1.9 \pm 0.2$, indicating a metal-poor but strongly ionised ISM, in line with other high-redshift observations \citep{bunker_gnz11,castellano+24,curti+25a}. However, no conclusion has been reached about the kinematic properties and the interactions between the two main components of this system, leaving open the question of whether they are distinct components of a merger or simply gas clumps in a rotating disc. In this work, we present NIRSpec/IFU medium-resolution spectroscopy of this target and argue that the former scenario is more likely. 

This paper is organised as follows. In Section \ref{sec:observations_data_redicton}, we discuss the data reduction pipeline, outlining the pre-processing applied to the raw data before using it in our analysis. Section \ref{sec:methods_overall} describes our point-spread-function (PSF) matching algorithm for spaxel spectra and the methods that we used to fit emission lines in the datacube, both spaxel by spaxel, as well as in spectra extracted from custom apertures. Section \ref{sec:results_general} also presents various components of our analysis and highlights the key results. Section \ref{sec:discussion_general} discusses the most probable nature of the MACS0647-JD system based on its resolved star formation histories, gas kinematics, and ISM properties. 
Our conclusions are outlined in Section \ref{sec:conclusions}. We assume a flat \textLambda CDM cosmology with $H_0 = 67.4$~\kms~Mpc$^{-1}$ and $\Omega_\mathrm{m}=0.315$ \citep{planck+2020}, giving a physical scale of 4.2~kpc~arcsec$^{-1}$ at a redshift of $z=10.17$ (before applying the lensing correction), corresponding to 1.5~kpc~arcsec$^{-1}$ in the source plane (for our assumed magnification of $\mu\sim8$). Stellar masses refer to the total stellar masses formed, assuming a \citet{chabrier2003} initial mass function (IMF), integrated between 0.1 and 120~\Msun. All magnitudes are in the AB system: $\rm AB \ mag=31.4-2.5\times\log F_{\rm nJy}$ \citep{oke+gunn1983}, and all EWs are in the rest frame, with negative EWs corresponding to line emission. All the logarithms in this paper are base-10 logarithms: $\log \equiv \log_{10}$.

\section{Data}

\label{sec:observations_data_redicton}

\begin{figure}
  \includegraphics[width=\columnwidth]{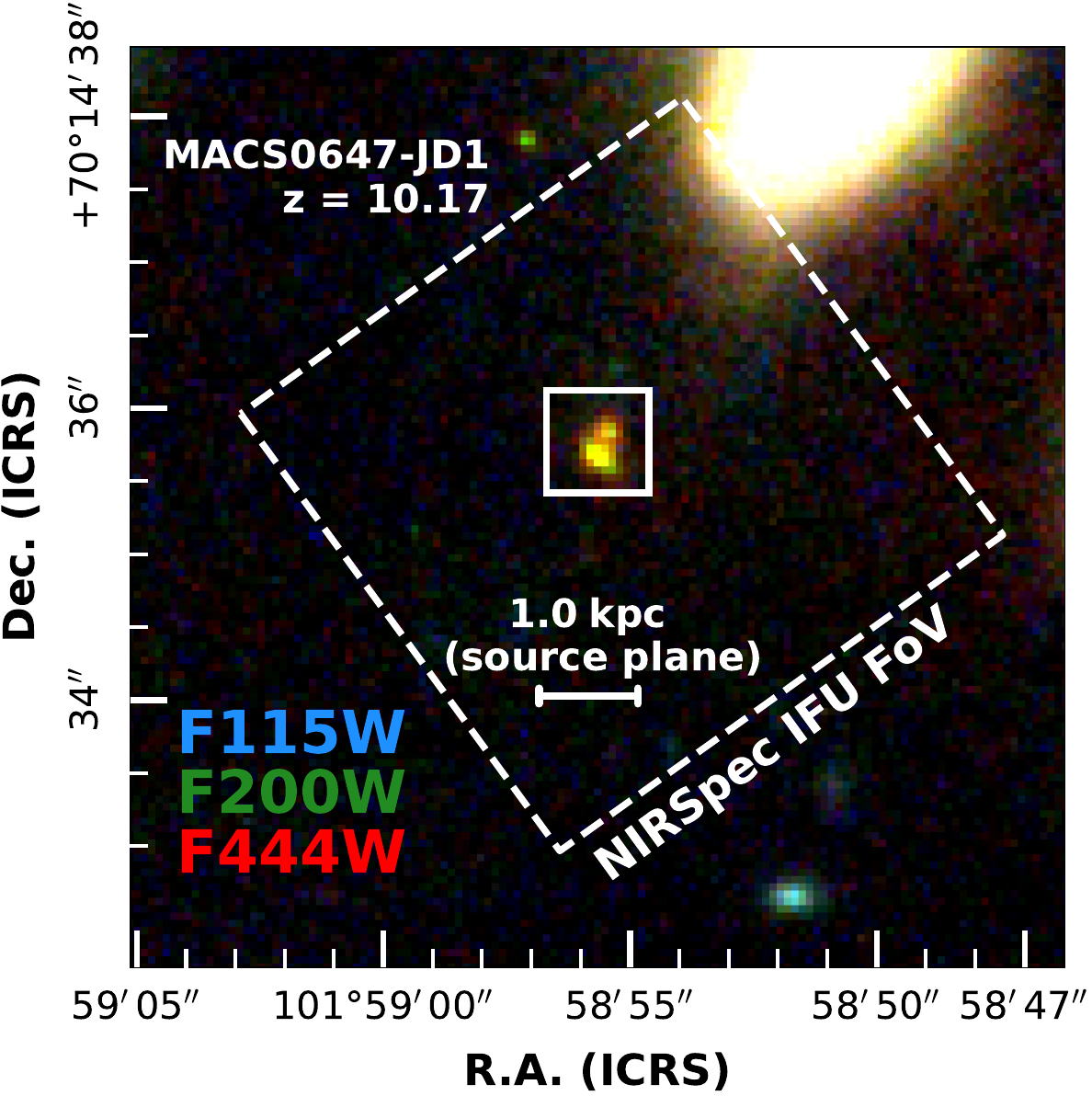}
  
  \caption{Composite RGB image of MACS0647-JD1 obtained by superposing F115W, F200W and F444W NIRCam photometry images. We also overlay the field-of-view of our IFU observations (dashed square), as well as the central $0.7\arcsec \times 0.7 \arcsec$ aperture (solid square), used for determining the integrated spectrum emission line fluxes and ISM properties. As discussed in Section \ref{sec:emission_lines}, we only fit the emission lines in the spectra of the spaxels contained within this aperture.}\label{f.rgb}
\end{figure}


\subsection{NIRSpec/IFU observations}
\label{sec:nirspec_obs} 

MACS0647-JD (ICRS coordinates of the MACS0647-JD1 image analysed in this paper: R.A. = $\mathrm{06^{h}47^{m}55.746^{s}}$; Dec. = $+70^{\circ}14 \arcmin 35.93 \arcsec$) was first observed with the Hubble Space Telescope using the Wide Field Camera 3 \citep[WFC3;][]{wfc3} and the Advanced Camera for Surveys \citep[ACS;][]{acs}; at the time, it was one of the most distant photometric candidates \citep{coe+2013}. Spectroscopic confirmation was obtained with \textit{JWST} ten years later \citep{hsiao_nirspec}. Our observations are part of the Galaxy Assembly with NIRSpec Integral Field Spectroscopy\footnote{\url{https://ga-nifs.github.io/}} (GA-NIFS; PIs: S.~Arribas and R.~Maiolino). GA-NIFS is a NIRSpec/IFU Guaranteed Time Observations programme, and our target is included in the \textit{JWST} Cycle 3 Program ID \#4528 (PI: K. Isaak). The observations used  the medium-resolution G395M/F290LP disperser/filter setup. 
We used a medium-cycling pattern 
with 8 dithers, 27 groups per integration, and one integration per exposure (with the NRSIRS2 readout mode; \citealt{rauscher+2017}), giving an on-source time of $4.4 \,\rm h$.

\subsection{NIRSpec data reduction and processing}
\label{sec:data_reduction}


The data reduction used the \textit{JWST} science calibration pipeline v1.17.1 with CRDS jwst\_1322.pmap. To improve the quality of the datacube, we performed a number of further data processing steps, in addition to the default pipeline. The customised procedures used for flagging cosmic rays snowballs and correcting pink noise (1/f noise) are described in detail in \citet{perna+23} and \citet{kashino+23}. These were applied to the Stage 1 count-rate products, before the Stage 2 and Stage 3 calibrations. For outlier removal, we used the outlier-detection algorithm described in \citet[][see also \citealt{ubler+23}]{deugenio+24_psb} and executed after Stage 2. 
This is adapted from the original \textsc{lacosmic} \citep{vandokkum+2001} algorithm. 
The background subtraction follows the procedure described in \citet{pascalau+26}, but here we used H$\gamma$ and [Ne \textsc{iii}]$\lambda3869$ to obtain the source mask. 
The datacube was obtained using the \textsc{cube\_build} algorithm with a pixel size of $0.05\arcsec$ and \textit{`drizzle'} weighting.

\subsection{Imaging and Photometry}
\label{sec:data_photom}

We retrieved NIRCam photometric data from the DAWN JWST Archive \textsc{v7.0} \citep[DJA;][]{heintz+2025}. We used the \textsc{v7.0} MACS0647 mosaics for the F115W, F150W, F200W, F277W, F356W and F444W filters. These imaging data were obtained as part of the \textit{JWST} Cycle 1 Program ID \#1433 (PI: D. Coe). Data reduction procedures have been described in detail by \citet{brammer+23} and \citet{valentino+23}. 
We produced $3\arcsec\times3\arcsec$ cutouts (similar to the IFU field-of-view) and applied background subtraction. We first removed bright sources using a $3\sigma$ mask produced with the \textsc{SigmaClip} algorithm from \textsc{astropy} \citep{astropy_collab}. We then used \textsc{Background2D} from \textsc{photutils} \citep{photutils_citation} to divide the masked cutouts into $0.35\arcsec\times0.35\arcsec$ regions. 
We used \textsc{MedianBackground} as our background estimator. The background values from neighbouring background-estimation regions are smoothed using a median filter and interpolated across the full image to produce a full 2-dimensional background image.

\subsection{Astrometry re-alignment of the datacube}\label{sec:astrometry}

\begin{figure}
  \centering
{\phantomsubcaption\label{fig:astrometry_a}
\phantomsubcaption\label{fig:astrometry_b}
\phantomsubcaption\label{fig:astrometry_c}
\phantomsubcaption\label{fig:astrometry_d}}
  \begin{subfigure}{\columnwidth}
    \centering
    \includegraphics[width=\linewidth]{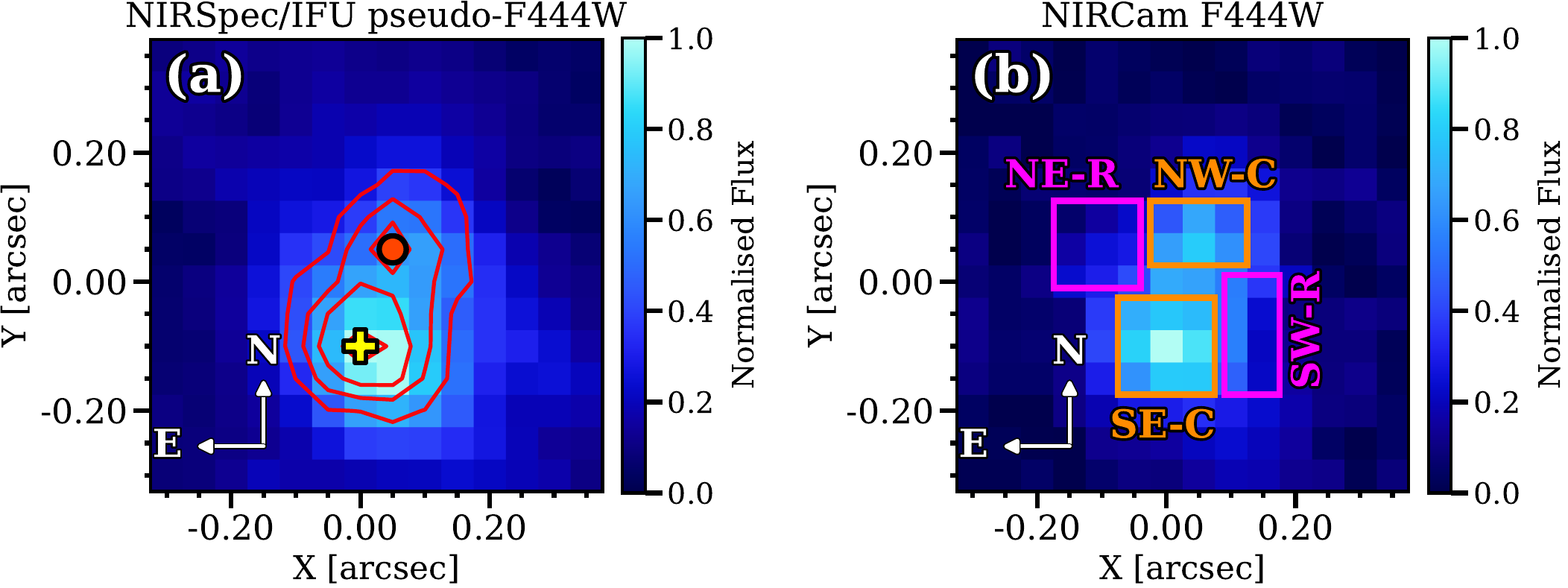}
    \label{f.astrometry.top}
  \end{subfigure}

  \vspace{-0.2em}

  \begin{subfigure}{\columnwidth}
    \centering
    \includegraphics[width=\linewidth]{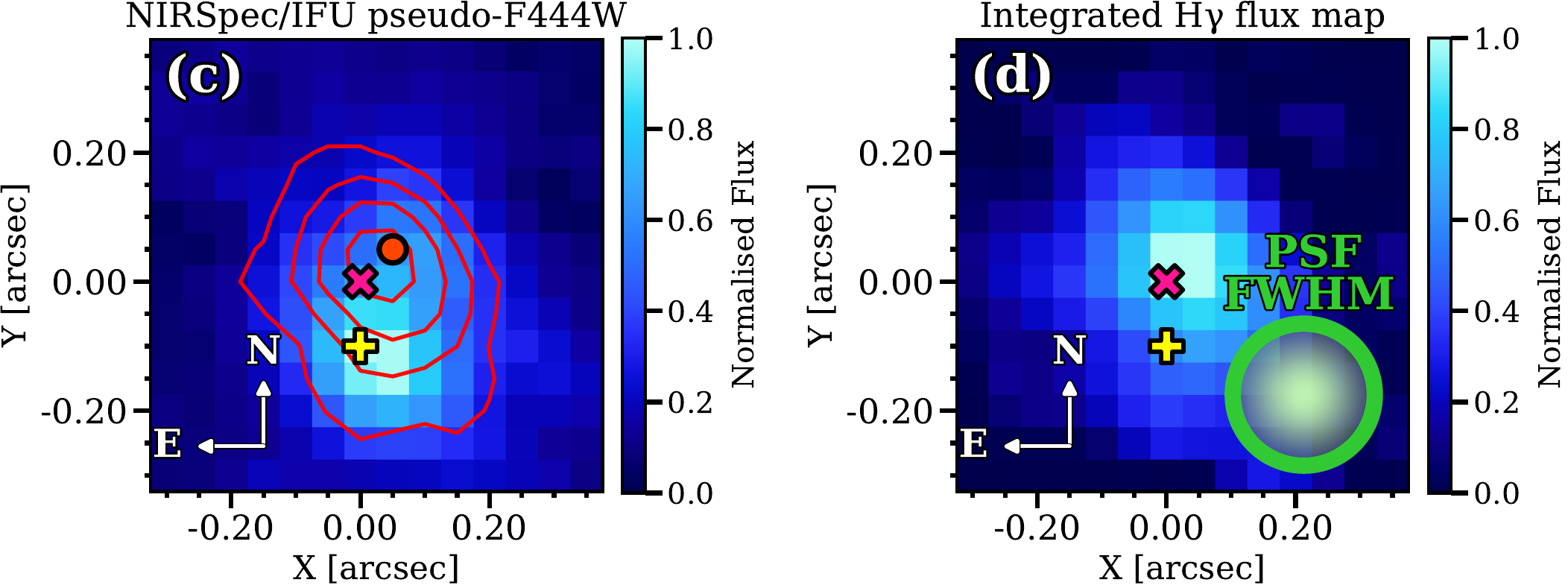}
    
    \label{f.astrometry.bottom}
  \end{subfigure}

  \caption{{\bf The top panels} show the astrometric alignment between the ICRS coordinates of the NIRSpec/IFU synthetic F444W image (\autoref{fig:astrometry_a}) and the NIRCam F444W photometry (\autoref{fig:astrometry_b}). In \autoref{fig:astrometry_a}, we overlay contours from the normalised NIRCam imaging data at 0.3, 0.5, 0.7, and 0.9 times the peak flux (defined by the stellar continuum centroid). We highlight the brightest spaxels in the two clumps in the system: south-east (yellow `+' cross) and north-west (red circle). The orange rectangles in \autoref{fig:astrometry_b} approximately mark the apertures corresponding to the SE and NW clumps, while the magenta rectangles indicate the regions outside the main stellar components. In \autoref{fig:astrometry_c}, we overlay the contours from the normalised H$\gamma$ integrated flux map (at the same fractional levels as in the top-left panel) on the synthetic F444W image from the datacube. We highlight the centroids of the south-east and north-west clumps (yellow `+' cross and red circle), together with the centroid of H$\gamma$ emission (pink `$\times$' cross). All the panels are shown in arcsecond offsets relative to the H$\gamma$ emission centroid. The same coordinate convention is adopted for all our subsequent maps in this paper. We also indicate the PSF FWHM (determined as described in Section~\ref{sec:psf}) with a circle of diameter equal to the FWHM, shown with a green edge and pale green Gaussian shading to illustrate our assumption of a two-dimensional Gaussian PSF.}
  \label{fig:astrometry.combined}
\end{figure}

The field of view 
of the NIRSpec/IFU pointing is $3.1\arcsec \times 3.1 \arcsec$, which is large enough to contain our target even without an accurate pointing procedure. 
The no-target acquisition mode provides an astrometric precision of $\sim 0.1\arcsec$ \citep{perna+23,ubler+23}. 
Because our study focuses on spatially resolved ISM properties, we refine the NIRSpec/IFU astrometry by matching the synthetic F444W image obtained from the datacube to the NIRCam F444W imaging data. This filter was chosen because its wavelength range covers the brightest part of the rest-frame optical continuum in young stellar populations, as well as key rest-frame optical emission lines in our spectra that we aim to map and study in detail. We find an offset of 0.19\arcsec\ between the NIRCam astrometry and our IFU data, comparable to those found in other studies \citep[e.g.,][]{arribas+24,lamperti+24,uebler+24,fujimoto25_alpine,fujimoto_grapes_u,parlanti+25_z55}. In \autoref{fig:astrometry.combined}, we mark the brightest spaxel in the synthetic F444W image with the yellow `+' cross mark. We regard this spaxel as the centroid of the stellar continuum light. However, the peak line emission centroid for the brightest emission lines in the system, H$\gamma$ and $\left[\rm Ne \ \textsc{iii}\right]\lambda3869$ (marked with a pink `$\times$' cross symbol) is offset by $0.1\arcsec$ from the stellar continuum centroid. We further discuss this spatial offset, together with the different morphologies of continuum and line emission in Section \ref{sec:discussion_general}. 



\subsection{Wavelength range extension and \texorpdfstring{$\boldsymbol{H\beta}$}{Hb} recovery}
\label{sec:hbeta_recovery}

The nominal wavelength range of NIRSpec G395M grating
reaches $\lambda=5.27~\mu \rm m$. However, wavelengths
$\lambda \lesssim 5.5~\mu \rm m$ are still recorded on the
detector, and can therefore be recovered, which is 
particularly useful for the brightest emission lines at high redshifts. The possibility of extending the recovered wavelength coverage beyond the nominal range has been demonstrated by several recent studies (\citealt{deugenio+25a_blackthunder_strikes_twice,torralba+25_warm+u} using high-resolution data; \citealt{parlanti+extension,valentino+25} using medium-resolution data). In
this work, we aim to measure H$\beta$, with a method similar to that implemented in \citet{deugenio+25a_blackthunder_strikes_twice}, but focusing on the
red end of the G395M grating. 

At $z\sim10.17$, H$\beta$ emission lies at $\lambda_{\rm obs}\approx5.43 \ \rm \mu m$, outside the nominal wavelength range covered by the G395M/F290LP disperser/filter setup. The line is well detected because the F290LP filter does not suppress longer wavelengths and the transmission of the grating does not drop sharply beyond $\lambda_{\rm obs}>5.27 \ \rm \mu m$ \citep{deugenio+25a_blackthunder_strikes_twice}. On the other hand, the $\left[\rm O \ \textsc{iii}\right]\lambda\lambda4959,5007$ doublet lines would fall at observed wavelengths of $\approx 5.54 \, \rm \mu m$ and $\approx 5.59 \ \rm \mu m$, respectively. At these wavelengths, the detector efficiency decreases, suppressing the $\left[\rm O \ \textsc{iii}\right]\lambda\lambda4959,5007$ emission, which is therefore not detected. 
We extract the spectra by extrapolating the flat-field curves and the wavelength calibration solution beyond the nominal range. There are only 89 spectral pixels (in the case of our medium-resolution grating) between the observed wavelength of H$\beta$ and the red end of the nominal spectral range of G395M. As a result of the well characterised behaviour of the grating, the error on the wavelength solution is small. 



\begin{figure}
  \includegraphics[width=\columnwidth]{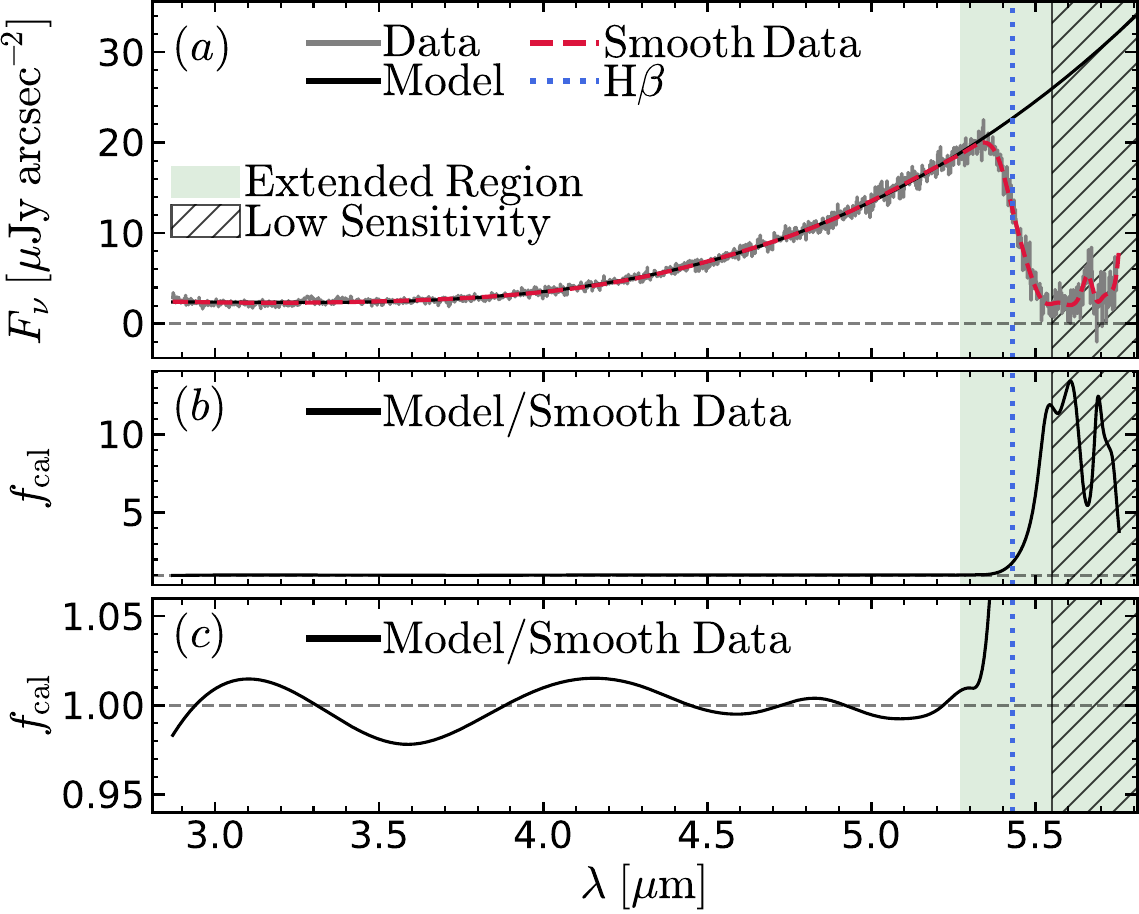}
  {\phantomsubcaption\label{f.fluxcal:a}
  \phantomsubcaption\label{f.fluxcal:b}
  \phantomsubcaption\label{f.fluxcal:c}}
  \caption{In \autoref{f.fluxcal:a} we display the raw and smoothed spectra from the circular aperture discussed in the main text in Section \ref{sec:emission_lines}, together with the extrapolation towards longer wavelengths, outside the nominal wavelength range, but where the detector is still sensitive enough for us to extract meaningful data. \autoref{f.fluxcal:b} illustrates the variation of the derived flux calibration factor $f_{\rm cal}$ across the wavelength range; we zoom in to $f_{\rm cal}\sim1$ in \autoref{f.fluxcal:c}.}\label{f.fluxcal}
\end{figure}

To derive the flux calibration of our extended data, we rely on extrapolating the background level.
We perform an initial data reduction in which we 
extend the existing calibration files for the flat-field correction and flux calibration. We initially use a linear extrapolation of the existing calibrations to the range 5.27--5.55~$\mu \rm m$. The calibrations consist of an
illumination flat, a detector flat, and a spectrograph flat.
After this initial step, the resulting datacube displays a sharp drop at $\lambda \gtrsim 5.35$~\mum. This is shown in \autoref{f.fluxcal:a}, where the grey spectrum was extracted in an empty
circular aperture from the initial datacube. Since the signal in this region is expected to increase
smoothly with wavelength, we use data in the well-calibrated spectral region $\lambda < 5.27$~\mum to model the background
(black line) and to derive an extended flux-calibration term, $f_\mathrm{cal}$, defined as the ratio between the
extrapolated model background and the smoothed data. This additional term rises rapidly at $\lambda
\gtrsim 5.35$~\mum (as shown in \autoref{f.fluxcal:c}), while at $\lambda \lesssim 5.27$~\mum it oscillates between $1.00\pm 0.02$ (\autoref{f.fluxcal:b}). We consider
the latter oscillations to be artefacts of our background model; hence, our additional flux-calibration
term is equal to unity at $\lambda \leq 5.27~\mum$ and is equal to the derived $f_\mathrm{cal}$ at
$\lambda > 5.27~\mum$.

\section{Analysis}
\label{sec:methods_overall}

\subsection{PSF matching for individual spaxels spectra}
\label{sec:psf}

In the case of individual spaxels ($0.05\arcsec \times 0.05 \arcsec$), it is crucial to take into account the size of the NIRSpec/IFU point-spread function (PSF). We assume the PSF to be a two-dimensional Gaussian, allowing for the observed asymmetry in the directions along and perpendicular to the IFU slices. The first such determination of the NIRSpec/IFU PSF size was performed by \citet{deugenio+24_psb} using three different methods that yielded consistent results. The authors report that the major axis of the 2-dimensional PSF is preferentially aligned with the cross-dispersion direction (corresponding to the long dimension of the slices). 
However, we can estimate the expected axial ratio of the PSF using the parametrisation in equations 3 and 4 from \citet{deugenio+24_psb}. We find that, for our wavelength range of interest $4\ {\rm \mu m}<\lambda_{\rm obs}<5.3 \ \rm\mu m$, the difference between the PSF sizes along and perpendicular to the direction of the slicers is less than 10\%. It is therefore reasonable to approximate the PSF shape as circular with the FWHM given by the geometric mean of the PSF FWHM values along and across the slicers. 

In a more recent work, \citet{jones2509_u} compared the wavelength-dependent PSF FWHM predicted by the \textsc{stpsf} software\footnote{\url{https://stpsf.readthedocs.io/en/latest/}} with PSF sizes estimates based on observations of two M-type stars and one white dwarf star in calibration programmes, as well as point-like quasi-stellar objects (QSOs) from \citet{marshall+23} and \citet{marshall+25_u}. \citet{jones2509_u} found a good agreement between these measurements and the \textsc{stpsf} calculations. Consequently, we use the PSF sizes determined by \textsc{stpsf} throughout our analysis. The emission line with the longest wavelength in our study is H$\beta$, and the PSF FWHM at this observed wavelength ($\sim 5.43 \ \mu\rm m$) is $0.22 \arcsec$. Thus, we chose to PSF-match our spectra to $\rm FWHM_{\rm PSF,max}=0.22\arcsec$ ($\sigma_{\rm PSF,max}=0.093 \arcsec$). First, we retrieve the PSF sizes at all wavelengths in our datacube, $\rm \sigma_{\rm PSF} \left(\lambda\right)$. The parametrisation we used was obtained by fitting the same functional form for $\sigma_{\rm PSF} \left(\lambda\right)$ as in \citet{deugenio+24_psb}:
\begin{equation}
    \sigma_{\rm PSF} \left[\rm arcsec\right]= 0.076 + 0.013\ \times \lambda \left[\mu {\rm m}\right] \times \exp \left(-7.6/\lambda\left[\mu {\rm m}\right]\right) .
\end{equation}



\noindent At each wavelength, we construct a 2-dimensional circular Gaussian kernel, $\mathcal{K}\left(\lambda\right)$, with a dispersion $\sigma_{K}\left(\lambda\right) =\sqrt{\rm \sigma_{\rm PSF,max}^{2}-\sigma_{\rm PSF}^{2}\left(\lambda\right)}$. 
We convolve the collapsed images of the datacube at each wavelength $\lambda$ with $\mathcal{K}\left(\lambda\right)$, and the collapsed variance datacube slices with $\mathcal{K}^{2}\left(\lambda\right)$ using the \textsc{convolve\_fft} routine from \textit{Astropy}.

\subsection{Emission lines fitting procedure}
\label{sec:emission_lines}






\begin{figure*}
{\phantomsubcaption \label{fig:fig_spectrum_a}
\phantomsubcaption \label{fig:fig_spectrum_b}}
  \begin{subfigure}[t]{0.58\linewidth}

  \includegraphics[width=\linewidth]{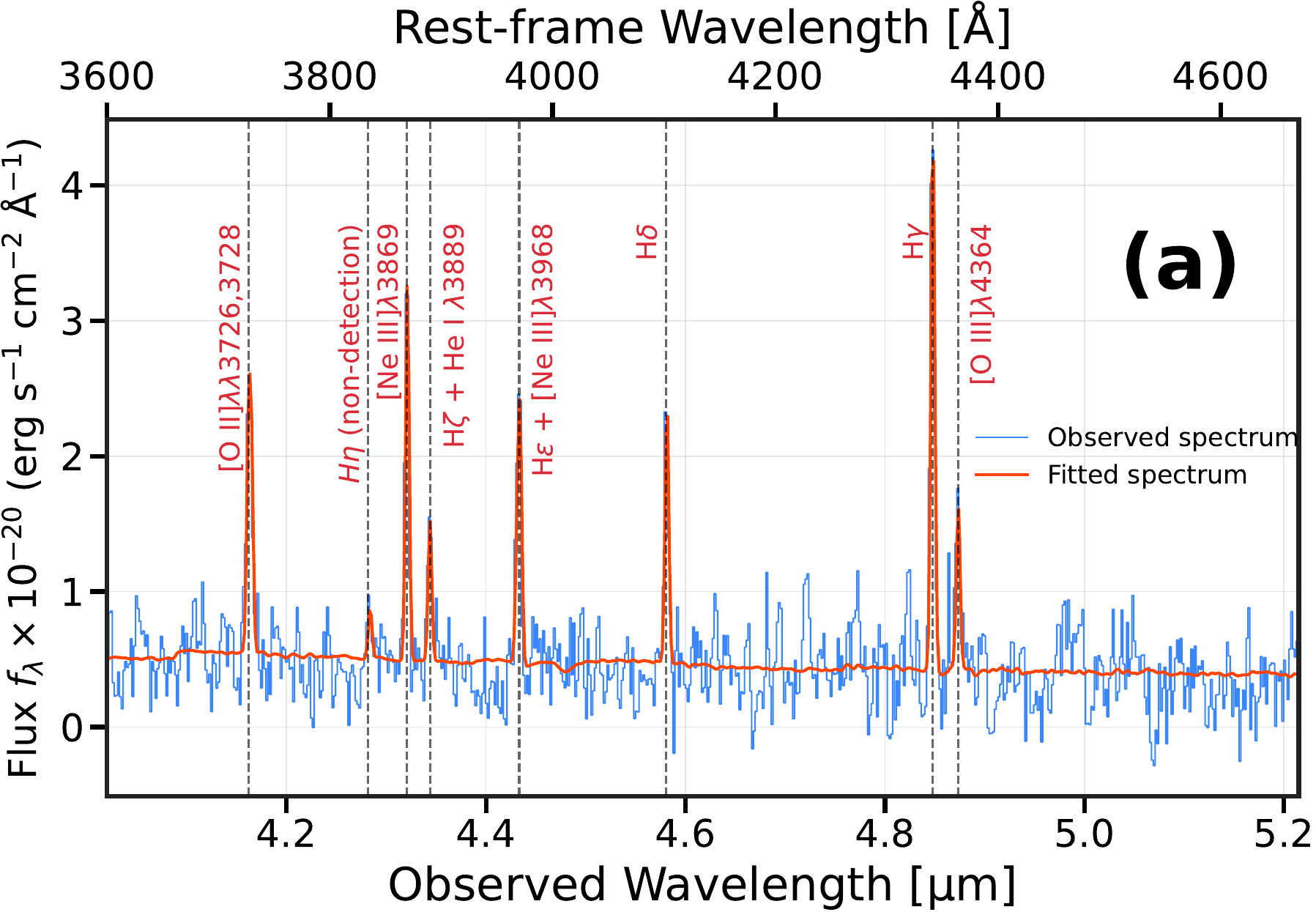}
  \end{subfigure}
  \hspace{0.02\linewidth}
    \begin{subfigure}[t]{0.39\linewidth}
    \includegraphics[width=\linewidth]{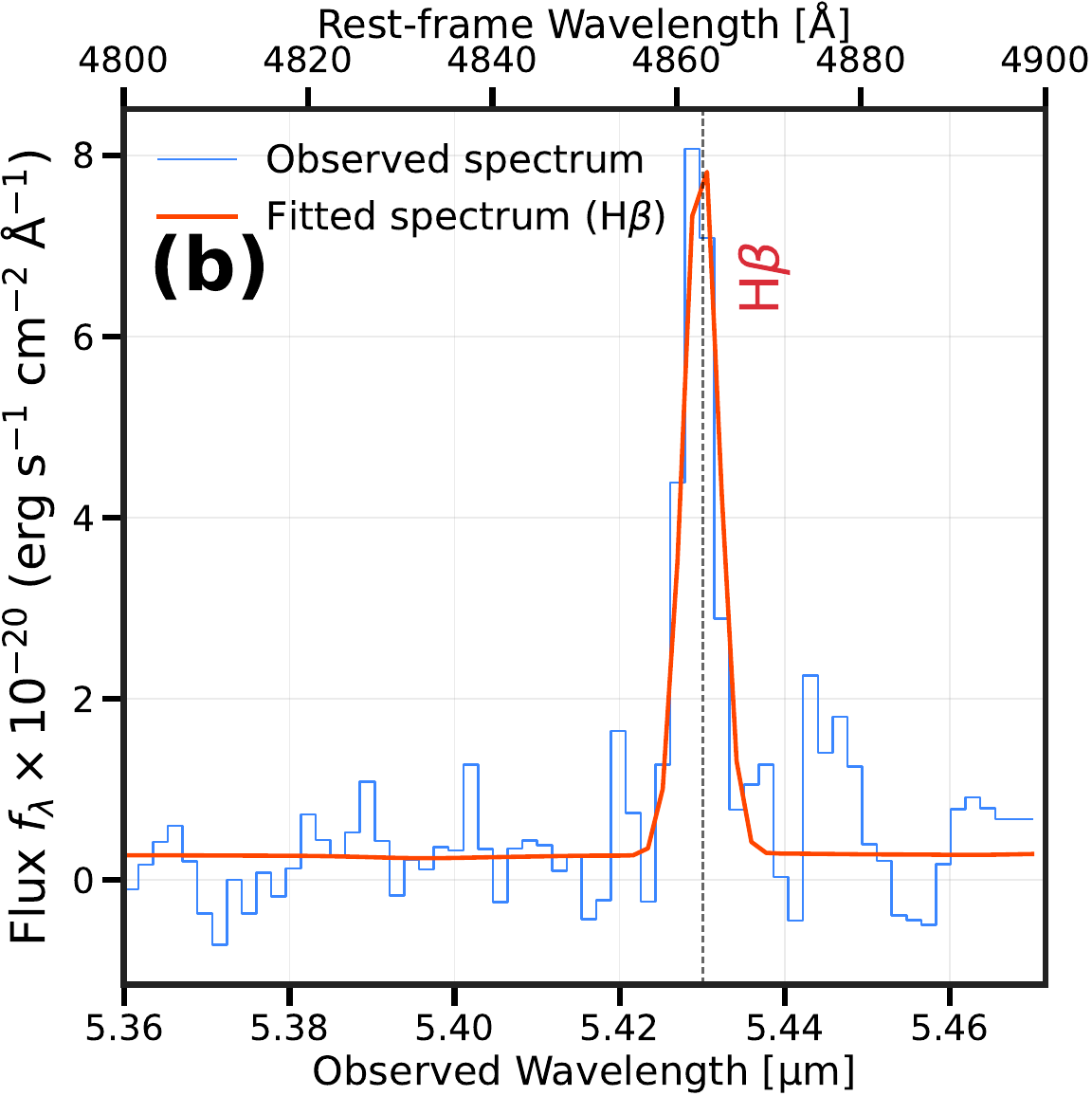}

  \end{subfigure}

  \caption{\autoref{fig:fig_spectrum_a} shows the observed integrated aperture spectrum in the rest-frame wavelength range between $\SI{3600}{\angstrom}$ and $\SI{4700}{\angstrom}$, highlighting the key emission lines discussed in the main text, including the $\left[\rm O \ \textsc{ii}\right]$ doublet and blended features. \autoref{fig:fig_spectrum_b} represents a zoom-in over the spectral wavelength range $\SI{4800}{\angstrom}<\lambda_{\rm rest}<\SI{4900}{\angstrom}$, which, for $z=10.17$, is redder than the nominal wavelength coverage of NIRSpec/G395M ($2.87-5.27~\mu \rm m$). The plot highlights the best-fit of the H$\beta$ line assuming only one Gaussian component, constrained by the kinematics of multiple emission lines shown in the left panel.}
  \label{fig:spectrum_integrated_figure}
\end{figure*}


We consider a square aperture of $0.7 \arcsec \times0.7 \arcsec$ centred on the H$\gamma$ emission centroid, and fit only the spaxels inside this aperture. This aperture contains most of the stellar continuum light (as shown in \autoref{f.rgb}) and the fluxes of key rest-frame optical lines (as shown in \autoref{fig:images_cube_vs_integrated}). We do not correct for aperture losses, as we estimate them to be $\lesssim10\%$ at $\lambda_{\rm obs}\sim4-5 \, \mu m$. This expectation is based on the calculations of \citet{bruno+2026} for a circular aperture of radius $r\sim0.4\arcsec$, which is comparable to the half-width of our square aperture and larger than the PSF FWHM at these wavelengths ($0.21-0.22\arcsec$, as argued in Section \ref{sec:psf}). We show the aperture spectrum in \autoref{fig:fig_spectrum_a}, covering the wavelength range $\SI{3600}{\angstrom}<\lambda_{\rm rest}<\SI{4700}{\angstrom}$. The key emission lines that we study are the $\left[\rm O \ \textsc{ii}\right]\lambda\lambda3726,3729$ doublet, the $\left[\rm Ne \ \textsc{iii}\right]\lambda\lambda3869,3968$ doublet, H$\delta$, H$\gamma$, the $\left[\rm O \ \textsc{iii} \right]\lambda 4363$ auroral line, and H$\beta$ (see Section \ref{sec:hbeta_recovery} for details on how it was extracted). 
H$\eta$ is not detected, as expected given the high order of this Balmer transition. H$\zeta$ and $\rm He \ \textsc{i}\ \lambda3889$ are blended so we perform a crude deblending by adopting a truncated Gaussian prior on the $H\delta/H\zeta$ ratio. The prior is centred on the case-B value of $H\delta/H\zeta=2.46$, computed using \textsc{PyNeb} \citep{pyneb}. This is a reasonable assumption because the system does not have significant dust attenuation (as discussed in Section \ref{sec:dust}). We choose a standard deviation of 0.3 for this Gaussian prior because it is sufficiently broad to account for our most pessimistic estimates of the flux calibration uncertainties ($\sim 10\% -15\%$). We selected H$\delta$ because it is the closest Balmer line to H$\zeta$ that is not blended. 
$\left[\rm Ne \ \textsc{iii}\right]\lambda3968$ and H$\epsilon$ are blended, but we can recover their fluxes by fixing $\left[\rm Ne \ \textsc{iii}\right]\lambda3968 / \left[\rm Ne \ \textsc{iii}\right]\lambda3869=0.301$ (e.g., \citealt{tripodi+24,gareth2024_z715_u}). All emission lines are fitted with a Gaussian profile; we tie the kinematics (the centroid and the intrinsic dispersion) of all lines. 
We fit the emission lines from $\left[\rm O \ \textsc{ii}\right]\lambda\lambda3726,3729$ to $\left[\rm O \ \textsc{iii}\right]\lambda 4363$. 
We clip and mask $>3\sigma$ outliers 
(in line free spectral regions) in the aperture and spaxel spectra, using the \textsc{sigma\_clip} procedure from \textsc{Astropy} \citep{astropy_collab}. 



We explored several parameterisations for the underlying continuum, including low-order polynomials and power-law models, but found strong degeneracies among the parameters governing the continuum shape in these approaches. These issues are avoided by adopting a \textsc{ppxf} \citep{cappellari2017ppxf,cappellari2023ppxf} fit for the continuum. 
In this work, we used templates built using \textit{FSPS v3.2}\footnote{\url{https://github.com/cconroy20/fsps}} \citep{fsps1, fsps2}, with the \textit{MILES} stellar library for the optical region \citep{sanchezblazquez2006, falcon2011} and the \textit{MIST} isochrones \citep{choi2016mist}. We adopted a \citet{chabrier2003} IMF. During the continuum fitting, spectral regions within $\pm\SI{10}{\angstrom}$ of the rest-frame wavelengths of the main emission lines indicated in \autoref{fig:spectrum_integrated_figure} were masked to prevent nebular emission from biasing the stellar continuum and absorption modelling. The continuum was fitted in the unmasked regions using both stellar and gas templates; the latter were kept enabled to account for possible residual contributions from faint nebular lines not included in the main emission-line mask. 
It is important to note, however, that the stellar continuum in our data is intrinsically very weak, and as a result the continuum fit does not strongly constrain detailed stellar population properties. Consistent with this, the underlying stellar Balmer absorption features are modest: in the full-aperture spectrum we measured rest-frame equivalent widths of $\mathrm{EW}(\mathrm{H}\gamma) \approx \SI{0.36}{\angstrom}$ and $\mathrm{EW}(\mathrm{H}\delta) \approx \SI{0.57}{\angstrom}$ in the best-fit stellar model. These values imply that uncertainties in the modelling of the continuum and stellar absorption features have a negligible impact on the emission line fluxes (which are calculated at a later step), which represent the core part of our analysis. 



To take into account the possible systematics in the errors measured within the pipeline, we rescaled the nominal flux uncertainties (extracted from the datacube `ERR' extension file) such that the median value of the noise in a given spaxel (measured across wavelength slices free of emission lines) matches the standard deviation of the residuals from the spectral fit in the line-free spectral regions. Because the noise rescaling factor is not known beforehand, we allowed it to be a free parameter in our fitting algorithm. A similar noise re-scaling approach was implemented by \citet{gareth_smouldering} in their study of a $z\sim4.26$ rotating disc star forming galaxy. For individual spaxels, rescaling factors vary between 1.9 and 2.5 \citep{ubler+23,uebler+24,venturi+25}. Finally, in the case of aperture spectra (obtained by adding the spectra of individual spaxels within each aperture), we obtain the uncertainties by adding in quadrature the re-scaled uncertainties in the spaxel spectra. 

We first identify the minimum $\chi^{2}$ solution of our emission line fitting algorithm using a least-squares procedure. To deconvolve the intrinsic line profiles from the observed features, we assume that the wavelength-dependent line spread function (LSF) of NIRSpec 
is given by the nominal, pre-launch resolution, $FWHM_\mathrm{LSF \ pre\text{-}launch}$ \citep{nirspec}, assumed to be Gaussian and improved by a factor of 0.8, following
\citet[][and roughly in agreement with e.g., \citealt{greene+2024}]{shajib+25_u}. The pre-launch estimates for the LSF width were too conservative because {\it JWST} provides a lower than expected wavefront error. The intrinsic FWHM$_\mathrm{intr}$ is thus: 


\begin{equation}
    \rm FWHM_{\rm intr}=\sqrt{\rm FWHM_{\rm obs}^{2}-\left(0.8 \times \rm FWHM_{\rm LSF \ pre\text{-}launch} \right)^{2}} \ .
\label{eq:LSF_correction}
\end{equation}


\noindent After this step, we use the least-squares solution 
to initialise the chains used to compute the posterior distributions of our models using a Markov chain Monte Carlo (MCMC) method and the software \textsc{emcee} \citep{emcee_paper2012}. 
The fiducial parameters are the median values of the marginalised posterior distributions, and the uncertainties are derived from the $16^{\rm th}$ and $84^{\rm th}$ percentiles. 
We assumed flat priors on all parameters and set the following prior bounds: $0<\log \left(F_{\rm int}/10^{-20} \ 
\rm erg \ s^{-1} cm^{-2}\right)<3$ for integrated fluxes, $10.15<z<10.20$ and $100 \ \rm km \ s^{-1}<\rm FWHM<500 \ \rm km \ s^{-1}$. For the ratio $\left[\rm O \ \textsc{ii}\right]\lambda3729$/$\left[\rm O  \ \textsc{ii}\right]\lambda3726$ we assume a range of $\left[0.34, 1.5\right]$ \citep{sanders+2016}. We fitted the two $\left[\rm O \ \textsc{ii}\right]\lambda\lambda3726,3729$ doublet components individually and tied their kinematics to those of other features (including the other, non-blended features). We note that, although we fitted the $\left[\rm O \ \textsc{ii}\right]\lambda\lambda3726,3729$ doublet components individually and tied their kinematics to those of other, non-blended features, our medium-resolution data cannot fully break the degeneracy between the red-to-blue line ratio $\left[\rm O \ \textsc{ii}\right]\lambda3729/\left[\rm O \ \textsc{ii}\right]\lambda3726$ and the flux of the blue component. However, the total $\left[\rm O \ \textsc{ii}\right]\lambda\lambda3726,3729$ flux is well constrained and we will focus on this parameter from now on.

Additionally, we tested an alternative fitting model for the integrated aperture spectrum in which we assumed that all emission lines are fitted with two Gaussians from two sets of kinematically tied components. The broader Gaussians are allowed to vary over a range of $100 \ \rm km \ s^{-1}<\rm FWHM<1000 \ \rm km \ s^{-1}$. However, a simple Bayesian Information Criterion (BIC; \citealt{bic}) test yields that $\Delta\rm BIC \equiv \rm BIC_{\rm 2 \ comp}-\rm BIC_{\rm 1 \ comp} \approx 11$. Hence, the two-Gaussians model is statistically disfavoured relative to the single-Gaussian model (even in the high-$\rm S/N$ aperture spectrum). We repeated this analysis for the spaxel spectra in the wavelength range $\SI{3600}{\angstrom}<\lambda_{\rm rest}<\SI{4700}{\angstrom}$, but found that no spaxel simultaneously satisfied both $\Delta \rm BIC<-10$ (which would indicate that a two-component fit is statistically favoured), and $\rm S/N>3$ for the secondary, high-FWHM component. Therefore, it is reasonable to adopt single-Gaussian fits for all spaxels. H$\beta$ is fitted separately from the other rest-frame optical emission lines, given the higher uncertainties of the continuum at $\lambda_{\rm obs}>5.27 \mu\rm m$, outside the nominal range covered by G395M/F290LP. The kinematics of H$\beta$ were tied to the lower-wavelength lines by imposing the posteriors of redshift and FWHM (derived from our $\SI{3600}{\angstrom}<\lambda_{\rm rest}<\SI{4700}{\angstrom}$ emission lines fit) as the priors on the H$\beta$ redshift and FWHM. This approach allows some flexibility in deriving the H$\beta$ kinematics and better accounts for possible wavelength calibration uncertainties than simply fixing $z_{H\beta}$ and $\rm FWHM_{H\beta}$ to the median values of the posterior distributions from the $\SI{3600}{\angstrom}<\lambda_{\rm rest}<\SI{4700}{\angstrom}$ fit. Only the H$\beta$ flux was fitted as a free parameter with a flat prior distribution.

Following these methods, we obtain the integrated aperture spectra over the $\SI{3600}{\angstrom}<\lambda_{\rm rest}<\SI{4700}{\angstrom}$ wavelength range (\autoref{fig:fig_spectrum_a}) and in the spectral region around H$\beta$ (\autoref{fig:fig_spectrum_b}). The latter allows us to estimate the uncertainties in our flux calibration outside the nominal wavelength range of the G395M/F290LP setup. We estimate an integrated H$\beta$ flux of $\left(3.69 \pm 0.23\right) \times 10^{-18} \rm erg \ s^{-1} \ cm^{-2}$ from the full-aperture spectrum (shown in \autoref{fig:fig_spectrum_b}). \citet{hsiao+2023b} presented MIRI/MRS IFU observations of the MACS0647-JD1 lensed image of our target but did not detect H$\beta$ directly. The authors estimated an H$\beta$ flux of $\left(3.29 \pm 0.37\right) \times 10^{-18} \ \rm erg \ s^{-1} \ cm^{-2}$ by extrapolating from the detected H$\alpha$ flux of $\left(9\pm 1 \right)\times 10^{-18} \rm \ erg \ s^{-1} \ cm^{-2}$ and assuming a case-B ratio of $H\alpha/H\beta\sim2.74$ at $T_{\rm e}=20 000\, \rm K$. While it is possible that a small amount of dust may be present, its effect on the inferred H$\beta$ flux would be much smaller than the error bars on the flux measurement itself (as we argue in more detail in Section \ref{sec:dust}). We therefore conclude that, within the associated uncertainties, the H$\beta$ flux inferred by \citet{hsiao+2023b} is consistent with the full-aperture H$\beta$ flux in our datacube. This suggests that the flux calibration uncertainties in the extended spectral region of the datacube are less than $\sim 15\%$.

\begin{figure}
{\phantomsubcaption \label{fig:fig4panels_a}
\phantomsubcaption \label{fig:fig4panels_b}
\phantomsubcaption \label{fig:fig4panels_c}
\phantomsubcaption \label{fig:fig4panels_d}}
	\includegraphics[width=0.99\columnwidth]{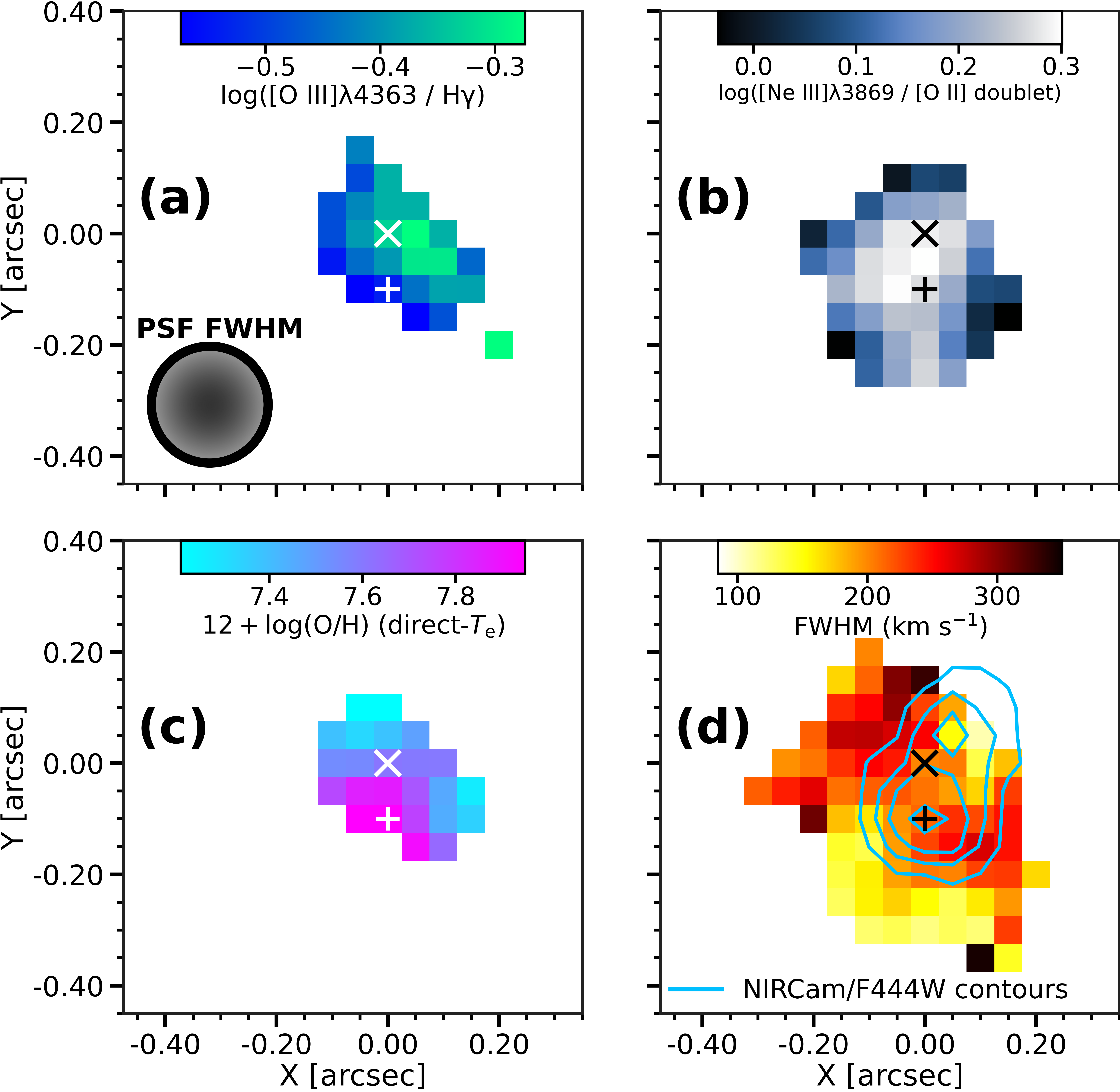}
    \caption{{\bf Top panels:} maps of the diagnostic ratios O3Hg and Ne3O2 (allowing only spaxels with $\rm S/N>3$ for each emission line involved). {\bf The bottom panels} show the spaxel-by-spaxel maps of gas-phase metallicity (\autoref{fig:fig4panels_c}; showing only the spaxels with $\rm S/N>3$ for $\left[\rm O \ \textsc{iii}\right]\lambda4363$, H$\gamma$ and inferred $\left[\rm O \ \textsc{iii}\right]\lambda5007$) and the intrinsic, deconvolved gas FWHM (\autoref{fig:fig4panels_d}; we only show spaxels with a $\rm S/N>5$ detection of H$\gamma$). The FWHM map shows two high-dispersion regions which coincide with the locations of the metal-poor gas. The `$\times$' and `+' crosses mark the H$\gamma$ emission-line and the F444W stellar continuum centroids, respectively (as in \autoref{fig:astrometry.combined}). We also show the PSF size on the top left panel for comparison.
	}\label{fig:maps}
\end{figure}

Using the same methods described above, we obtained the line ratios maps for $\rm O3Hg \equiv \log \left( \left[\rm O \ \textsc{iii}\right]\lambda4363/H\gamma\right)$ and
$\rm Ne3O2 \equiv \log \left( \left[\rm Ne \ \textsc{iii}\right]\lambda3869 / \left[\rm O \ \textsc{ii}\right]\lambda\lambda3726,3729\right)$. 
These are displayed in \autoref{fig:fig4panels_a} and \autoref{fig:fig4panels_b}, respectively, together with the gas FWHM map in \autoref{fig:fig4panels_d}.

\subsection{Combined SED fitting of photometry and spectroscopy}
\label{sec:sed}

To study the link between the spatially resolved ISM properties and star formation histories, we measure the stellar-population properties of the MACS0647-JD system using \textsc{prospector}, a Bayesian SED modelling tool \citep{johnson+21}, built around the Flexible Stellar Population Synthesis (\textsc{fsps}; \citealt{fsps1,fsps2}) tool. We fit the NIRCam photometry and NIRSpec/G395M spectroscopy of MACS0647-JD1 simultaneously for all spaxels in the cutout described in Section~\ref{sec:emission_lines}. We use the \textit{MILES} stellar spectral library \citep{sanchezblazquez2006,falcon2011} and the \textit{MIST} isochrones \citep{choi2016mist}. Our stellar mass prior is informed by the observed stellar mass function \citep{wang_b2023,qiao_newest}, and the composite stellar populations are constructed as in the \textsc{prospector}-$\beta$ model \citep{wang_b2023}. The stellar metallicity follows a truncated Gaussian prior, $\log \left(Z_{\ast}/Z_{\odot}\right)\sim\mathcal{N}\left(-1.0;\ 0.5\right)$, truncated to $\left[-2.0;\ 0.0\right]$. We adopt a flexible non-parametric SFH continuity prior \citep{leja+2019} with eight age bins. Nebular emission is modelled using \texttt{Cloudy} (v13.03; \citealt{ferland+2013, ferland+2017}) as implemented in \texttt{FSPS} \citep{byler2017_fsps}. Dust attenuation is modelled using the two-component model of \citet{charlot_fall+2000}, with a diffuse dust component optical depth $\tau_{2} \sim \mathcal{N}\left(0.3;\ 1\right)$, truncated to $\left[0;\ 4\right]$, and a birth-cloud-to-diffuse dust ratio $\tau_{1}/\tau_{2} \sim \mathcal{N}\left(1;\ 0.3\right)$, truncated to $\left[0;\ 2\right]$. The attenuation curve slope $n_{\rm dust}$ is drawn from a uniform prior over $\left[-1.2;\ 0.4\right]$ \citep{noll2009}. We do not include dust emission in our model, as it is negligible at the rest-frame wavelengths $\lesssim \SI{5000}{\angstrom}$ probed here. We also account for absorption by the intergalactic medium following the prescription of \citet{inoue2014}. To account for flux calibration and slit-loss effects, we model the spectral response using an eighth-order polynomial distortion that allows the observed spectrum to match the continuum shape predicted by the SED model. We additionally include a single jitter parameter that rescales the spectral uncertainties, capturing residual calibration systematics and ensuring a statistically consistent fit. Overall, the model configuration and prior choices are motivated by a broad range of previous studies of high-redshift galaxies \citep[e.g.,][]{leja+2019,tacchella2022_apj1,tacchella2022_apj2,tacchella+2023,duan2024_addingvalue,Harvey2025}.

We present our results in Section \ref{sec:sed_stellar}, where we discuss the implications of the spatial distributions of stellar mass and star formation burstiness, defined as $\rm SFR_{10}/SFR_{100}$, the ratio of the star formation rates averaged over lookback timescales of 10 and 100 Myr. We report the star formation histories and the corner plots showing the posterior distributions and the correlations between various SED parameters for the two main stellar components (\autoref{fig:prospector_full_big} and \autoref{fig:prospector_full_small} in Appendix \ref{sec:prospector_results_appendix}).

\section{Results}\label{sec:results_general}

\subsection{Identification of AGN dominated spaxels}
\label{sec:agn_identification}

\begin{figure*}
\centering
  \includegraphics[width=\linewidth]{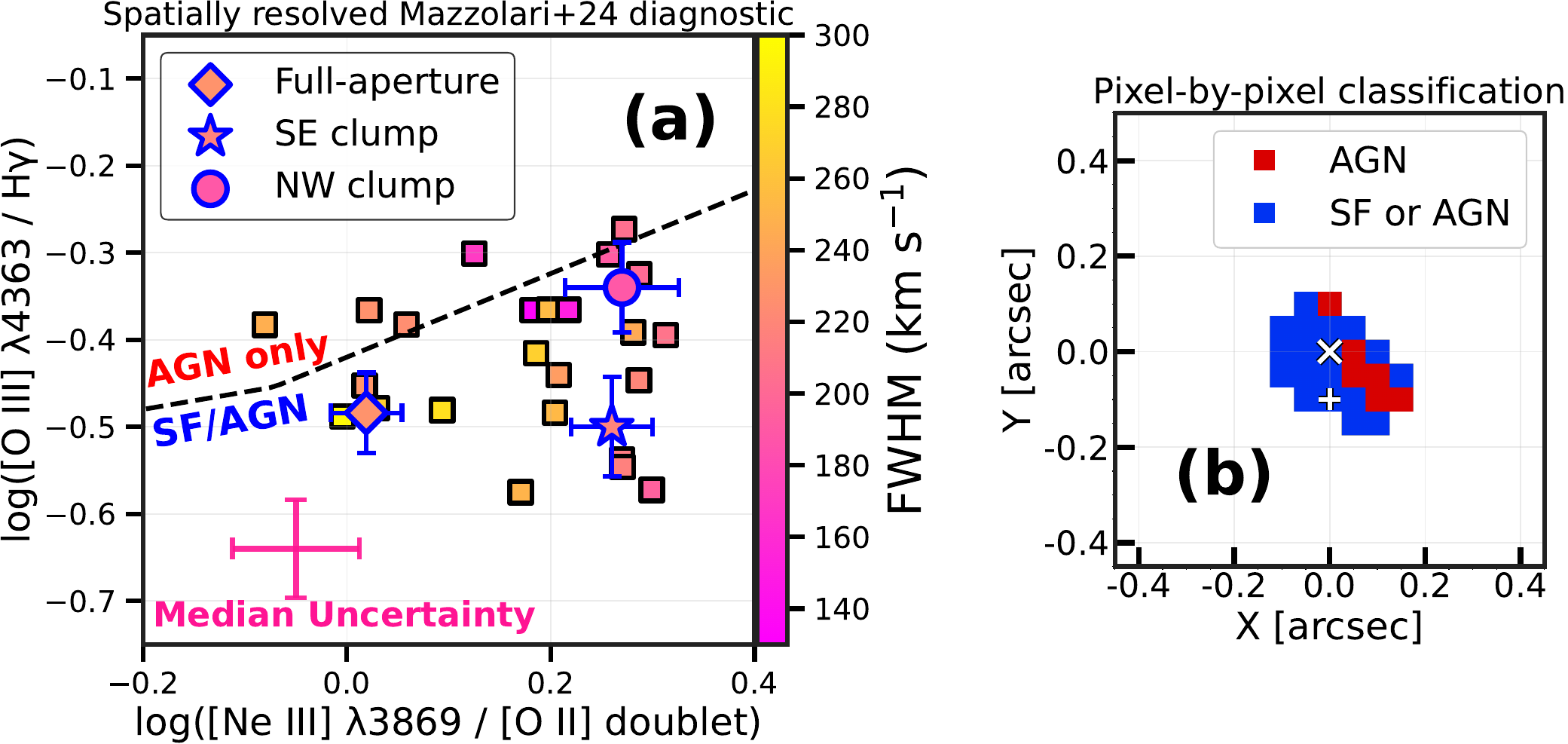}
  {\phantomsubcaption\label{fig:fig_mazzolari_a}
  \phantomsubcaption\label{fig:fig_mazzolari_b}}
  
  \caption{\autoref{fig:fig_mazzolari_a} illustrates the position of individual spaxels on the O3Hg vs Ne3O2 diagnostic diagram from \citet{mazzolari+24}. The dashed demarcation line separates regimes that are fully dominated by AGN photoionisation (AGN only) from regions in which both AGN and/or star formation may have similar contributions. Squares are individual spaxels, colour coded by the FWHM from the single Gaussian component fit. We only show the pixels in which $\rm S/N>5$ for H$\gamma$ and $\left[\rm Ne \ \textsc{iii}\right]\lambda3869$ and $\rm S/N>3$ for $\left[\rm O \ \textsc{iii}\right]\lambda4363$ and for $\left[\rm O \ \textsc{ii}\right]$. The pink error-bars are the median uncertainties from spaxels measurements. The ratios from the full-aperture, NW and SE clump spectra lie below the demarcation line. On the spatially resolved map (\autoref{fig:fig_mazzolari_b}), we identify the spaxels with either AGN or SF/AGN classification.} 
  \label{fig:mazzolari_main}
\end{figure*}

To constrain the dominant ISM excitation mechanism, we use the star-forming/AGN diagnostic diagram from \citet{mazzolari+24}, which can classify spectra as AGN-only, or consistent with both AGN and star formation. These diagnostics are commonly used in studies of high-redshift AGN (e.g., \citealt{mazzolari+25,geris+25_u,ivey+26,juodzbalis+25u}). We focus on the emission line ratios O3Hg and Ne3O2. 
According to equation (5) of \citet{mazzolari+24}, the `AGN-only' scenario is disfavoured for the integrated spectrum of MACS0647-JD1 and the SE clump, but it is consistent within the $1\sigma$ uncertainties for the NW clump. A number of spaxels in the south-west region are identified as AGN dominated, lying above the demarcation line from \autoref{fig:fig_mazzolari_b}, and others lie below the demarcation line, but still close to it (\autoref{fig:fig_mazzolari_a}). Given the large error bars and the predominant location of `AGN-only' spaxels away from the centre, we disfavour an AGN interpretation. 

Supporting this interpretation, no spaxel is identified as purely dominated by AGN if we instead use the $\rm O3Hg- O32$ or $\rm O3Hg - O33$ ($\equiv \log \left(\left[\rm O \ \textsc{iii}\right]\lambda5007\right/\left[\rm O \ \textsc{iii}\right]\lambda4363$) diagnostics (as shown in \autoref{fig:mazzolari_more} in Appendix \ref{sec:appendix_spatial_maps_O33_O32}). However, in this case, the uncertainties are larger than for the previous diagram because: {\it i)} both the O32 and O33 ratios require a reddening correction, and {\it ii)} our spatially resolved $\left[\rm O \ \textsc{iii}\right]\lambda5007$ fluxes are obtained using the extrapolation method described in Section \ref{sec:full_spectrum}. In summary, we disfavour, although do not completely rule out, an AGN interpretation.

\subsection{Direct-method gas-phase metallicities}
\label{sec:full_spectrum}

\begin{table}
\centering
\caption{{\bf Physical parameters of the MACS0647-JD1 system and of its main stellar components.} The top rows in this table present our results for some properties derived directly from large integrated aperture spectra and from the spectra corresponding to the SE and NW clumps. The bottom rows, on the other hand, give the values of various parameters resulting from spectro-photometric SED modelling (see Section \ref{sec:sed} and Section \ref{sec:sed_stellar}).}

\begin{tabular}{ c c c c }
\hline
Parameter & Full Aperture & SE Clump & NW Clump  \\ 
\hline
gas FWHM ($\rm km \ s^{-1}$)  & $228 \pm 19 $ & $215\pm34$ & $189 \pm 41$ \\
$12 + \log\left(\rm O/H\right)$ (dir) & $7.59 \pm 0.17$ & $7.89 \pm 0.16$ & $7.47 \pm0.20$\\
$12 + \log\left(\rm O/H\right)$ (str) & $7.66 \pm 0.14$ & $7.71 \pm 0.13$ & $7.49 \pm 0.15$\\
$T_{\rm e} \left[\rm O \ \textsc{iii}\right] /10^{4} \ \rm K$ & $1.90 \pm 0.37$ & $1.52 \pm 0.29$ & $2.38^{+0.32}_{-0.40}$ \\
$\log U$ & $-2.05 \pm 0.07$ & $-1.84 \pm 0.06$ & $-1.84 \pm 0.09 $ \\
$\rm SFR_{H\gamma}$ ($\rm M_{\odot} \ \rm yr^{-1}$) & $5.5 \pm 1.0$ & $1.0 \pm 0.2$ & $0.50 \pm 0.15$ \\
\hline

$\rm SFR_{100}$ ($\rm M_{\odot} \ yr^{-1}$) & $2.9 \pm 0.4$ & $0.56\pm 0.10$ & $0.24 \pm 0.05$ \\
$\log\left(M_{\ast}/M_{\odot}\right)$ & $8.38 \pm 0.21$ & $7.78 \pm 0.07$ & $7.41 \pm 0.08$ \\
$A_{\rm V,cont}$ (\textsc{Prospector}) & $0.26 \pm 0.15$ & $0.13 \pm 0.05$ & $0.18 \pm 0.07$ \\

\hline

\end{tabular}

\label{tab:table_parameters_integrated}
\end{table}

To calculate the electron temperatures from the spectra of individual spaxels (which are required to infer the direct-$T_{\rm e}$ metallicities), we need to estimate the $\left[\rm O \ \textsc{iii}\right]\lambda5007$ fluxes. We use equation (1) from \citet{joris+2021}, which connects the observed $\left[\rm O \ \textsc{iii}\right]\lambda5007 \, /\, \left[\rm O \ \textsc{ii}\right]\lambda\lambda3726,3729$ and $\left[\rm Ne \ \textsc{iii}\right]\lambda3869 \, /\, \left[\rm O \ \textsc{ii}\right]\lambda\lambda3726,3729$ ratios for star forming galaxies: 

\begin{equation}
\rm \log \left(\frac{\left[\rm O \, \textsc{iii}\right]\lambda5007}{\left[\rm O \, \textsc{ii}\right]\lambda\lambda3726,3729}\right) = 0.905  \log \left(\frac{\left[\rm Ne \, \textsc{iii}\right]\lambda3869} { \left[\rm O \, \textsc{ii}\right]\lambda\lambda3726,3729}\right) +1.078 .
\label{eq:joris}
\end{equation}

\noindent In this paragraph, we discuss possible caveats. This empirical conversion relation has a mean scatter of 0.22 dex, calculated based on the full sample of {\it SDSS} SFGs \citep{sdss}. However, the subset of galaxies with a higher ionisation parameters ($ \log \left( \left[\rm Ne \ \textsc{iii}\right]\lambda3869 / \left[\rm O \ \textsc{ii}\right]\lambda\lambda3726,3729\right)>-1$; a regime that is more common in the case of high-$z$, low metallicity galaxies) have a scatter of 0.12 dex around the \citet{joris+2021} equation. As a conservative approach, we propagate this source of uncertainty by adding 0.22 dex in quadrature to the numerical uncertainties associated with the $\left[\rm O \, \textsc{iii}\right]\lambda5007$ fluxes (for integrated-aperture or individual spaxels spectra). Another possible caveat is that Equation \ref{eq:joris} is mainly valid for SF-dominated (rather than AGN-dominated) spaxels. However, we notice (based on \autoref{fig:fig_mazzolari_b} and \autoref{fig:fig4panels_b}) that the AGN-dominated spaxels all have $\rm Ne3O2>0.08$, a regime in which the $\left[\rm O \, \textsc{iii}\right]\lambda 5007 \, / \, \left[\rm O \, \textsc{ii}\right]\lambda\lambda3726,3729 - \left[\rm Ne \, \textsc{iii}\right]\lambda 3869 \, / \, \left[\rm O \, \textsc{ii}\right]\lambda\lambda3726,3729$ empirical conversion relations for SFGs and Seyfert-1 galaxies are offset by $\lesssim0.03 \, \rm dex$ (Figure 4 of \citealt{joris+2021}). This source of systematic uncertainty is therefore negligible compared to, for example, the mean scatter of the empirical relation itself. We can therefore reasonably adopt the $\left[\rm O \, \textsc{iii}\right]\lambda5007$ fluxes predicted by equation \ref{eq:joris}. This method of estimating the flux of $\left[\rm O \, \textsc{iii}\right]\lambda 5007$ has also been used by several studies of the physical properties of the ISM (including the metallicity and ionisation parameter) in $z>10$ galaxies \citep{bunker_gnz11,jan_carbon_z112u}. Although the empirical equation \ref{eq:joris} was derived using $z\sim0$ galaxies from \textit{SDSS}, such a correlation between $\left[\rm O \ \textsc{iii}\right]\lambda5007\,/\,\left[\rm O \ \textsc{ii}\right]\lambda3726,3729$ and $\left[\rm Ne \ \textsc{iii}\right]\lambda3869\,/\,\left[\rm O \ \textsc{ii}\right]\lambda3726,3729$ has recently been shown to exist at least up to $z\sim 9.5$, as illustrated in figure 6 of \citet{zamora+25b}, based on a compilation of $z>5$ \textit{JWST} observations \citep{arellano-cordova+22,cameron+23,curti+2023,sanders+23a,heintz+24,schaerer+24,topping+24,curti+25a}. 
For the integrated-aperture spectrum of MACS0647-JD1, the inferred $\left[\rm O \ \textsc{iii}\right]\lambda5007$ line flux is $\left(2.24 \pm 0.52 \right) \times 10^{-17} \ \rm erg \ s^{-1} \ cm^{-2}$, consistent with the value obtained from MIRI medium-resolution spectroscopy of this target \citep{hsiao+2023b}. This result demonstrates the potential of the method proposed by \citet{joris+2021} to extrapolate $\left[\rm O \ \textsc{iii} \right]\lambda5007$ based on $\left[\rm Ne \ \textsc{iii}\right]\lambda 3869$ at $z>10$, where $\left[\rm O \ \textsc{iii} \right]\lambda5007$ falls outside the wavelength range of NIRSpec. 





\begin{figure}
{\phantomsubcaption{\label{fig:te_map_a}
\phantomsubcaption{\label{fig:te_map_b}}}}
\centering
  \includegraphics[width=\linewidth]{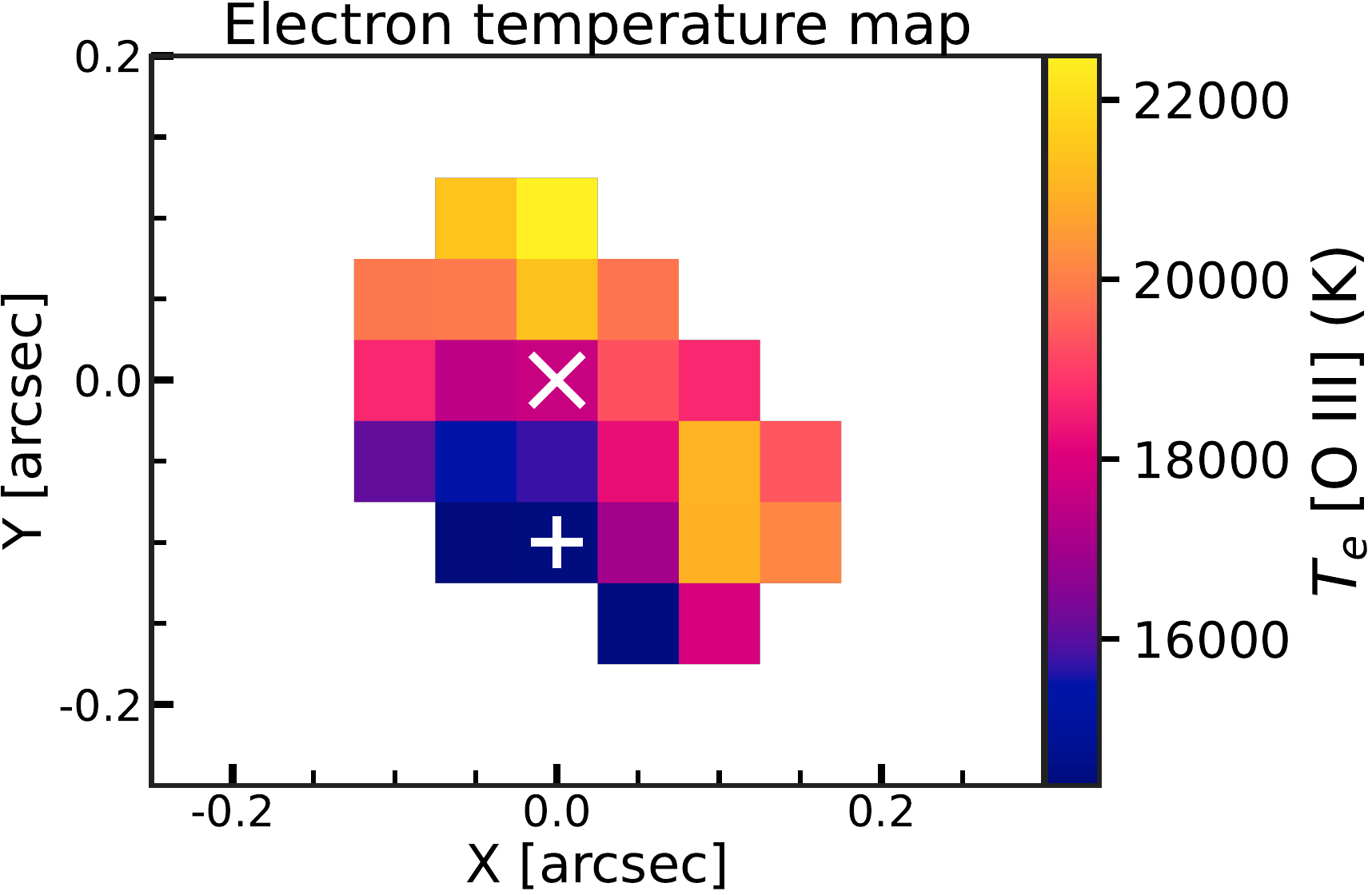}
  \caption{
  This figure shows the map of the electron temperature of the $\left[\rm O \ \textsc{iii}\right]$ emitting gas in the central region of the system. We only show spaxels with $\rm S/N>3$ for the auroral $\left[\rm O \ \textsc{iii}\right]\lambda4363$ line. The white `+' and `$\times$' crosses mark the centroids of stellar continuum and H$\gamma$ line emission, respectively.}
  \label{fig:te_map}
\end{figure}

The electron temperature is calculated using $\left[\rm O \  \textsc{iii} \right]\lambda4363/\left[\rm O \  \textsc{iii} \right]\lambda5007$. 
We first assume that dust attenuation is negligible ($A_{\rm V,cont}\lesssim 0.1 \ \rm mag$; \citealt{hsiao_nirspec}). For the full aperture spectrum, \textsc{PyNeb} \citep{pyneb} predicts a mean temperature of the $\left[\rm O \ \textsc{iii}\right]$-emitting gas $T_\mathrm{e} \left[\rm O \ \textsc{iii}\right] =\left(1.9 \pm 0.4\right)\times10^{4}  \ \rm K$ for an auroral-to-strong line ratio of $0.031 \pm 0.009$, assuming an electron density of $2000 \, \rm cm^{-3}$ (as determined by \citealt{abdurrouf+2024}, using high-resolution spectroscopic data for our target). We also tested a wide range of electron densities $100 \ \rm cm^{-3}<n_\mathrm{e}<10000 \ \rm cm^{-3}$ which extends beyond the typical range of $\left[\rm O \, \textsc{ii}\right]\lambda\lambda3726,3729$
based electron densities in high-redshift galaxies \citep{isobe+23,marconcini+2024,jan_cos3018} and found that the dependence of our $T_\mathrm{e} \left[\rm O \ \textsc{iii}\right]$ estimate on the assumed $n_\mathrm{e}$ is negligible. 
A similar procedure is applied to compute the electron temperatures from the spectra of individual spaxels, and our results are displayed in \autoref{fig:te_map_b}. 
Once we calculate our $T_{\rm e} \left[\rm O \ \textsc{iii}\right]$ values, we can determine the electron temperature of the $\left[\rm O \ \textsc{ii}\right]$ emitting gas using equation (14) from \citet{izotov+2006}. 
We then compute the gas-phase metallicities (as given by the oxygen abundances). 
\citet{dors+2020} showed that the contribution of O$^{3+}$ is negligible even in highly AGN-dominated regimes. Thus, we can assume all oxygen atoms are either singly or doubly ionised in these conditions. 
We use $\left[\rm O \ \textsc{ii}\right]\lambda\lambda3726,3729/H\beta$ and $\left[\rm O \ \textsc{iii}\right]\lambda 5007/H\beta$ to compute the abundances of the O$^{+}$ and O$^{++}$ ions using \textsc{PyNeb}. For the aperture spectrum, we obtain $12+\log\left(\rm O/H\right)=7.59 \pm0.17$, in agreement with the result of \citet{hsiao+2023b} and similar to the direct-$T_{\rm e}$ metallicities reported for a number of other SFGs at $z\sim 4-9$ \citep[e.g.,][]{curti+2023,heintz+23b,nakajima+23,morishita+24}. To further test the consistency of our method, (given the indirect approach used to extract H$\beta$ fluxes), we repeat our calculation but using the ratios $\left[\rm O \ \textsc{ii}\right]\lambda\lambda3726,3729/H\gamma$ and $\left[\rm O \ \textsc{iii}\right]\lambda 5007/H\gamma$ (instead of their counterparts involving H$\beta$). We obtain $12 + \log\left(\rm O/H\right)=7.52 \pm 0.18$ in agreement with our previous value, and acting as a cross-validation of both the method and the flux calibration. For individual spaxels, we additionally perform the dust attenuation corrections described in Section \ref{sec:dust}. The resulting metallicity map is shown in \autoref{fig:fig4panels_c}.

To assess the impact of dust reddening on our calculated parameters, we adopted a stellar continuum dust attenuation following the Calzetti law \citep{Calzetti2000} with $A_{\rm V,cont}=1\ \rm mag$ for the full-aperture spectrum (although this value is unreasonably high from a physical point of view; \citealt{ferrara+25}). This would imply a gas-phase metallicity of $12+\log\left(\rm O/H\right)=7.71$, which is within the quoted uncertainties of our $A_{\rm V,cont}=0$ estimate. For an assumed $A_{\rm V,cont}=2 \ \rm mag$, we obtain $12+\log\left(\rm O/H\right)=7.90$. Thus, as a first-order approximation, the change in calculated metallicity and the change in assumed dust attenuation are related by $\Delta \log \left(\rm O/H\right)/\Delta A_{\rm V,cont} \approx0.18 \ \rm dex \ mag^{-1}$.



\subsection{Dust attenuation corrections}
\label{sec:dust}

Various recent studies use the Balmer decrement ratios to compute the nebular dust attenuation for individual spaxels \citep{venturi+24,ivey+26}. In our case, thanks to our custom H$\beta$ extraction procedure (Section \ref{sec:hbeta_recovery}), the lowest order Balmer lines are H$\beta$ and H$\gamma$. We use \textsc{PyNeb} to compute the theoretical case-B Balmer ratio $\left(H\beta/H\gamma\right)_{0}= 2.105$ (assuming $T_{\rm e}= 19000 \ \rm K$ as derived in Section \ref{sec:full_spectrum}). The stellar continuum dust attenuation is: 


\begin{equation}
    A_{\rm V,cont}= -2.5 \times0.44 \times \log \left(\phi_{\rm obs}/\phi_{0}\right) \times R_{V}/\left(\rm k_{\lambda,H\beta}-k_{\lambda,H\gamma} \right)\ ,
\label{eq:dust_correction}
\end{equation}


\noindent where the factor of 0.44 comes from the relation between stellar and nebular dust attenuation \citep{Calzetti2000}, and we define $\phi \equiv H\beta/H\gamma$. We assume $R_{V}=4.05$ \citep{Calzetti2000}. For the dust attenuation law, we assume a slightly modified version of the Calzetti law that is suitable for low-mass, low-metallicity galaxies \citep{shivaei+20}. We find that, for all spaxels considered in computing the direct-method metallicity map, except for three outliers, $A_{\rm V,cont}\lesssim0.3 \ \rm mag$ (and consistent with 0 within the uncertainties) whereas the highest outlier has $A_{\rm V,cont} \approx0.6 \ \rm mag$. According to our previously calculated $\Delta \log \left(\rm O/H\right)/\Delta A_{\rm V,cont}\approx 0.18 \ \rm \ dex \ mag^{-1}$ (Section \ref{sec:full_spectrum}), this implies that the effect of dust attenuation correction on the derived spaxels metallicities is less than 0.11 dex. We will therefore, as a first-order approximation, correct the line ratios from spaxels spectra assuming the Calzetti dust attenuation curve with the $A_{\rm V,cont}$ derived using this method.

We now separately discuss individual spaxels with $H\beta/H\gamma < \left(H\beta/H\gamma\right)_{0}$ (which cannot be explained by dust attenuation assuming case B recombination). In all but four cases of spaxels with below case-B $H\beta/H\gamma$ decrement ratios and $\rm S/N>3$ for H$\beta$, H$\gamma$ and H$\delta$, we notice that the $H\gamma/H\delta$ remains above the corresponding case-B Balmer ratio. In principle, stronger stellar absorption in H$\beta$ could be responsible for this behaviour \citep{groves+2012,tacchella+2023}; while our \textsc{ppxf} continuum modelling allows for stellar absorption (as described in Section \ref{sec:emission_lines}), the weakness of the continuum of our $z\sim10.17$ target implies that any stellar absorption has a negligible impact on the measured emission line fluxes. 
Moreover, potential calibration uncertainties in the extrapolated wavelength range ($\lambda_{\rm obs}>5.27 \mu\rm m$) could also contribute to artificially low H$\beta$ fluxes. Because of the wavelength proximity of H$\gamma$ and H$\delta$, this Balmer decrement ratio is not suitable for accurately calculating the dust attenuation, especially in low-$\rm S/N$ targets. As a result, we assume that $A_{\rm V,cont}=0 \ \rm mag$ for these spaxels.

There are only four spaxels for which both $H\beta/H\gamma$ and $H\gamma/H\delta$ are more than 1$\sigma$ below their expected case-B values and all are located west of the south-east clump, in the region where we expect a non-negligible contribution from AGN (as discussed in Section \ref{sec:agn_identification}). Indeed, below case-B Balmer ratios, including $H\beta/H\gamma$ and $H\alpha/H\beta$, have been reported for stacked Type-2 AGN \citep{jan_t2}. Case-B assumptions may not be fulfilled under optically thin or non-equilibrium conditions \citep{scarlata+24}. While such conditions typically require high electron temperatures and/or low electron densities, which are not indicated by our measurements, we cannot exclude localised departures from Case-B assumptions in these regions. We adopt a similar approach of assuming $A_{\rm V,cont}=0 \ \rm mag$ for these spaxels. We expect this assumption to have a negligible impact on the derived gas-phase metallicities, given the findings of \citet{hsiao_nirspec}, who report $A_{\rm V,cont} \lesssim 0.3 \ \rm mag$ for the integrated aperture spectrum (covering the south-east clump within the slit).

As an additional sanity check, we estimate $A_{\rm V,cont}$ based on our combined spaxel-by-spaxel \textsc{prospector} SED fits (described in Section \ref{sec:sed}). Taking into account that both diffuse dust (optical depth $\tau_{2}$) and dust from the birth clouds (optical depth $\tau_{1}$) contribute to the total dust attenuation, we approximate $A_{\rm V,cont} \approx 2.5 \times 0.44 \times \log_{10} \left(e\right) \times \left(\tau_{1}+\tau_{2}\right)$. In reality, this formula gives an upper limit on $A_{\rm V,cont}$, with the actual value depending on the exact fraction of light coming from young stars ($<10 \ \rm Myr$) relative to the total. Constraining this parameter is beyond the scope of this paper. From this calculation, we find that, in the case of spaxels with $\rm S/N_{\left[\rm O \, \textsc{iii}\right]\lambda4363} >3$, the highest $A_{\rm V,cont}$ is $\sim0.5 \ \rm mag$, although most spaxels have $A_{\rm V,cont} \sim 0.0-0.4 \ \rm mag$. The values obtained for the two clumps and for the full cutout aperture are reported in \autoref{tab:table_parameters_integrated}. Overall, the results obtained with these methods are in agreement, illustrating the difficulty of providing accurate dust corrections on spatially resolved scales \citep[as also mentioned by][]{waz_arc}. However, we demonstrate that, regardless of the dust correction method implemented, the effect on the inferred metallicities is expected to be smaller than the measurement uncertainties. 


\subsection{Gas-phase metallicities (strong-line method)}
\label{sec:gas_strong_line}

\begin{figure}
\centering

  \includegraphics[width=\linewidth]{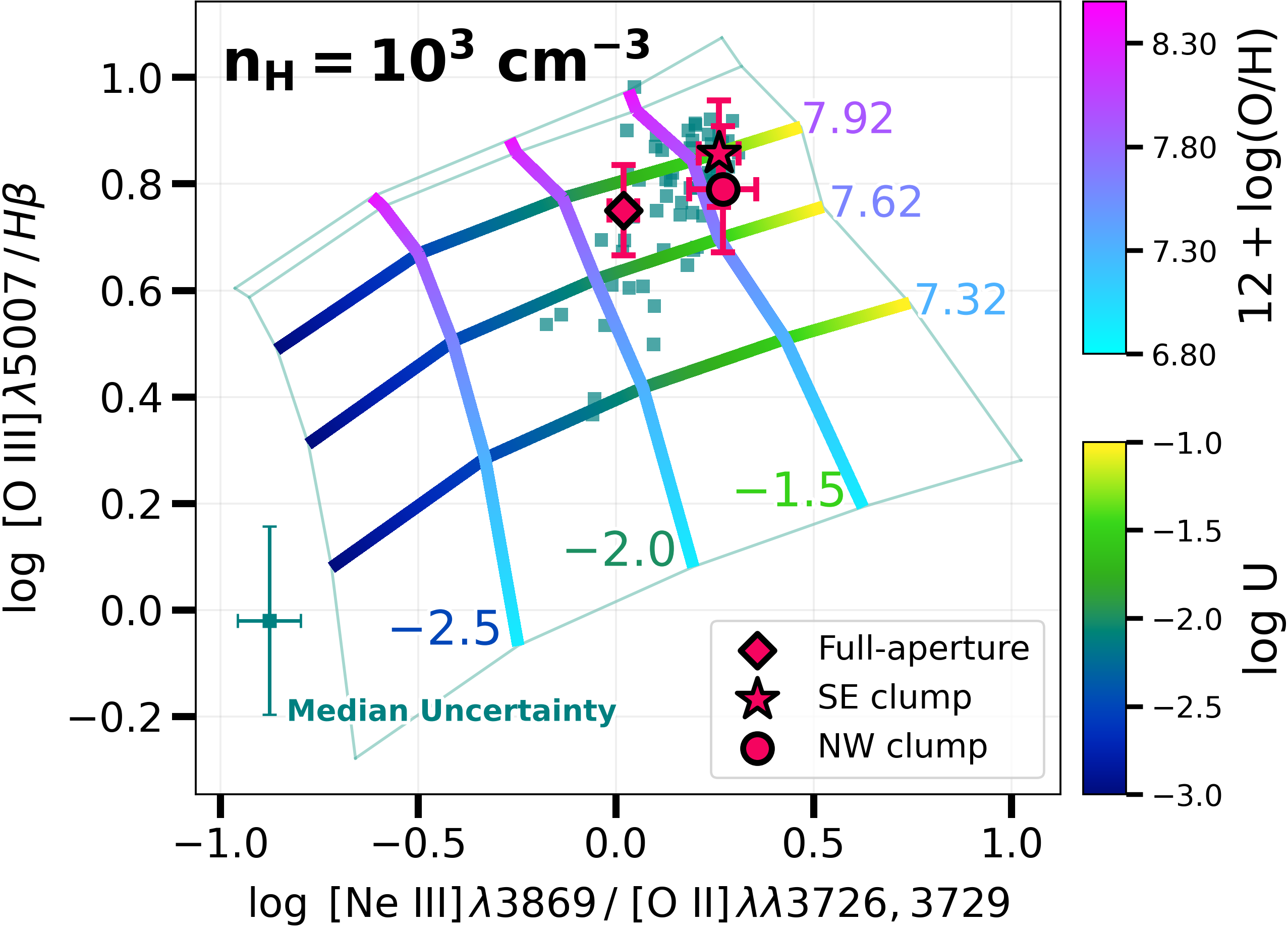}
  
  \caption{This figure shows the R3-Ne3O2 photoionisation grid (with varying metallicity and ionisation parameter) of SFGs from \citet{gutkin+2016} based on \textit{CLOUDY} models \citep{ferland+2013,ferland+2017}. The choice of the grid is motivated in the main text of Section \ref{sec:gas_strong_line}. Individual spaxels (with $\rm S/N>5$ on $\left[\rm Ne \, \textsc{iii}\right]\lambda 3869$, $\left[\rm O \, \textsc{ii}\right]\lambda3726,3729$ and H$\beta$ lines) are plotted as dark-cyan small squares and their median uncertainties are shown in the bottom left part. Blue symbols denote the results from the full-aperture and the two main SE and NW components. These two line ratios also offer the advantage that they do not require dust reddening correction.} 
  \label{fig:strong_another}
\end{figure}

The strong nebular emission lines provide an alternative method for estimating gas-phase metallicities in high-redshift SFGs (e.g., \citealt{nakajima_empress5,curti+24,cataldi+25_u,scholte+25,sanders_newest_u}). These calibrations are valid only for SFGs and link strong emission-line ratios to metallicities based on the more reliable and more general `direct' method. However, it is important to note that these studies use integrated aperture spectra, so they do not account for spatial variations in the ISM conditions and/or systematic physical effects (e.g. localised shocks, AGN) that can affect the metallicity measurements on resolved scales. In this work, we choose to use the robust SFG $\rm R3 \ \rm vs \ Ne3O2$ photoionisation grid from \citet{gutkin+2016}, as shown in \autoref{fig:strong_another}. 
We use $\rm R3 \equiv \log \left(\left[\rm O \ \textsc{iii}\right]\lambda5007/H\beta\right)$ as a primary tracer of metallicity, with 
$\left[\rm O \ \textsc{iii}\right]\lambda5007$ fluxes calculated based on the \citet{joris+2021} equation (equation \ref{eq:joris}). To break the degeneracy with the ionisation parameter, we additionally use Ne3O2 \citep{maiolino_manucci+2019}. This photoionisation grid method has the further advantage that both line ratios are insensitive to dust attenuation, since all the emission lines involved in our calculations have similar wavelengths.


\citet{hsiao_carbon_u} measure a carbon abundance of $\log\left(\rm C/O\right)=-0.44\pm0.07$ for MACS0647-JD1. Given the value of $\left(\rm C/O\right)_{\odot}=0.44$ assumed by \citet{gutkin+2016}, we select the grid with a carbon abundance of $\rm \left(\rm C/O\right)= 0.72 \left(\rm C/O\right)_{\odot}$ which is the closest available value to  $0.82^{+0.13}_{-0.09} \left(\rm C/O\right)_{\odot}$ measured by \citet{hsiao_carbon_u}. We choose an upper stellar mass cutoff for the IMF of $100 \ \rm M_{\odot}$ for our grid and a fiducial hydrogen gas density of $n_{H}=10^{3} \ \rm cm^{-3}$, similar to the average electron density reported by \citet{abdurrouf+2024} for this system. However, no significant changes in the grid positions occur even if we instead choose $n_{H}=10^{2} \ \rm or \ 10^{4} \ \rm cm^{-3}$. We assume a low dust-to-metal mass ratio of $\xi_{d}=0.1$, suitable for $z>10$ galaxies \citep{ferrara+25}. The results are shown in \autoref{fig:strong_main}, where we illustrate the positions of individual spaxels spectra on the chosen photoionisation grid. To convert from $Z_{\odot}$ to oxygen abundance, we use $12+\log\left(\rm O/H\right)=8.80+\log\left(Z/Z_{\odot}\right)$, where 8.80 is the solar oxygen abundance computed by \citet{gutkin+2016} for $\xi_{\rm d}=0.1$. We obtain an excellent agreement between the results of the strong-line and direct-$T_{\rm e}$ methods, except for the south-western region of the system. This is where line ratios show tentative evidence of photoionisation influenced by other processes (\autoref{fig:fig_mazzolari_b}), in addition to star formation (such as AGN, shocks or diffuse ionised gas; the latter two will be discussed in Sections \ref{sec:shocks} and \ref{sec:gas_discussion}, respectively). 

Indeed, previous studies have suggested that such abundance discrepancy factors (ADFs) between the two gas-phase metallicity calculation methods may be caused by ISM inhomogeneities \citep{stasinska+2007,chen_y+2023}. We also note that, using multiple emission lines and assuming a stratified ISM with multiple density zones, numerical models indicate that the contribution of the dense gas component to $\left[\rm O \, \textsc{iii}\right]\lambda4363$ flux can increase the inferred total metallicity \citep[e.g.,][]{marconi+2024}. As a result, we provide a cautionary note regarding the interpretation of strong-line metallicities in the south-western region.

\subsection{Ionisation parameter}
\label{sec:ionisation_param}

For the ionisation parameter, one standard approach is to use the diagnostic $\rm O32 \equiv \log \left[\rm O \ \textsc{iii}\right]\lambda5007/\left[\rm O \ \textsc{ii}\right]\lambda\lambda3726,3729$ as these lines trace different ionisation stages of the same atomic species \citep{diaz+2000,Kewley_2002,papovich+22}. $\left[\rm O \ \textsc{iii}\right]\lambda5007$ is unavailable in our spectra, despite our recovery of spectral data outside the nominal range of G395M grating. We therefore use $\rm Ne3O2$ which closely tracks $\rm O32$ and has the advantage that the wavelengths of these lines are close to each other, so their ratio is insensitive to dust reddening \citep{nagao+2006,perez-montero+2007,levesque_richardson+2014,maiolino_manucci+2019}. To determine the ionisation parameter, we use equation (3) from \citet[][based on the original equation by \citealt{diaz+2000} from single-star photoionisation models]{joris+2021}.

With this standard approach, we obtain $\log  U = -2.05 \pm 0.07$ for the full aperture spectrum and $\log U= -1.84 \pm 0.06$ for both the south-east and the north-west components, consistent with the result obtained by \citet{hsiao_nirspec}. However, given the non-negligible spatial variations in gas-phase metallicity between different regions of the system (notably, between the SE and the NW clumps), an alternative, more robust calculation of $\log U$ should take this effect into account. The most straightforward way to perform this calculation is to use the \citet{gutkin+2016} metallicity$-$ionisation parameter photoionisation grid (\autoref{fig:strong_another}; also described in Section \ref{sec:gas_strong_line}). By interpolating the measured positions of the data points corresponding to the full-aperture, SE, and NW clumps spectra on the $\rm R3-Ne3O2$ plane onto the grid, we obtain $\log U =-1.79 \pm 0.15$, $-1.41 \pm 0.19$, and $-1.47 \pm 0.24$ for the full-aperture, SE, and NW clumps, respectively. The values calculated using this alternative method are higher than those determined using the standard, $\rm Ne3O2$-based method that does not treat the metallicity dependence of this line ratio explicitly. This implies that, in the case of a metal-poor ISM, the usual $\rm Ne3O2-\log U$ calibrations may substantially underestimate the ionisation state. The calculated $\log U$ values are also higher than those of other star-forming galaxies at $3.5<z<10$ \citep{christensen+2012,troncoso+2014,shapley+2017,marconcini+2024,bruno+2024,hayes+2025,parlanti+25_z55}, suggesting a more intense radiation field. Our values are instead similar to other $z>10$ galaxies such as GHZ2 \citep{castellano+24,zavala+24} and GN-z11 \citep{bunker_gnz11}. On spatially resolved scales, \autoref{fig:fig4panels_b} shows a relatively uniform Ne3O2 (and, by extrapolation, $\log U$) in the central regions of MACS0647-JD1, with a radial decrease of $0.33 \pm 0.14$ dex from the centre of the system towards its outer regions, with substantially lower ionisation in the south-western region, a feature possibly indicating ionisation stratification, which we discuss in greater detail in Section \ref{sec:the_nature_lines_morph}.

\section{The nature of the MACS0647-JD system}
\label{sec:discussion_general}


\subsection{Emission lines spatial distribution}
\label{sec:the_nature_lines_morph}

\begin{figure}
\centering
  \includegraphics[width=\linewidth]{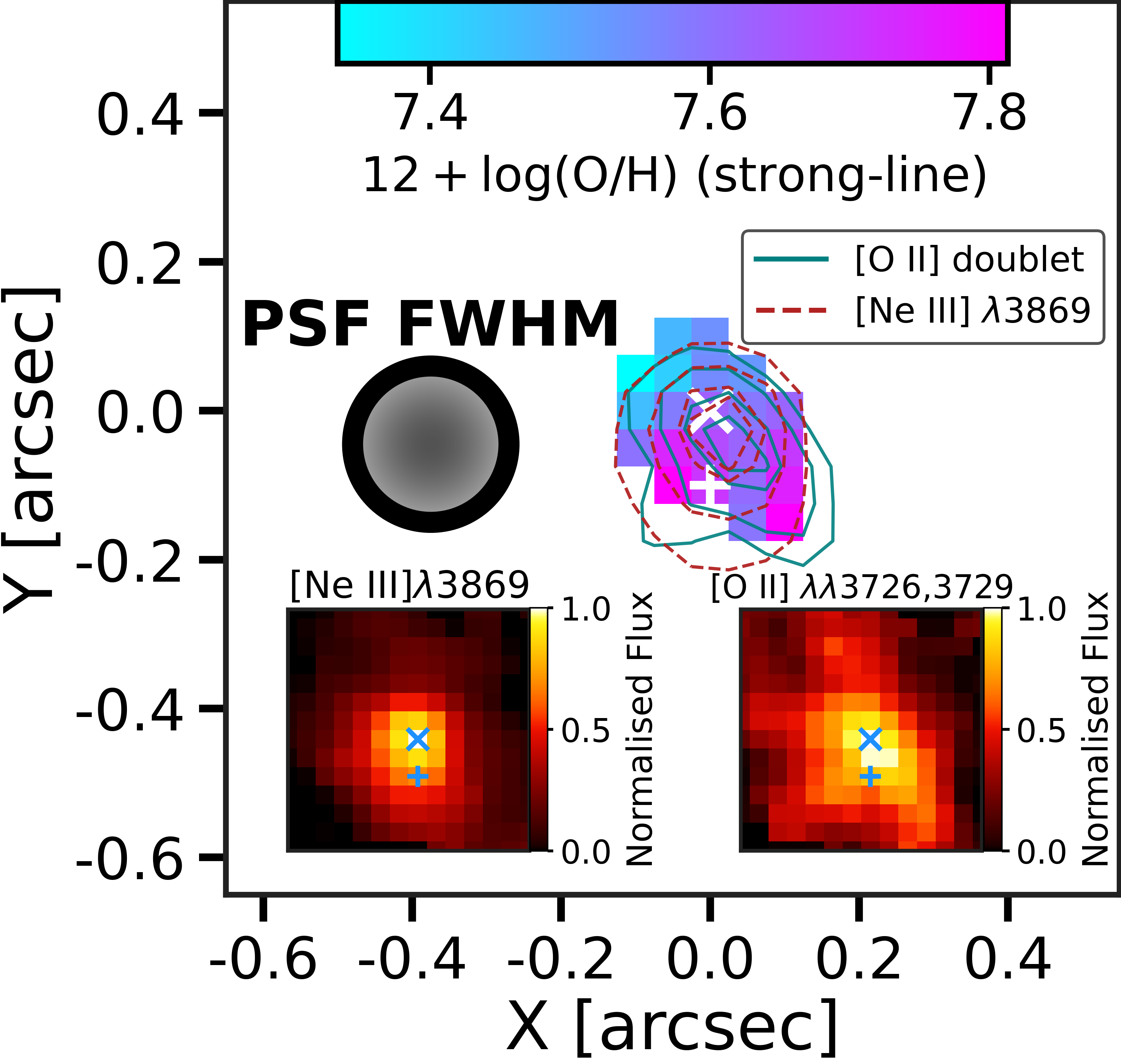}
  \caption{Metallicity map based on R3-Ne3O2 line ratios (Section \ref{sec:gas_strong_line}). Here, we only highlight the spaxels that have $\rm S/N>5$ for all lines involved and additionally $\rm S/N>3$ for $\left[\rm O \ \textsc{iii}\right]\lambda4363$ enabling a direct comparison to the direct method metallicity map (\autoref{fig:maps}). We overplot the contours of the integrated fluxes of $\left[\rm O \ \textsc{ii}\right]\lambda\lambda3726,3729$ (teal) and $\left[\rm Ne \ \textsc{iii}\right]\lambda3869$ (fire-brick). Both the contours and the inset panels show that $[\mathrm{Ne}\,\textsc{iii}]$ has a more compact distribution while $\left[\rm O \, \textsc{ii}\right]$ is more extended. The blue `$\times$' and `+' crosses in the inset panels are the centroids of peak H$\gamma$ (and $\left[\rm Ne \ \textsc{iii}\right]\lambda3869$) emission and of stellar continuum, respectively.}
  \label{fig:strong_main}
\end{figure}

To characterise the morphology of the ISM within the MACS0647-JD1 system, we first identify the brightest spaxel in the emission-line maps of H$\gamma$ and $\left[\rm Ne \ \textsc{iii}\right]\lambda3869$. This spaxel is shifted by two spaxels (corresponding to a physical, delensed distance of $\sim 150 \ \rm pc$, as shown in \autoref{fig:astrometry.combined}) relative to the centroid of the synthetic F444W image obtained by convolving the datacube with the response function of the NIRCam/F444W bandpass. Such a shift could arise from differential dust attenuation between the emission lines and the stellar continuum. However, MACS0647-JD is dust poor (as shown in Section \ref{sec:dust} and by \citealt{hsiao_nirspec}), making this possibility unlikely. Thus, the observed offset likely implies a different spatial distribution of the line-emitting gas relative to the stars. Since the instantaneous SFR directly determines the strength of the Balmer lines \citep{kennicutt+2012}, it means that the central region between the two stellar clumps (\autoref{fig:astrometry_b}) hosts the most intense recent star formation activity, while the two stellar components are dominated by older stellar populations. This could favour a merger scenario for this system, in agreement with the findings of \citet{qiao4585_u} and \citet{puskas+25_lya}, who report that, on average, high-redshift mergers trigger short-lived star formation enhancements if the components become separated by a proper distance of less than 20 kpc.

There is a striking difference between the spatial distributions of the $\left[\rm O \ \textsc{ii}\right]\lambda\lambda3726,3729$ and $\left[\rm Ne \ \textsc{iii}\right]\lambda3869$ emission. The latter is a representative tracer of the high-ionisation compact central region of the system (with very similar morphologies revealed by both Balmer recombination lines and $\left[\rm O \ \textsc{iii}\right]\lambda4363$, all of which are shown in Appendix \ref{sec:appendix_images}). In contrast, the $\left[\rm O \ \textsc{ii}\right]\lambda\lambda3726,3729$ emission is both clumpier and more spatially extended, primarily tracing the low ionisation gas in the system. Since we observe that $\log (\rm Ne3O2)>0$ even in the outskirts, the most plausible explanation is a classical, inside-out stratification of the system, although we cannot completely rule out more exotic scenarios such as the presence of diffuse ionised gas (DIG; e.g. \citealt{zhang+2017}), which typically boosts $\left[\rm O \ \textsc{ii}\right]\lambda\lambda3726,3729$ emission \citep{hunter+90,kennicutt+98} and metal lines strength \citep{garg+2024,will2024,clarendon+2025}. A recent NIRSpec/IFU study by \citet{waz_arc} tentatively identified DIG in a highly lensed $z\sim 5$ massive post-starburst galaxy. However, while $\left[\rm O \ \textsc{ii}\right]\lambda\lambda3726,3729$ emission truly appears enhanced relative to hydrogen lines in the southern and southwestern regions of our target (compared to the central regions), \citet{hsiao_carbon_u} report non-detections of $\left[\rm N \ \textsc{ii}\right]\lambda6548$ and $\left[\rm N \ \textsc{ii}\right]\lambda6583$ (these lines are expected to be prominent in regions with DIG) in the MIRI spectra. This challenges the interpretation that DIG is present in the southern-southwestern region of the MACS0647-JD1 system. Of course, at $z>10$, intrinsically low $N/O$ abundance
ratios could also explain the undetected $\left[\rm N\,\textsc{ii}\right]$ doublet in the extended region, given
that N/O-enhanced systems tend to be very compact
\citep[e.g.,][]{xihan+2025_NO}.

\subsection{Shock-driven photoionisation modelling}
\label{sec:shocks}

\begin{figure*}
{\phantomsubcaption\label{fig:shocks_a}
\phantomsubcaption\label{fig:shocks_b}
\phantomsubcaption\label{fig:shocks_c}
\phantomsubcaption\label{fig:shocks_d}
\phantomsubcaption\label{fig:shocks_e}
\phantomsubcaption\label{fig:shocks_f}}
\centering
  \includegraphics[width=\linewidth]{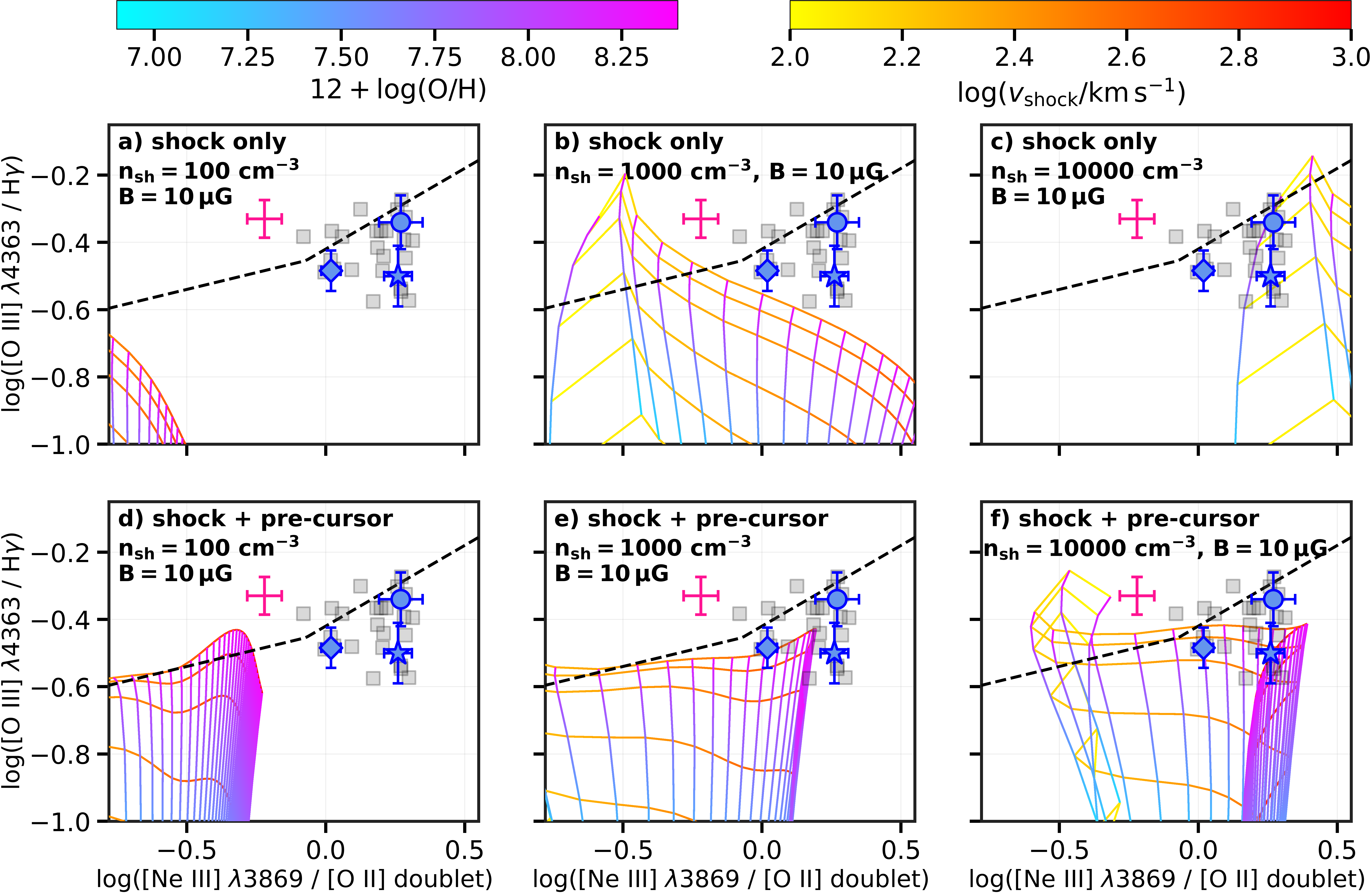}
  \caption{In this figure, we illustrate the various shock models overlaid on the \citet{mazzolari+24} diagnostic diagram, in order to compare with our data-points. Individual spaxels are marked with grey squares (with the pink error bars showing their median uncertainties), whereas the blue symbols have similar meanings to \autoref{fig:fig_mazzolari_a}: the diamond represent the full-aperture, the star is the SE clump and the circle is the NW clump. The black, dashed line is the demarcation between the AGN-dominated regime and the regions where SF and AGN have similar contributions to photoionisation (SF or AGN regime). We display a number of shock and shock $+$ precursor grids computed using \textsc{mappings v} \citep{mappings5} at fixed pre-shock density, $n_{\rm sh}$, fixed magnetic field, $\rm B= 10 \ \mu G$ and varying metallicity $7<12+\log\left(\rm O/H\right)<8.4$ and shock velocity $2<\log \left(v_{\rm shock} \, / \, \rm km \ s^{-1}\right)<3$. Our observations can only be reproduced with either high metallicities or high pre-shock densities, which, although not completely ruled out locally, are difficult to reconcile with the gas-phase metallicities (\autoref{fig:fig4panels_c}) and electron densities inferred by \citet{abdurrouf+2024} for our system.}
  \label{fig:shocks}
\end{figure*}

\begin{figure*}
\centering
{\phantomsubcaption \label{fig:correlations_a}
\phantomsubcaption \label{fig:correlations_b}
\phantomsubcaption \label{fig:correlations_c}}
  \includegraphics[width=\linewidth]{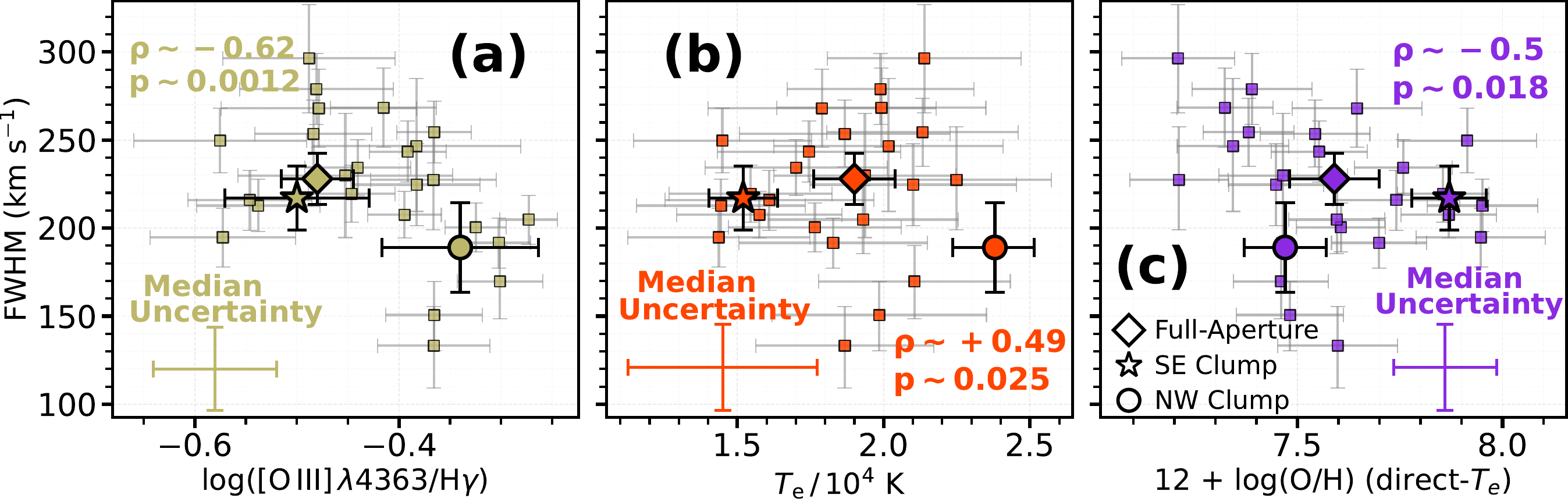}
  \caption{This figure shows the variation of gas velocity dispersion (FWHM) as a function of O3Hg line ratio (\autoref{fig:correlations_a}), electron temperature (\autoref{fig:correlations_b}) or gas-phase metallicity calculated using the direct method (\autoref{fig:correlations_c}). As before, we only show the spaxels with $\rm S/N_{\left[\rm O \ \textsc{iii}\right]\lambda4363}>3$, but we also show the values corresponding to the SE and NW clump and for the full-aperture spectra. We display the individual and the typical median uncertainties for the per-spaxel measurements. The bootstrapped median Spearman correlation coefficient $\rho$ and the p-value are reported for the set of correlations considered.}\label{fig:correlations}
\end{figure*}

Given the tentative position of the spaxels in the south-western region of the system in the AGN dominated region of the \citet{mazzolari+24} diagnostic diagram, as well as the subtle discrepancy between the gas-phase metallicities calculated using the direct-$T_e$ and the strong-line methods, we test the possibility that shocks may have a significant contribution to the photoionisation. We use an approach similar to that of \citet{deugenio+2024_wide}, who report shock-driven photoionisation in a SFG at $z\sim4.6$. We use the shock models from the 3MdB database\footnote{\url{https://sites.google.com/site/mexicanmillionmodels/}}. The grids are computed using the \textsc{mappings v} code \citep{mappings5}. We tested both shock only and shock with precursor models \citep{allen+2008,alarie+2019} spanning a large parameter space (the {\it BnZv space}): magnetic field $10^{-4} \ \mu {\rm G} < B < 10 \ \mu {\rm G}$, shock velocity $100 {\rm \ km \ s^{-1}} <v < 1000 \ \rm km \ s^{-1}$, pre-shock (upstream) density $1 \ {\rm cm^{-3}} < n_{\rm sh} < 10000 \ \rm cm^{-3}$ and metallicity $10^{-4} <Z < 0.04$. The metallicities are converted to oxygen abundance values assuming $Z_{\odot}=0.01524$ and $12+\log\left(\rm O/H\right)_{\odot} =8.8$ (using the same chemical abundances from \citealt{gutkin+2016} as used by \citealt{alarie+2019} in the shock grid calculations and adopting a dust depletion $\xi_{d}=0.1$, similar to Section \ref{sec:gas_strong_line}). As before, we choose the $\rm C/O = 0.72 \left(\rm C/O\right)_{\odot}$ photoionisation parameters from \citet{gutkin+2016}. 

In our particular case, we select a fixed magnetic field strength of $\rm B= 10 \ \mu G$ for all models displayed in \autoref{fig:shocks}. This value is comparable to the magnetic field in the solar neighbourhood ($2-25 \rm \mu G$; \citealt{vallee+2022}). However, there is plausible evidence that the galactic magnetic fields are likely larger in starburst galaxies ($50-100 \  \mu G$; \citealt{beck_mag}), and at earlier cosmic times \citep{84mug_field}. Because $10 \ \mu G$ is the highest value available in the computed shock grids, we regard this value as the most sensible choice. The diagnostic line ratios considered here (O3Hg and Ne3O2) are only weakly sensitive to the magnetic field strength over the range of parameter space probed by our data. In particular, magnetic-field-dependent effects become significant only at very low O3Hg ratios, $\log \left(\left[\rm O \ \textsc{iii}\right]\lambda4363 \, / \, H\gamma\right)<-1.5$, where model tracks corresponding to different magnetic fields diverge.

We explore shock model grids with pre-shock densities, $n_{\rm sh} =100, \, 1000 \, \rm \, and \  10000 \ \rm cm^{-3}$. Models with $n_{\rm sh}\lesssim100 \ \rm cm^{-3}$ are unable to reproduce the observed O3Hg and Ne3O2 measurements in any configuration, so they are strongly disfavoured. For $n_{\rm sh}=10^{3}-10^{4} \ \rm cm^{-3}$, shock-plus-precursor grids (\autoref{fig:shocks_e} and \ref{fig:shocks_f}) intersect the data only for parameter combinations that imply oxygen abundances of $12+\log\left(\rm O/H\right) \sim 8.2-8.4$, exceeding the metallicities inferred from spaxel or clump aperture spectra ($12+\log\left(\rm O/H\right) \sim 7.2-8.0$ using either the direct or the strong-line method). It is generally true that the metallicities calculated with the direct method might be biased low if shocks are present and not accounted for. This is because shocks alter the ionisation structure of the ISM and can create large temperature variations on small spatial scales. As a result, the electron temperature and oxygen abundance may be decoupled, causing metallicities to be underestimated in shock-heated regions. However, the shock grid displayed in \autoref{fig:shocks_e} further requires high velocity shocks ($\sim 500-1000 \ \rm km \ s^{-1}$) to match our data. Such velocities are substantially larger than the characteristic ionised gas kinematic widths in our system ($\rm FWHM \sim 100-350 \ \rm km \ s^{-1}$). Additionally, our spectral fits show no statistical requirement for an additional broader component, together suggesting that these shock-plus-precursor models are unlikely to dominate over star-formation driven photoionisation on large scales, but they may be important locally. 

A shock-only model with $B=10 \ \mu G$ and $n_{\rm sh}=10^{4} \ \rm cm^{-3}$ (\autoref{fig:shocks_c}) can reproduce the observed line ratios for the SE and NW clump apertures at lower metallicities and lower shock velocities than the shock-plus-precursor cases. However, such a model requires an upstream density of $10^{4} \ \rm cm^{-3}$. Moreover, shocks compress the gas by at least a factor of four in the strong adiabatic shock limit (and potentially more if radiative), implying post-shock densities $\gtrsim 4\times 10^{4} \ \rm cm^{-3}$. While these required densities are at least one order of magnitude above the average electron density of $\log \left(n_{\rm e} \rm /cm^{-3}\right)=3.3^{+0.7}_{-0.6}$, derived from the full-aperture spectrum of the MACS0647-JD1 image \citep{abdurrouf+2024}, this direct comparison is not fully conclusive for ruling out shock-driven photoionisation processes. The reported electron density was obtained by \citet{abdurrouf+2024} by spectrally resolving the $\left[\rm O \, \textsc{ii}\right]\lambda3726,3729$ doublet with high-resolution {\it JWST} spectroscopy. \textsc{PyNeb} predicts critical densities of $\sim 1500 \ \rm cm^{-3}$ and  $\sim 5000 \ \rm cm^{-3}$ (assuming an electron temperature of $1.9\times 10^{4} \ \rm K$, similar to the one derived from the full aperture spectrum) for $\left[\rm O \ \textsc{ii}\right]\lambda3729$ (first upper level) and $\left[\rm O \ \textsc{ii}\right]\lambda3726$ (second upper level), respectively. These values are in agreement with those obtained by \citet{wang+2004}. As a result, this doublet primarily traces low-ionisation and lower density phases of the ISM. In a multi-zone ISM,  the $\left[\rm O \, \textsc{ii}\right]$-based density may not fully capture denser gas that might produce the bulk of $\left[\rm O \, \textsc{iii}\right]\lambda4363$ ($n_{\rm crit}\sim 3 \times 10^7 \, \rm cm^{-3}$; \citealt{Moreschini2026}) emission or shocks. Recent observational and ISM modelling studies have further revealed that metallicities inferred under a low-density ISM assumption may be underestimated \citep{marconi+2024,harikane_zones,martinez_zones,Moreschini2026}. This is because auroral lines such as $\left[\rm O \, \textsc{iii}\right]\lambda4363$, commonly used in direct-$T_e$ method calculations, typically probe higher density regions \citep{jones_t2020,yang_lidz2020,yang2021,nakazato2023}. In high density regimes, collisional de-excitation and radiative decays have comparable contributions to de-populating the atomic upper levels associated with the key emission lines, which could bias our estimated abundances. Invoking a substantial fraction of the ionised gas mass with a density of $\sim 10^5-10^6 \, \rm cm^{-3}$ (to accommodate the high pre-shock densities required to match the observed emission line ratios; \autoref{fig:shocks}) would be difficult to reconcile with a starburst-dominated system exhibiting only marginal and localised evidence for non-SF processes. A complete picture would require high-resolution {\it JWST} IFU observations of the system using the G235H or G395H gratings. Such observations would provide spatially resolved maps of both low- and high-density gas (by spectrally resolving the $\left[\rm O \, \textsc{ii}\right]\lambda3726,3729$ and the far-ultraviolet $\rm  C \, \textsc{iii}]\lambda\lambda1907,1909$ doublets, respectively). 

Furthermore, as a simplified, illustrative approach, under the assumption that shocks are the main excitation mechanism of the ISM in MACS0647-JD1, we can estimate the average photoionisation mass rate associated with this process. If shocks rather than star formation dominate the emission lines production, it means that at least half of the $\left[\rm O \, \textsc{iii}\right]\lambda5007$ flux (total of $\left(2.24 \pm 0.52\right)\times 10^{-17} \, \rm erg \, s^{-1} \, cm^{-2}$) would arise from the shock line component. Taking into account the luminosity distance and the magnification of the system, this would imply $L_{\left[\rm O \, \textsc{iii}\right]\lambda 5007, \, \rm shocks}\gtrsim 2 \times 10^{42} \, \rm erg \, s^{-1}$. According to the first equation from \citet{laor+98}, and adopting a typical maximum shock velocity of $v \sim 1000 \, \rm km \, s^{-1}$ (similar to the maximum value within the shock grids from the {\it 3MdB} database), we would obtain $M_{\rm shocked \, gas}\gtrsim 100 \, \rm M_{\odot} \, yr^{-1}$. This value is significantly larger than the inferred H$\gamma$-based SFR of $\sim 5.5 \pm 1.0 \, \rm M_{\odot} \, yr^{-1}$, consistent with our starting assumption. If the gas mass in the system is comparable to the stellar mass (see the discussion in Section \ref{sec:scenarios_merger_or_disc}), such a high rate of shock-driven photoionisation of the cold gas would imply that this reservoir would be consumed on $\lesssim1-5 \, \rm Myr$ timescales, shorter than the stellar ages of $\sim 20-30 \, \rm Myr$ of the two clumps derived from SED fitting. Overall, while we cannot firmly rule out shocks affecting the structure and ISM properties of MACS0647-JD1, the existing observations favour a starburst-dominated ISM.


\subsection{Spatially resolved properties of the ISM gas}
\label{sec:gas_discussion}

\begin{figure}
\centering
  \includegraphics[width=\linewidth]{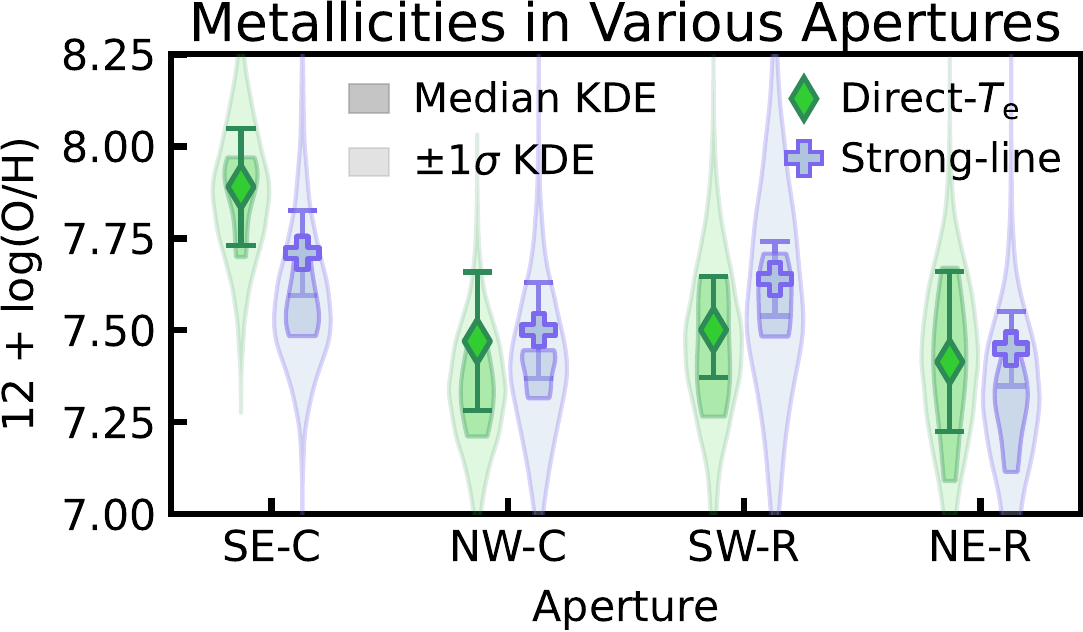}
  \caption{This figure shows the calculated gas-phase metallicities (using both the direct-$T_{\rm e}$ and the strong-line methods) and the uncertainties for the spectra of the four apertures identified in \autoref{fig:astrometry_b}. For each aperture and method, violin plots illustrate the ranges of metallicities of spaxels within the said aperture. Darker-shade violins show the kernel density estimate (KDE) of median metallicities, while lighter-shaded violins include $\pm1\sigma$ spaxel-level uncertainties.}
  \label{fig:violins}
\end{figure}

Based on our methods for calculating the strong-line metallicity (discussed in Section \ref{sec:gas_strong_line}), we obtain the map shown in \autoref{fig:strong_main}. We applied similar spatial cut of the map in order to facilitate the comparison with the direct metallicity map (\autoref{fig:fig4panels_c}). We find that our results are consistent between these methods. Both the direct-$T_{\rm e}$ and the strong-line methods reveal that the SE clump is more enriched than both the NW clump and the turbulent, metal-poor gas located in the NE region of the system (i.e. between the two clumps). 

To provide quantitative evidence of the different nature of the ISM at various locations, we divide the observed system within the field of view in four regions, as shown in \autoref{fig:astrometry.combined}: the north-west and the south-east clumps (NW-C and SE-C), together with the north-east and south-west regions (NE-R and SW-R). We illustrate the resulting direct-$T_{\rm e}$ and strong-line metallicities from the integrated spectra of these four apertures (obtained by adding the spaxels within each aperture) in \autoref{fig:violins}. We find that, for each aperture, the two methods of calculating the metallicity give consistent results. Moreover, \autoref{fig:violins} shows that, although some non-negligible spaxel-to-spaxel variations may occur in our maps, our conclusion that the SE clump is more metal-rich than its surrounding environment is statistically robust. 

\autoref{fig:fig4panels_a}, \ref{fig:fig4panels_c}, and \ref{fig:fig4panels_d} reveal spatial variations in gas kinematics, excitation, and chemical properties across the system, suggesting possible connections between gas velocity dispersion, excitation-sensitive line ratios (particularly O3Hg), and gas-phase metallicity on spatially resolved scales. We quantify these trends by testing for spaxel-scale correlations between gas velocity dispersion (FWHM) and ISM properties. We find a significant anti-correlation between FWHM and the O3Hg line ratio (with a p-value of $1.2 \times10^{-3}$ and a Spearman correlation coefficient $\rho\sim-0.62$; \autoref{fig:correlations_a}). While this correlation is intriguing, its physical origin is difficult to pinpoint because the $\left[\rm O \, \textsc{iii}\right]\lambda4363 /H\gamma$ ratio depends simultaneously on multiple ISM properties, including metallicity, electron temperature, ionisation parameter, and the local SFR. We find only a marginal anti-correlation between FWHM and metallicity, and a tentative positive correlation between FWHM and electron temperature (as shown in \autoref{fig:correlations_b} and \autoref{fig:correlations_c}). Our analysis does not find any correlation between FWHM and Ne3O2 ($\rm p\sim0.38$; Spearman correlation coefficient of $\rho\sim -0.12$), although central spaxels tend to have both high O3Hg and high Ne3O2 ratios. Similar to our calculation in Section \ref{sec:ionisation_param} for the integrated-aperture spectrum, we compute the ionisation parameters for the spectra of all individual spaxels, based on the \citet{gutkin+2016} photoionisation grid. We find only a weak, statistically insignificant correlation between FWHM and $\log U$ of individual spaxels ($\rm p \sim 0.08; \, \rho \sim +0.24$). Remarkably, we also obtain a correlation of higher significance between FWHM and the star formation burstiness ($\rm SFR_{10}/SFR_{100}$), with $\rm \rho \sim +0.39; \, p \sim 10^{-3}$. We note that, in this case, the total number of spaxels considered is larger than for the anti-correlation reported between FWHM and O3Hg, and the tentative $\rm FWHM-T_e$ and FWHM-metallicity correlations. This is because in the latter case, we take into account all spaxels with $\rm S/N_{H\gamma}>5$, whereas in the former case, we are restricted to the central region of the system for which we imposed $\rm S/N_{\left[\rm O \, \textsc{iii}\right]\lambda4363}>3$ for a statistically significant detection of this auroral line. We caution that the reported Spearman coefficients and p-values are based on median estimates using face-value measurements in the individual spaxels considered for each pair of physical parameters or observables. A bootstrap analysis accounting for per-spaxel uncertainties yields p-values exceeding 0.05 for the FWHM–$T_{\rm e}$ and FWHM–metallicity relations, indicating that these trends are not robust. Only the correlation between FWHM and O3Hg remains significant, with an 84th-percentile p-value of 0.028. 

The differing strengths of these correlations likely reflect the fact that O3Hg is a direct and unbiased observable that is sensitive to both the local excitation and thermal conditions of the ionised gas in the ISM, and also to the local instantaneous SFR. By contrast, the electron temperature and the inferred metallicity are derived quantities subject to additional assumptions and associated systematic uncertainties. The most important of these include a secondary dependence on the ionisation parameter, the possibility of a multi-zone ISM, as well as possible uncertain dust corrections in calculations of metallicities and electron temperatures. Importantly, a locally elevated temperature of the $O^{++}$ gas may not necessarily imply a higher $\left[\rm O \ \textsc{iii}\right]\lambda4363 \, / \, H\gamma$ ratio. While the effective H$\gamma$ recombination rate has a weak inverse dependence on electron temperature, this effect can be easily outweighed by spatial variations in electron and ion densities or gas filling factor, particularly in a complex, multi-zone ISM. The gas FWHM traces emission-line broadening due to both thermal and non-thermal processes. The presence of dynamically turbulent and hot gas (as shown in \autoref{fig:fig4panels_d} and \ref{fig:te_map_b}) between the clumps may indicate enhanced energy injection into the ISM. This may arise from localised shocks (see the discussion in Section \ref{sec:shocks} for arguments disfavouring large-scale shock-driven photoionisation), or from a combination of turbulent star formation-driven feedback and complex gravitational interactions between the clumps and the surrounding metal-poor gas. Our observational findings agree with recent theoretical models, which posit that turbulent gas in low-mass bursty galaxies at high redshift, being highly dynamically unstable, favours clumpy star formation, and an ISM with large density and temperature variations, even on relatively small spatial scales \citep{fire_turbulent}, providing a possible explanation for why there is a statistically significant correlation between FWHM and star formation burstiness. 


Our results disagree with results from the local Universe and lower-redshift studies, which find positive correlations between gas-phase metallicity and (stellar) velocity dispersion for massive, evolved galaxies (e.g., \citealt{li+2018vc,cappellari2023ppxf}). Such a positive correlation is also seen in individual galaxies within the SPT0311-58 protocluster at $z\sim6.9$ \citep{arribas+24}. The possible anti-correlation observed in the case of MACS0647-JD is likely not the result of secular, interconnected processes shaping the co-evolution of various galaxy properties. Such processes have not yet been established in low-mass systems at $z>10$. Instead, our findings suggest that our target is observed in an early, short-lived evolutionary phase. 

\subsection{Spatially resolved stellar populations: bursty star formation in the north-eastern region of the system}
\label{sec:sed_stellar}

\begin{figure*}
\centering
{\phantomsubcaption \label{fig:bursty_a}
\phantomsubcaption \label{fig:bursty_b}
\phantomsubcaption \label{fig:bursty_c}}
  \includegraphics[width=\linewidth]{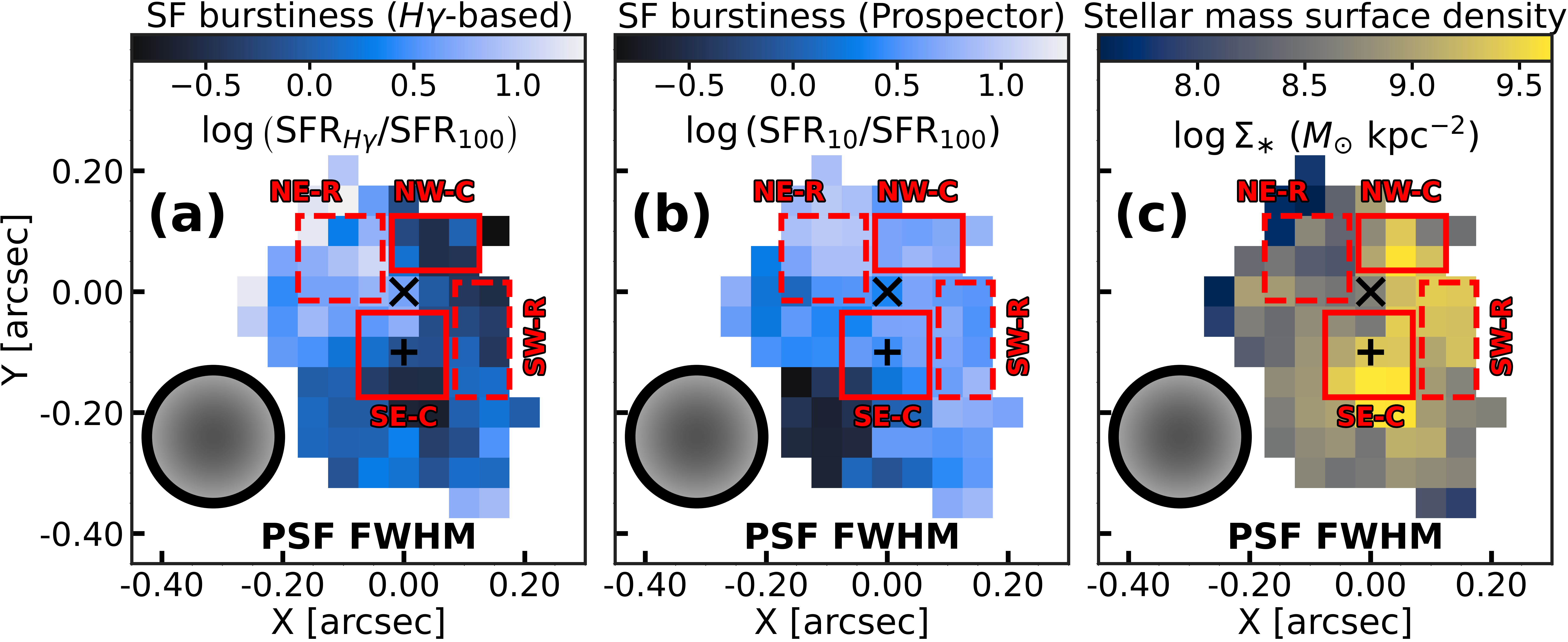}
  \caption{In this figure, we display the spatially resolved star formation burstiness in MACS0647-JD1 system. The burstiness probes the recent versus prolonged star formation activity in our target, and is defined in two alternative ways: $\rm SFR_{H\gamma}/SFR_{100}$ (\autoref{fig:bursty_a}) and $\rm SFR_{10}/SFR_{100}$ (\autoref{fig:bursty_b}). The red rectangles are the apertures corresponding to the two clumps (SE and NW, as identified from NIRCam F444W imaging; see \autoref{fig:astrometry.combined}). Both maps consistently show a remarkable recent star formation enhancement in the north-east region and a less bursty SFH in the region of the more massive SE clump. \autoref{fig:bursty_c} shows the stellar mass density map of this system and confirms the more massive nature of the two clumps, compared to other regions. On all panels, we only show the spaxels with a $\rm S/N>5$ detection of H$\gamma$ from our analysis of NIRSpec/IFU data. We also show the PSF size in all panels, together with the centroid of stellar continuum (`+' cross) and of the H$\gamma$ emission line map (`$\times$' cross).}\label{fig:bursty}
\end{figure*}

We compute the burstiness of star formation on spaxel-by-spaxel scales using the SFR averaged over the last 100 Myr in lookback time ($\rm SFR_{100}$ from our \textsc{prospector} SED fits) as a probe of the long-timescale SFR. For instantaneous or short-timescale SFR, we use two alternative tracers: {\it a)} $\rm SFR_{10}$ (similar to $\rm SFR_{100}$ but it probes only the last 10 Myr in lookback time; this is also computed by \textsc{prospector}); and {\it b)} $\rm SFR_{\rm H\gamma}$ inferred from the $\rm H\gamma$ integrated line flux. To estimate $\rm SFR_{H\gamma}$, we first correct the H$\gamma$ fluxes of each spaxel based on the observed $\rm H\beta/H\gamma$ Balmer decrement in that spaxel (we assume no dust correction in the cases where the $\rm H\beta/H\gamma$ ratio from the spectrum is lower than the case-B value of 2.105). We then assume an intrinsic, case-B $\rm H\alpha/H\gamma=5.79$ ratio, valid for an electron temperature $T_{\rm e}\sim20000 \ \rm K$ \citep{groves+2012} and we adapt the relation between H$\alpha$ intrinsic luminosity and SFR from equation (12) of \citet{kennicutt+2012} as follows:

\begin{equation}
\log \rm SFR_{H\gamma} \left(M_{\odot} \ yr^{-1}\right)= \log L_{H\gamma} \left(erg \ s^{-1}\right) -40.5 -\log \mu -0.3.
\end{equation}

\noindent Here, for delensing, we assume a lensing magnification of $\mu=8\pm1$ \citep{chan+2017}. However, the SFR-H$\alpha$ equation derived by \citet{kennicutt+2012} is calibrated in the local Universe, assuming a solar metallicity and a \citet{Calzetti2000} dust attenuation law. At higher redshifts, galaxies typically have different properties from local targets, so a more suitable set of assumptions is a different dust attenuation law and a lower metallicity. \citet{reddy+22} and \citet{shapley+23} showed that the correction term is around $-0.3 \ \rm dex$ if we instead assume a Small Magellanic Cloud dust attenuation law and a metallicity of $0.27 \ \rm Z_{\odot}$.

We show our results in \autoref{fig:bursty_a} and \autoref{fig:bursty_b}. Both panels indicate a significant enhancement of recent star formation in the NE region, the same location where the gas is both more turbulent (\autoref{fig:fig4panels_d}) and metal-poor (\autoref{fig:fig4panels_c}). By contrast, the more massive, more metal-enriched SE clump does not undergo a noticeable starburst episode within the timescales probed. As shown in \autoref{fig:bursty_b}, we find only moderate burstiness in the two clumps (more pronounced for the NW clump). \autoref{fig:bursty_c} confirms the more massive nature of the two clumps compared to the surrounding regions. The stellar mass densities of the spaxels in the NE region are typically $\sim 1 \ \rm dex$ below the values within the two clumps, which implies that this region may be associated with a metal poor gas inflow. Another plausible scenario is that this region contains gas stripped from the NW clump by gravitational interactions with the massive SE clump.  Indeed, the metallicities of NE-R and NW-C are consistent within the uncertainties as shown in \autoref{fig:violins}. 

More puzzlingly, the SW region of the system, which is intrinsically fainter than either of the clumps, appears to be substantially more massive than the bursty, low-$M_{\ast}$ NE region. In addition, $\rm SFR_{H\gamma}$ indicates negligible star formation enhancement (\autoref{fig:bursty_a}). When recent star formation is instead probed via $\rm SFR_{10}$, a starburst effect is observed in the SW region, although it is less intense than in the NE region. Physically, the SW region could represent gas and stars that originated in the SE clump but were redistributed during a merger event.

\subsection{Merging galaxies or clumps in a disc?}
\label{sec:scenarios_merger_or_disc}



The observed properties of MACS0647-JD1 allow, in principle, two broad interpretations: either the system represents a gas-rich rotating disc hosting multiple clumps, or it is caught during an early-stage merger between two distinct components. Disc galaxies are generally characterised by a ratio of ionised gas velocity to velocity dispersion of $\rm V_{\rm gas}/\sigma\gtrsim1$ and by ordered rotation on spatially resolved scales \citep{forster_schreiber_review,danhaive_kin}. In Appendix \ref{sec:velocity_problem}, we show that the observed rotational pattern in the velocity field of the MACS0647-JD1 system may be dominated by instrumental systematics, making it difficult to recover the intrinsic velocity field. One reason why we cannot firmly dismiss the hypothesis of a real rotating structure is the fact that the rotational gradient is spatially more extended than the FWHM of the PSF (the latter being $\sim 4.4$ spatial pixels). In addition, as we argue in Appendix \ref{sec:velocity_problem}, the amplitude of the artificial rotation pattern produced by the slit-width effect (described in \citealt{isobe_empress} and Appendix \ref{sec:velocity_problem}) is noticeably smaller (2.6 times) than the observed velocity gradient in our data. It is, however, crucial to note that we only provide a simplified calculation that illustrates that the slit-width effect could potentially mimic a rotational pattern in {\it JWST/NIRSpec} IFU data. Disentangling instrumental effects from genuine galaxy kinematics would require additional IFU observations obtained at a position angle rotated by $90^{\circ}$ relative to the current setup (\autoref{fig:fake_a}).

In this section, we argue that the morphological, chemical and kinematic properties of this system are more naturally explained by a gas-rich merger between the NW and the SE clumps. First and foremost, the two main components exhibit markedly different gas-phase metallicities (see Section \ref{sec:gas_discussion}), with the SE clump being significantly more enriched than both the NW clump and the surrounding gas. Using the values in \autoref{tab:table_parameters_integrated}, we obtain a metallicity difference of $0.42 \pm 0.18 \, \rm dex$ (using direct-method metallicities; or $0.22 \pm 0.12 \, \rm dex$ using the metallicities derived based on the strong-line method) between the two main components of the system. Such metallicity contrasts over short projected physical scales ($\sim 300 \, \rm pc$ in the source frame, assuming a magnification of $\mu \sim 8$) are difficult to reconcile with in-situ clump formation within a single disc, where clumps primarily form from the existing ISM gas reservoir. While several studies have revealed metallicity inhomogeneities in proto-disc systems at $z>6$ \citep{arribas+24,venturi+24,ivey+26}, the reported metallicity gradients are $\left | \nabla_{r} \log\left(\rm O/H\right) \right| \lesssim 0.2 \, \rm dex \ kpc^{-1}$. The amplitude of the variations we measure in the MACS0647-JD system suggests that additional processes beyond simple in-situ fragmentation (such as external fresh gas accretion or mergers) may be involved.

Second, the highest gas velocity dispersions are observed outside the two massive clumps (\autoref{fig:fig4panels_d}) and are anti-correlated with excitation-sensitive line ratios (\autoref{fig:correlations_a}). This suggests that turbulence is not primarily driven by star formation feedback, as expected in clumpy discs, but by gravitational interactions possibly leading to a large-scale circulation of the gas, consistent with a merger scenario. The observation of this dynamically hot gas in the regions outside the main components is facilitated by the lack of strong stellar sources in these locations, allowing the emission lines to probe large path lengths along the line of sight and therefore trace large-scale gas motions. In addition, the star formation burstiness maps show a clear asymmetry (\autoref{fig:bursty}): the most intense enhancement of recent star formation is in the north-east region, co-spatial with confidently identified metal-poor, highly turbulent gas.

Third, the centroid of nebular emission is offset from the stellar continuum peak by at least two spatial pixels ($\sim 0.1 \arcsec$, measured using the $\rm H\gamma$ emission map; see \autoref{fig:astrometry.combined}). The spatial offset is larger if we consider, for instance, $\rm H\beta$, $\rm H\delta$ or $\left[\rm Ne \ \textsc{iii}\right]\lambda3869$ centroids. While instrument systematics can sometimes create an artificial shift of the centroid of point sources as a function of wavelength, the magnitude of this effect is too small ($\sim 0.02\arcsec$; \citealt{ifu_report}) to justify the different spatial distributions of stellar continuum and emission lines. In a clumpy disc, the most intense line emission is generally co-spatial with the dominant stellar clumps \citep[e.g.,][]{romeo+2017}, whereas star formation between stellar components is more naturally explained by merger-driven gas instability.

Although an accurate estimate of the gas mass in the system requires the completion of the existing allocated {\it NOEMA} observations, we can provide a crude estimate using a method similar to \citet{nakajima+25u}. The angular separation between the two clumps is approximately $0.15 \arcsec$ (as shown in \autoref{fig:astrometry_a} and \autoref{fig:bursty_c}), corresponding to $\sim 220 \, \rm pc$ in the source plane. As a result, given our delensed value of the $\rm SFR_{H\gamma}$ estimated for the full-aperture spectrum, $\rm SFR_{H\gamma}=5.5\pm1.0 \ \rm M_{\odot} \ yr^{-1}$ (\autoref{tab:table_parameters_integrated}), and assuming an approximate radius of $\sim 200 \pm 50 \ \rm pc$ for the gas reservoir in the inter-clump region, we obtained a SFR density of $\log \left(\Sigma_{\rm SFR_{H\gamma}}/\rm M_{\odot} \ yr^{-1} \ kpc^{-2}\right)=1.35 \pm 0.30$. We use equation (2) from \citet{bouche+2007} to estimate the gas mass surface density. In this equation, $\Sigma_{\rm SFR} \propto \Sigma_{\rm gas}^{1.71 \pm 0.05}$ and the power-law index of $1.71 \pm 0.05$ is significantly larger than the $1.4 \pm 0.15$ value given in \citet{kennicutt+98}. This substantial difference arises from collisionally enhanced star formation efficiency in high-density gas ($\sim 1000 \, \rm cm^{-3}$), a regime which is more commonly observed in the ISM of high-redshift galaxies. With our previous value of $\log \Sigma_{\rm SFR_{H\gamma}}$, we obtain $\log \left(\Sigma_{\rm gas} \,/\,\rm M_{\odot} \ pc^{-2}\right)=3.15 \pm 0.20$, implying $\log \left(M_{\rm gas}\,/\,M_{\odot}\right)=8.6 \pm 0.3 $. This indicates that the gas mass in MACS0647-JD1 is likely larger than the stellar mass within the SE and NW clumps. 

If the system is a rotating disc, we can estimate its dynamical mass using the following equation (adapted from \citealt{sauron4}):

\begin{equation}
\ M_{\rm dyn} \approx \frac{5   R_{\rm eff}  \ \sigma_{\ast}^{2}}{G} \ .
\label{eq:dynamical_mass}
\end{equation}

\noindent We assume an effective radius $R_{\rm eff}=70 \pm 24 \ \rm pc$ \citep{hsiao_nirspec} and estimate the stellar velocity dispersion from the gas velocity dispersion, $\sigma_{\rm gas}= \rm FWHM_{\rm gas}\,/\,2.355=97 \pm 8 \ \rm km \ s^{-1}$ from the full-aperture spectrum. Adopting the empirical relation $\sigma_{\ast} \approx 1.26 \times \sigma_{\rm gas}$ \citep{bezanson2018}, we obtain $\sigma_{\ast}=122 \pm 10 \ \rm km \ s^{-1}$ and therefore $\log(M_{\rm dyn}/M_{\odot}) = 9.1 \pm 0.5$. The quoted uncertainty accounts for the standard scatter of the virial mass estimator equation and the systematics associated with choosing a factor of 5 in Equation \ref{eq:dynamical_mass}, instead of using equation (7) from \citet{vanderWel2022}. The latter equation accounts more accurately for variations due to the object's axial ratio and S\'{e}rsic index. However, given the two-clump morphology of our target (\autoref{fig:astrometry_b}), a simple single S\'{e}rsic profile fit to the F444W image of MACS0647-JD1 would not be suitable. This caveat prevents us from applying morphology-dependent virial corrections such as those discussed in \citet{vanderWel2022}.

The estimated dynamical mass of the system is statistically consistent (within the $1\sigma$ uncertainties) with the combined mass of the stars (see \autoref{tab:table_parameters_integrated}) and gas. However, this comparison does not provide a decisive diagnostic of the dynamical state of the system. The dark matter (DM) mass, on the other hand, is even more difficult to accurately constrain at such high redshifts. Galaxies at $z>4$ and $\log\left(M_{\ast}/M_{\odot}\right)<9$ are expected to be strongly dark matter dominated, as revealed by recent {\it JWST} spectroscopic observations \citep{degraaff_ionised,danhaive_dm_u} and simulations \citep{degraaff_dm,will_dm}. We can provide a simple estimate of the DM mass in the central region between the two clumps as follows. The total stellar mass of the two clumps is $\log \left(M_{\ast,\,\rm NW+SE}/M_{\odot}\right)=7.94 \pm 0.09$. We assume that the dark matter density follows the classical NFW profile (with the inner slope $\gamma = -1$) satisfying the equation \citep{nfw1996,lenses2001general}: 

\begin{equation}
\mathrm{\rho_{\rm DM} \left ( r \right ) \propto \frac{1}{\left (r  /   r_{b} \right )^{-\gamma} \cdot \left (1 + r  /  r_{b} \right ) \ ^{3+\gamma}}}.
\label{eq:nfw_profile}
\end{equation}

\noindent Here, $\mathrm{r_{b}}$ is the so-called ``break" radius of the dark matter halo of MACS0647-JD1. Using the stellar mass $-$ halo mass (SMHM) relation from \citet{behroozi+2019}, we obtain a virial halo mass of $\log \left(M_h/M_{\odot}\right)\approx 10.5\pm 0.3$. Here, the quoted uncertainty combines the intrinsic scatter of the SMHM parametrisation at $z\sim10$ and the systematic uncertainty introduced by summing the stellar masses within the two clumps (rather than including all the spaxels in \autoref{fig:bursty_c}). The virial radius of the halo can be estimated using: 

\begin{equation}
\mathrm{M_{h} = \frac{4 \pi}{3} \Delta_{vir} \rho_{m} \left ( z \right )  R_{vir}^{3}} .
\label{eq:mvir_rvir}
\end{equation}

\noindent In this equation, we assume that the density contrast is $\Delta_{\rm vir} = 200$, obtaining $R_{\rm vir} \approx 8.8 \pm 2.2$ kpc. Next, we estimate the concentration parameter $\mathrm{c_{200} = R_{vir}  /  r_{b}}=4.2 \pm 0.2$ from Figure 7 of \citet{som_character}. This implies $r_b=2.1 \pm 0.5 \, \rm kpc$. Under the spherical symmetry assumption, integrating the DM density profile would yield the following formula for the DM mass enclosed within a radius $r$:

\begin{equation}
M_{\rm DM}\left(<r\right) =
M_{\rm vir}
\,
\frac{
\ln\!\left(1+\dfrac{r}{r_b}\right)
-
\dfrac{r/r_b}{1+r/r_b}
}{
\ln\!\left(1+c_{\rm vir}\right)
-
\dfrac{c_{\rm vir}}{1+c_{\rm vir}}
}
.
\end{equation}

\noindent Within the innermost $r=200 \, \rm pc$ corresponding to the two clumps, we thus obtain $\log \left(M_{\rm DM}/M_{\odot}\right)=8.2 \pm 0.3$. These findings suggest that stars, gas and dark matter have similar contributions to the total mass budget of the system. Therefore, it is currently difficult to estimate the spatial distribution of gas or dark matter in this system. 
\autoref{fig:bursty_c} illustrates a clear double-peaked stellar-mass density distribution, with two spatially distinct mass concentrations corresponding to the south-eastern and north-western stellar components. For a disc-like galaxy to maintain long-term gravitational stability, however, the total mass (including stars, gas and dark matter) profile should be centrally peaked \citep{romeo+2018}. Because a detailed analysis of the distributions of gas and dark matter within this system is currently impossible, the spatially resolved map of the total mass cannot be determined accurately. In conclusion, it is not possible to infer the dynamical nature of the system based entirely on the available $M_{\ast}$ spatial distribution. 

\section{Summary and Conclusions}
\label{sec:conclusions}


In this paper, we present a spatially resolved analysis of the ISM in MACS0647-JD1, the brightest galaxy currently known at $z>10$ (owing to gravitational lensing), using medium-resolution NIRSpec/IFU G395M observations from the GA-NIFS GTO programme. We present a novel method for extrapolating the {\it JWST} spectra outside the nominal wavelength range of the G395M grating, enabling us to obtain the H$\beta$-wavelength collapsed flux map for a target at $z\sim10.2$. By combining these spectroscopic data with archival {\it JWST} imaging, we infer star formation histories across different regions of the system, enabling a detailed investigation of the physical processes governing its early assembly. This approach provides new insight into the complex interplay between star formation activity and chemical enrichment in one of the earliest known starburst systems. We summarise our conclusions as follows:

\begin{itemize}

\item The system consists of two bright components detected in NIRSpec/IFU and NIRCam imaging. The stellar continuum centroid coincides with the more massive component, which is spatially offset relative to the emission-line centroid by 150 pc (in the source plane). This is most likely a signpost of recent star formation activity in between the two components, potentially triggered by a pre-coalescence merger event.

\item Our target provides the highest-redshift direct-metallicity measurement obtained with NIRSpec, and the second highest-redshift measurement with any instrument, surpassed only by GN-z11 \citep{bunker_gnz11,alvarez+2025}. We compute metallicities from spaxel spectra using two independent approaches: the direct method (based on the electron temperature) and a customised strong-line calibration. Both methods yield consistent results within the uncertainties. We find that the brighter south-eastern clump is more metal-rich ($12+\log(\rm O/H)_{\rm direct}=7.89 \pm 0.11$), while the north-western clump is less enriched ($12+\log(\rm O/H)_{\rm direct}=7.47 \pm 0.14$). Such a pronounced asymmetry would be difficult to explain if the two components were clumps within a single rotating disc. Although we attempted to reconstruct the velocity field (see Appendix \ref{sec:velocity_problem}), the current data do not provide conclusive evidence for ordered rotation.


\item The strong-line metallicities are less reliable in the south-western part, where the unusual morphology of the $\left[\rm O \ \textsc{ii}\right]\lambda\lambda3726,3729$ emission (as compared to high-ionisation or hydrogen recombination lines), together with tentative AGN-like signatures on one of the \citet{mazzolari+24} diagnostic diagrams, may imply the presence of diffuse ionised gas or a complex stratification structure. We have additionally shown that shocks are strongly disfavoured, although not completely ruled out by our observed emission line ratios.

\item We analyse the ISM conditions on spatially resolved scales by focusing on various diagnostic ratios. Analysing the spectra of individual spaxels, we find a meaningful anti-correlation between the gas velocity dispersion (FWHM; a tracer of turbulence) and $\left[\rm O \ \textsc{iii}\right]\lambda4363/H\gamma$ line ratio (a proxy for the excitation state of the ISM), but statistically weaker trends between FWHM and gas-phase metallicity or electron temperature, both of which are likely influenced by a secondary dependence on the ionisation parameter.


\item Combining our spectroscopic and stellar populations analyses, we demonstrate that, remarkably, the north-eastern region has a small stellar mass and predominantly contains metal-poor gas, while its star formation burstiness is significantly elevated compared to the surroundings. This suggests that a very recent star formation event has been initiated in that region, which could be either associated with a fresh gas inflow or a redistribution of the metal-poor gas already present in the north-western component, as a result of gravitational interaction. These two scenarios could explain the metallicity difference between the south-eastern and north-western components, which are massive gas clumps, with only low to moderate burstiness.


\end{itemize}

\section*{Acknowledgements}

We are grateful to the referee for insightful and constructive feedback that substantially improved the quality of this manuscript. We thank Dr Xihan Ji, Prof. Alessandro Romeo, and Dr. Guochao Sun for insightful comments that led to a better interpretation of our results in the context of recent developments in the theoretical models of bursty star formation at high redshift. RGP acknowledges funding support from a STFC PhD studentship. RGP, FDE, RM, GCJ, JS and LRI acknowledge support by the Science and Technology Facilities Council (STFC), by the ERC through Advanced Grant 695671 ``QUENCH'', and by the UKRI Frontier Research grant RISEandFALL. RM also acknowledges funding from a research professorship from the Royal Society. QD acknowledges a PhD studentship from Trinity College, Cambridge. YI is supported by JSPS KAKENHI Grant No. 24KJ0202 and 24KJ1160. H{\"U} acknowledges funding by the European Union (ERC APEX 101164796. Views and opinions expressed are however those of the authors only and do not necessarily reflect those of the European Union or the European Research Council Executive Agency. Neither the European Union nor the granting authority can be held responsible for them. SZ, SC and GV acknowledge support from the European Union's HE ERC Starting Grant No. 101040227 - WINGS. IL acknowledges support from PRIN-MUR project ``PROMETEUS" financed by the European Union - Next Generation EU, Mission 4 Component 1 CUP B53D23004750006. AJB and GCJ acknowledge funding from the ``FirstGalaxies'' Advanced Grant from the European Research Council (ERC) under the European Union's Horizon 2020 research and innovation program (Grant Agreement No. 789056). BRP, SA and MP acknowledge grant PID2021-127718NB-I00 funded by the Spanish Ministry of Science and Innovation/State Agency of Research (MICIN/AEI/ 10.13039/501100011033); MP also acknowledges the grants RYC2023-044853-I and PID2024-159902NA-I00, funded by  MICIU/AEI/10.13039/501100011033 and European Social Fund Plus (FSE+). BRP also acknowledges support by the grant PID2024-158856NA-I00. MC acknowledges support from ESO via the ESO Fellowship Europe. EB and GC acknowledge the INAF GO grant ``A JWST/MIRI MIRACLE: Mid-IR Activity of Circumnuclear Line Emission'' and the “Ricerca Fondamentale 2024” INAF program (mini-grant 1.05.24.07.01). TH was supported by JSPS KAKENHI 25K00020. ST acknowledges support from the Royal Society Research Grant G125142. JW acknowledges support from the Cosmic Dawn Center through the DAWN Fellowship; the Cosmic Dawn Center (DAWN) is funded by the Danish National Research Foundation under grant No. 140.

\section*{Data Availability}

This work is based on observations with the NASA/ESA/CSA James Webb Space Telescope. The data were obtained as part of the JWST program ID 4528 (Cycle 3, PI: K. Isaak). These data are available from \href{https://mast.stsci.edu/portal/Mashup/Clients/Mast/Portal.html}{Mikulski Archive for Space Telescopes} at the Space Telescope Science Institute, which is operated by the Association of Universities for Research in Astronomy, Inc., under NASA contract NAS 5-03127 for JWST. The reduced datacube is available upon reasonable request. The photometry data are available from the public DAWN JWST Archive (DJA) data \footnote{Available at \url{https://s3.amazonaws.com/grizli-v2/JwstMosaics/v7/index.html}}. 


\bibliographystyle{config/mnras}
\bibliography{biblio} 




\section*{}
\noindent
\textit{\small{
$^{1}$ Kavli Institute for Cosmology, University of Cambridge, Madingley Road, Cambridge CB3 0HA, UK \\
$^{2}$ Cavendish Laboratory – Astrophysics Group, University of Cambridge, 19 JJ Thomson Avenue, Cambridge CB3 0HE, UK \\
$^{3}$ Department of Physics and Astronomy, University College London, Gower Street, London WC1E 6BT, UK \\
$^{4}$ Waseda Research Institute for Science and Engineering, Faculty of Science and Engineering, Waseda University, 3-4-1, Okubo, Shinjuku, Tokyo 169-8555, Japan \\
$^{5}$ Centro de Astrobiolog\'{i}a (CAB), CSIC-INTA, Ctra. de Ajalvir km 4, Torrej\'on de Ardoz, E-28850, Madrid, Spain\\
$^{6}$ Sub-Department of Astrophysics, Department of Physics, University of Oxford, Denys Wilkinson Building, Keble Road, Oxford, OX1 3RH, UK \\
$^{7}$ Sorbonne Universit\'e, CNRS, UMR 7095, Institut d'Astrophysique de Paris, 98 bis bd Arago, 75014 Paris, France \\
$^{8}$ Max-Planck-Institut f{\"u}r extraterrestrische Physik (MPE), Gie\ss{}enbachstra\ss{}e 1, 85748 Garching, Germany
$^{9}$ INAF - Osservatorio Astrofisco di Arcetri, largo E. Fermi 5, 50127 Firenze, Italy\\
$^{10}$ European Space Agency, c/o Space Telescope Science Institute, 3700 San Martin Drive, Baltimore MD 21218, USA\\
$^{11}$ Scuola Normale Superiore, Piazza dei Cavalieri 7, I-56126 Pisa, Italy\\
$^{12}$ Association of Universities for Research in Astronomy (AURA) Inc. for the European Space Agency (ESA)\\
$^{13}$ Center for Astrophysical Sciences, Department of Physics and Astronomy, The Johns Hopkins University, 3400 N Charles St. Baltimore, MD 21218, USA\\
$^{14}$ European Southern Observatory, Karl Schwarzschild Stra{\ss}e 2, D-85748 Garching bei M{\"u}nchen, Germany\\
$^{15}$ Department of Astronomy, University of Texas, Austin TX 78712, USA\\
$^{16}$ Dipartimento di Fisica e Astronomia, Universit\`a di Firenze, Via G. Sansone 1, 50019, Sesto F.no (Firenze), Italy\\
$^{17}$ Cosmic Dawn Center (DAWN), Copenhagen, Denmark\\
$^{18}$ Niels Bohr Institute, University of Copenhagen, Jagtvej 128 DK-2200, Copenhagen, Denmark\\
}
}
\FloatBarrier

\appendix

\section{Emission line maps}
\label{sec:appendix_images}

This appendix presents images of integrated fluxes maps for a number of key rest-frame optical emission lines, together with the corresponding wavelength slices of the datacube. All these maps show that high-ionisation and Balmer recombination emission are mostly constrained to the very central part of MACS0647-JD1. The central region between the two stellar clumps (\autoref{fig:astrometry_b}) hosts the most intense recent star-formation activity, and, consequently, the most intense line emission. In contrast, the outskirts of the system show more pronounced emission from the low-ionisation gas traced via $\left[\rm O \ \textsc{ii}\right]\lambda3726,3729$ emission.

\begin{figure*}
\centering
  \includegraphics[width=0.8\linewidth]{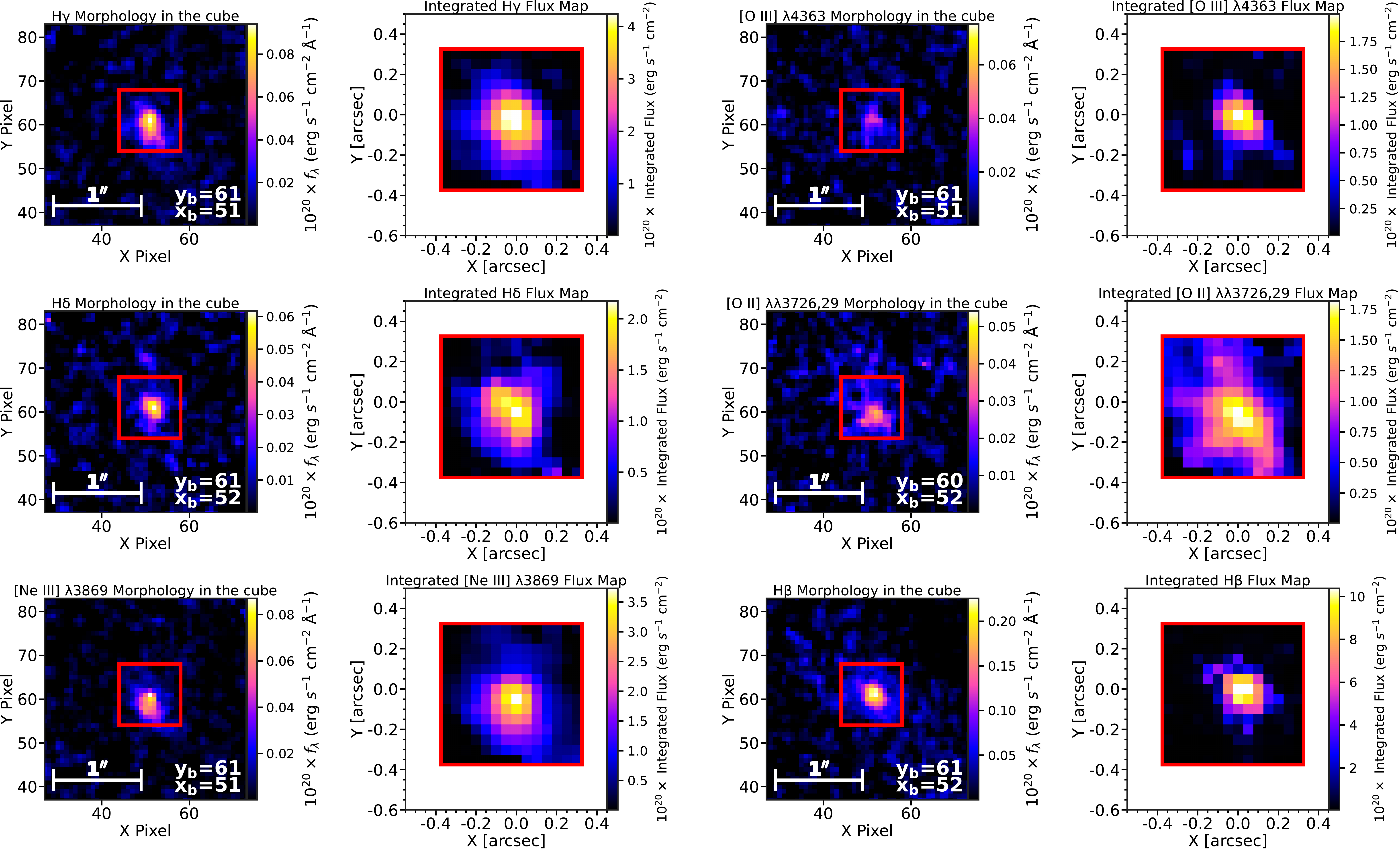}
  \caption{This figure displays a comparison between the images obtained by collapsing the PSF-matched datacube (see Sec. \ref{sec:psf} for details of the PSF matching method) along the wavelength slices corresponding to our key emission lines and the images resulting from mapping out the integrated fluxes of these emission lines which were determined by our MCMC spectral fitting analysis. We remark on small shifts between the centroids of different emission lines (identified by the coordinates $y_{\rm b}$ and $x_{\rm b}$ of the brightest spaxel for each emission line map) but this is not worrying given that these differences are on the order of $\sim 1 \ \rm spaxel$. This centroid drift as a function of wavelength is likely to be caused by a combination of filter transmission effects and imperfect NIRSpec/IFU calibrations.}\label{fig:images_cube_vs_integrated}
\end{figure*}



\section{Further spatially resolved emission lines diagnostics}
\label{sec:appendix_spatial_maps_O33_O32}

\begin{figure*}
\centering
  \includegraphics[width=0.74\linewidth]{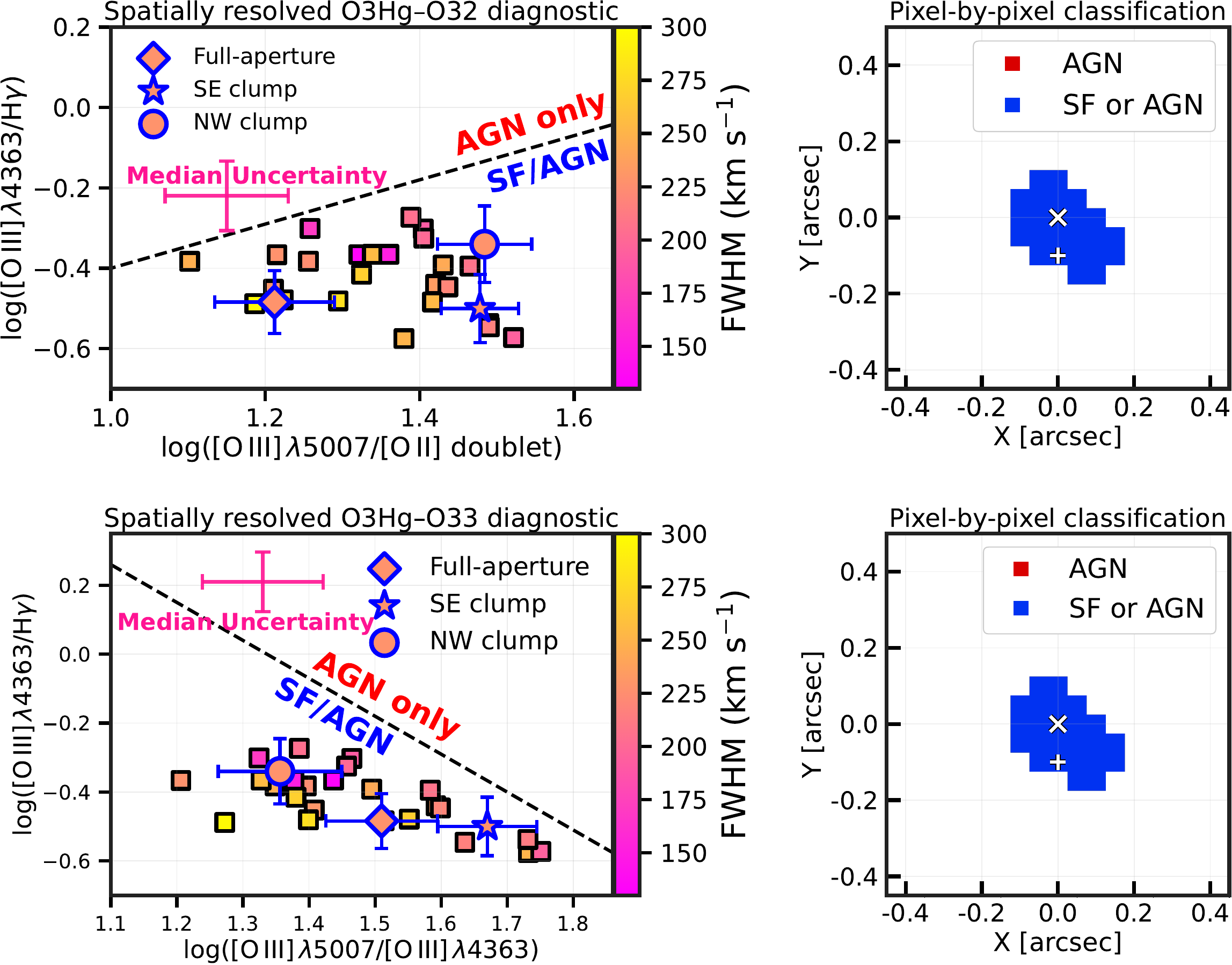}
  \caption{This diagram is equivalent (in terms of labelling and the meaning of various symbols) to \autoref{fig:mazzolari_main}, which presents the O3Hg vs Ne3O2 diagnostic diagram on a spatially resolved spaxel-by-spaxel scale. However, alternative AGN vs SFG diagnostics \citep{mazzolari+24} are illustrated: O3Hg-O32 \textbf{(top panels)} and O3Hg-O33 \textbf{(bottom panels)}. We can remark that, in contrast to the results based on the diagnostic $\rm O3Hg-Ne3O2$ (from \autoref{fig:mazzolari_main}, showing the presence of a south-southwest AGN-dominated region), if we use either $\rm O3Hg-O32$ or $\rm O3Hg-O33$ diagnostics, we would observe that all regions of the MACS0647-JD1 system are either dominated by star-formation driven photoionisation or the contribution of SF and AGN are similar.}\label{fig:mazzolari_more}
\end{figure*}



In \autoref{fig:mazzolari_more}, we report spaxel-by-spaxel maps of O33 and O32 diagnostic ratios \citep[as defined in the main text of Sec. \ref{sec:agn_identification} and following the criteria and demarcation lines identified by][]{mazzolari+24}. In the case of both diagnostics, all spaxels reside below the `AGN only' regime, revealing that AGN is likely not the main excitation source of the ISM in neither region of this galaxy. 

\section{On the velocity field of MACS0647-JD1}
\label{sec:velocity_problem}

\begin{figure}
\centering
  \includegraphics[width=\linewidth]{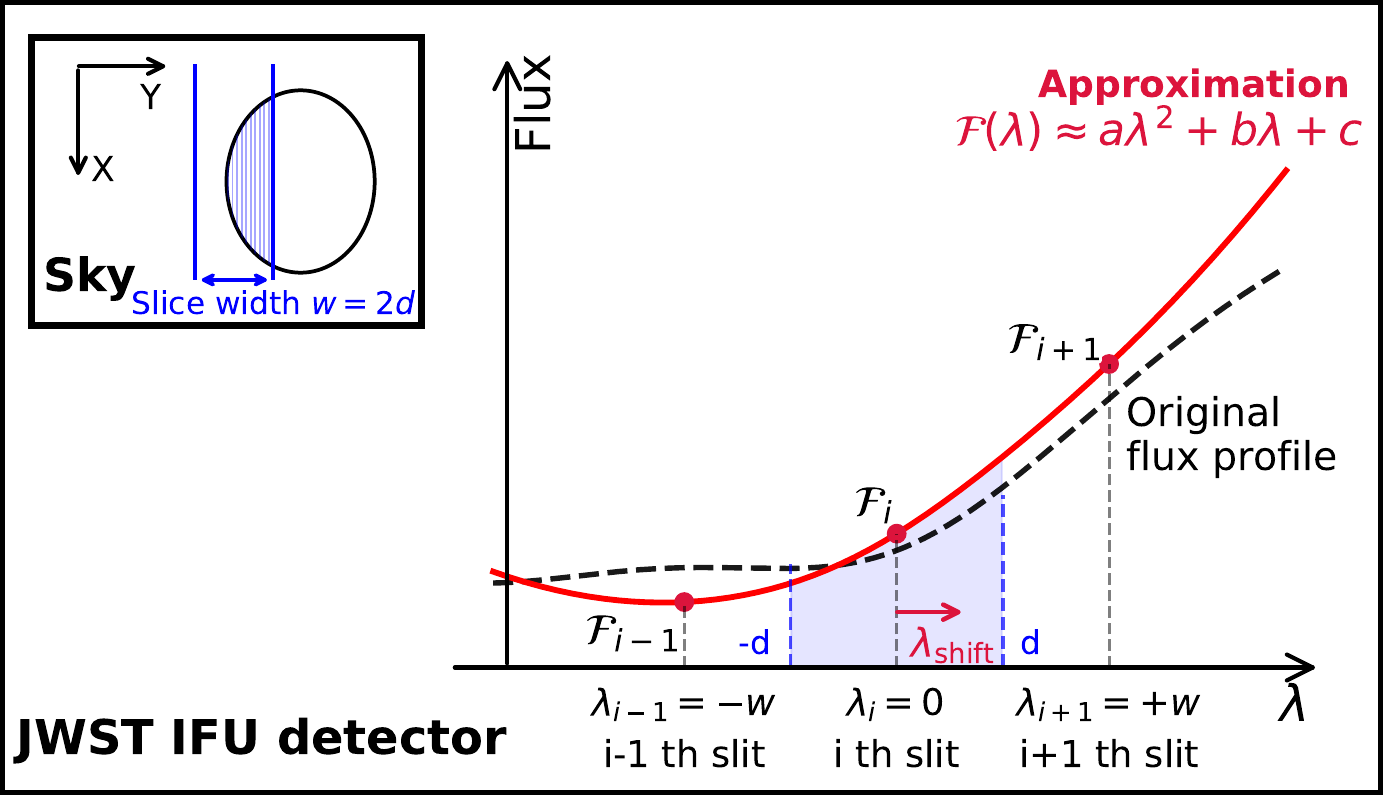}

  \caption{In this figure, we present a simplified explanation of the slit width effect. The main panel shows the flux profile projected onto the {\it JWST} NIRSpec/IFU detector (black dashed curve) and the measured fluxes $\mathcal{F}_{i-1}$, $\mathcal{F}_{i}$ and $\mathcal{F}_{i+1}$ on three adjacent slices. In this approach, the original flux profile is approximated with a quadratic function (the red solid curve) within a wavelength range $-w<\lambda<w$. The inset panel schematically shows an hypothetic circular object on the sky. The X (Y) direction denotes the direction of the slice length (width). The blue lines show the position of the edges of the central slice. The light coming into the slice (blue hatched) is projected onto the detector, and the Y direction in the sky is parallel to the dispersion (or wavelength) direction on the detector.}
  \label{fig:yuki_empress}
\end{figure}

\begin{figure}
\centering
  \includegraphics[width=\linewidth]{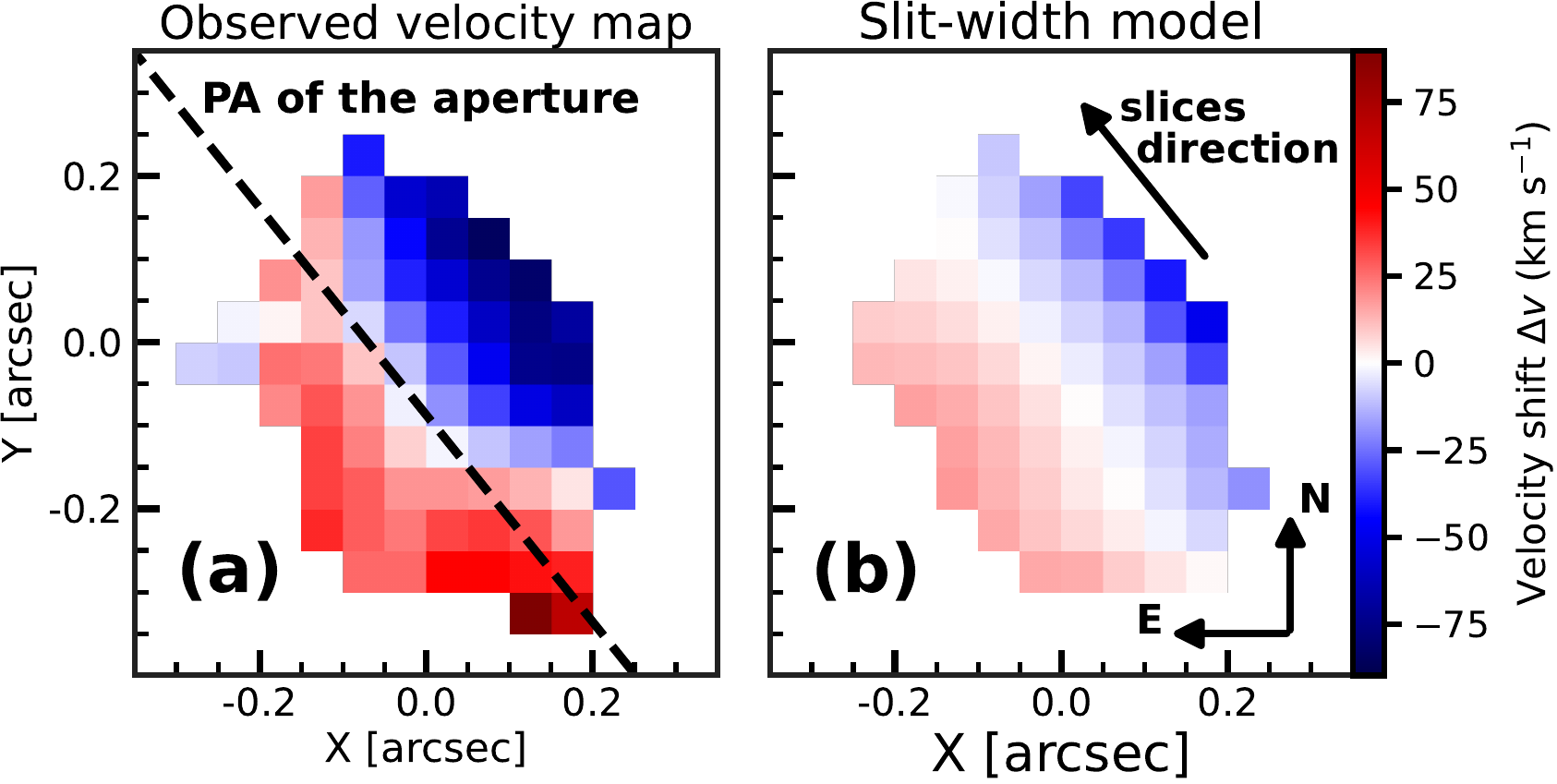}

  {\phantomsubcaption\label{fig:fake_a}
  \phantomsubcaption\label{fig:fake_b}}
  \caption{This figure shows the spatially resolved $\Delta v$ (the velocity field) map derived from single-Gaussian component spaxel-by-spaxel spectral fits (\autoref{fig:fake_a}). We only show the spaxels with a solid detection of H$\gamma$ ($\rm S/N>5$). We also indicate the position angle of the aperture of these observations, $\rm PA =38.83^{\circ}$, i.e. the direction of the NIRSpec/IFU slices. The velocity gradient of the observed rotational pattern is perpendicular to the direction of the slices. In \autoref{fig:fake_b}, we show a mock velocity pattern resulting from the IFU slices geometrical model described in the text.}
  \label{fig:fake_rotation}
\end{figure}

In \autoref{fig:fake_a}, we present the spatially resolved maps of $\Delta v$ (velocity shift of a particular spaxel with respect to galaxy's systemic velocity). We obtain $\Delta v$ from the MCMC fitted redshift corresponding to each spaxel spectrum:

\begin{equation}
    \left(1+z_{\rm pix}\right)=\left(1+z_{\rm pix,med}\right)\times\left(1+\Delta v/c\right) \ .
\end{equation}

\noindent Here, the median redshift of the spaxels is $z_{\rm pix,med}=10.1672$. The fact that the direction of the rotational gradient approximately coincides with the direction of IFU slicers, together with the relatively small amplitude of 160 $\rm km \ s^{-1}$ ($\sim 1.45$ spectral pixels, peak-to-peak) makes it uncertain whether the observed rotation is a real feature. This issue has been identified and documented (symptom NS-IFU07\footnote{\url{https://jwst-docs.stsci.edu/known-issues-with-jwst-data/nirspec-known-issues/nirspec-ifu-known-issues/}}) but no official solution has been proposed to date. Similar artificial rotation patterns have been reported in ground-based IFU observations obtained with the Subaru/FOCAS instrument \citep{kashikawa+2002,ozaki+2020} and, more recently, in high spectral resolution IFU studies of high-redshift galaxies \citep{ivey_lrd_u}. In particular, \citet{isobe_empress} demonstrated that a differential effective slit width across the IFU slices can introduce spurious velocity gradients aligned with the instrumental geometry, thereby mimicking ordered rotation in emission-line velocity fields. 

While the geometrical model implemented within the current pipeline typically succeeds in providing an accurate wavelength calibration \citep{2016_cali,2022_cali}, a recent report by \citet{ifu_report} argues that some imperfections in the current IFU spectral distortion model remain. Artificial kinematics shifts are observed for multiple grating/filter combinations and regardless of the cube weighting algorithms (`drizzle', `EMSM' or `MSM'). In the particular case of G395M grating that we used in our observations, \citet{ifu_report} reports a peak-to-peak amplitude of the artificial kinematic shift of $\sim 1.6$ spectral pixels in the case of a planetary nebula with substantially lower intrinsic kinematics. This is similar to the amplitude of the velocity gradient in \autoref{fig:fake_a}. However, it should be noted that the rotational pattern observed in our case is both more spatially extended (and larger than the PSF size), but also significantly less noisy (as there are less large kinematic variations in adjacent spaxels) than the pattern reported by \citet{ifu_report}. 

In \autoref{fig:fake_b}, we show that a simple geometrical model of the IFU slicer configuration (that is described and explained schematically in \autoref{fig:yuki_empress}) can reproduce an artificial velocity pattern similar in amplitude and orientation to that observed in our data. 
The underlying effect arises from spatial variations in the emission-line surface brightness across the finite width of an IFU slice \citep{bacon+95}. If we have a slice of projected width \emph{w} on the IFU detector and with a flux profile $\mathcal{F} \left(\lambda\right)$ along the dispersion (wavelength) direction coordinate $\lambda$ (perpendicular to the slit width direction), the effective centroid of the detected emission line can be shifted relative to the true wavelength. The method to quantify the magnitude of this wavelength shift is thoroughly described in \citet{isobe_empress}, but here we give an outline of the numerical calculations. We can estimate the wavelength shift caused by the slit-width effect from a barycenter of the flux following a $\mathcal{F}\left(\lambda\right)$ profile intercepted by the slice:

\begin{equation}
    \lambda_{\rm shift} = \frac{\int \limits_{-w/2}^{+w/2} \lambda\times \mathcal{F}\left(\lambda\right) \ \rm d\lambda}{\int \limits_{-w/2}^{+w/2}\mathcal{F}\left(\lambda\right) \ \rm d\lambda} \ .
    \label{eq:lambda_fake_shift}
\end{equation}

\noindent where $\lambda$ denotes the dispersion-axis coordinate across the projected slice width rather than the wavelength. This expression captures the fact that asymmetric illumination of the slice can bias the measured wavelength centroid. Following \citet{isobe_empress}, and assuming that $\mathcal{F} \left(\lambda\right)$ can be locally approximated by a second-order polynomial ($\mathcal{F} \left(\lambda\right) \approx a\lambda^{2} + b\lambda +c$), then Equation (\ref{eq:lambda_fake_shift}) can be simplified by performing the integration:

\begin{equation}
\lambda_{\rm shift} = w \times \frac{bw}{aw^{2}+12c}
\label{eq:shift_v1}
\end{equation}

\noindent We then consider three adjacent IFU slices on the detector, at the dispersion axis coordinates $\lambda_{i-1}=-w$, $\lambda_{i}=0$ and $\lambda_{i+1}=+w$, such that the central wavelength slice `i' contains the peak of H$\gamma$ line emission. The fluxes measured in each of these slices are therefore:

\begin{equation}
\mathcal{F}_{i+1}\equiv \mathcal{F} \left(\lambda_{i+1}\right) = aw^2 +bw+c
\end{equation}

\begin{equation}
\mathcal{F}_{i-1}\equiv \mathcal{F} \left(\lambda_{i-1}\right) = aw^2 -bw+c
\end{equation}

\begin{equation}
\mathcal{F}_{i}\equiv \mathcal{F} \left(\lambda_{i}\right) = c
\end{equation}

\noindent Using these three parametrisation equations for the fluxes observed on the three adjacent slices, we can further simplify Equation (\ref{eq:shift_v1}):

\begin{equation}
    \lambda_{\rm shift} =  w \times \frac{\mathcal{F}_{i+1}-\mathcal{F}_{i-1}}{\mathcal{F}_{i+1}+22\times \mathcal{F}_{i}+\mathcal{F}_{i-1}} \ ,
    \label{eq:lambda_fake_shift2}
\end{equation}


\noindent This formulation provides an estimate of the wavelength bias induced by slice-to-slice flux gradients. For JWST/NIRSpec, the undispersed IFU slices are 2 pixels wide according to JWST documentation\footnote{\url{https://jwst-docs.stsci.edu/jwst-near-infrared-spectrograph/nirspec-observing-modes/nirspec-ifu-spectroscopy}}. Using the spectral dispersion and line-spread function appropriate for the observed H$\gamma$ wavelength, this corresponds to an effective projected width of $w \approx \SI{35.8}{\angstrom}$. The peak-to-peak amplitude of the rotational pattern resulting from our slit geometry model is $\approx 55 \ \rm km \ s^{-1}$, $\sim 3.2$ times smaller than in the observed data ($\approx 175 \ \rm km \ s^{-1}$). However, if we compare the $5^{\rm th}-95^{\rm th}$ percentiles amplitudes, the discrepancy is reduced ($114 \ \rm km \ s^{-1}$ in the observed data and $44 \rm \ km \ s^{-1}$ in our simple model). We emphasize that our implementation represents a highly simplified, first-order description of the IFU slit geometry and does not capture more subtle instrumental effects or deviations of the slit geometry from a simple, parametric form. In conclusion, while instrumental systematics may plausibly account for a substantial fraction of the observed velocity gradient, we are not able to fully rule out the hypothesis that this kinematic pattern might be real, which could imply that the MACS0647-JD1 system would consist of two bright clumps within the same rotating disc.

\section{\textsc{prospector} SED fits of spectroscopy and photometry}
\label{sec:prospector_results_appendix}

\begin{figure*}
\centering
  \includegraphics[width=\linewidth]{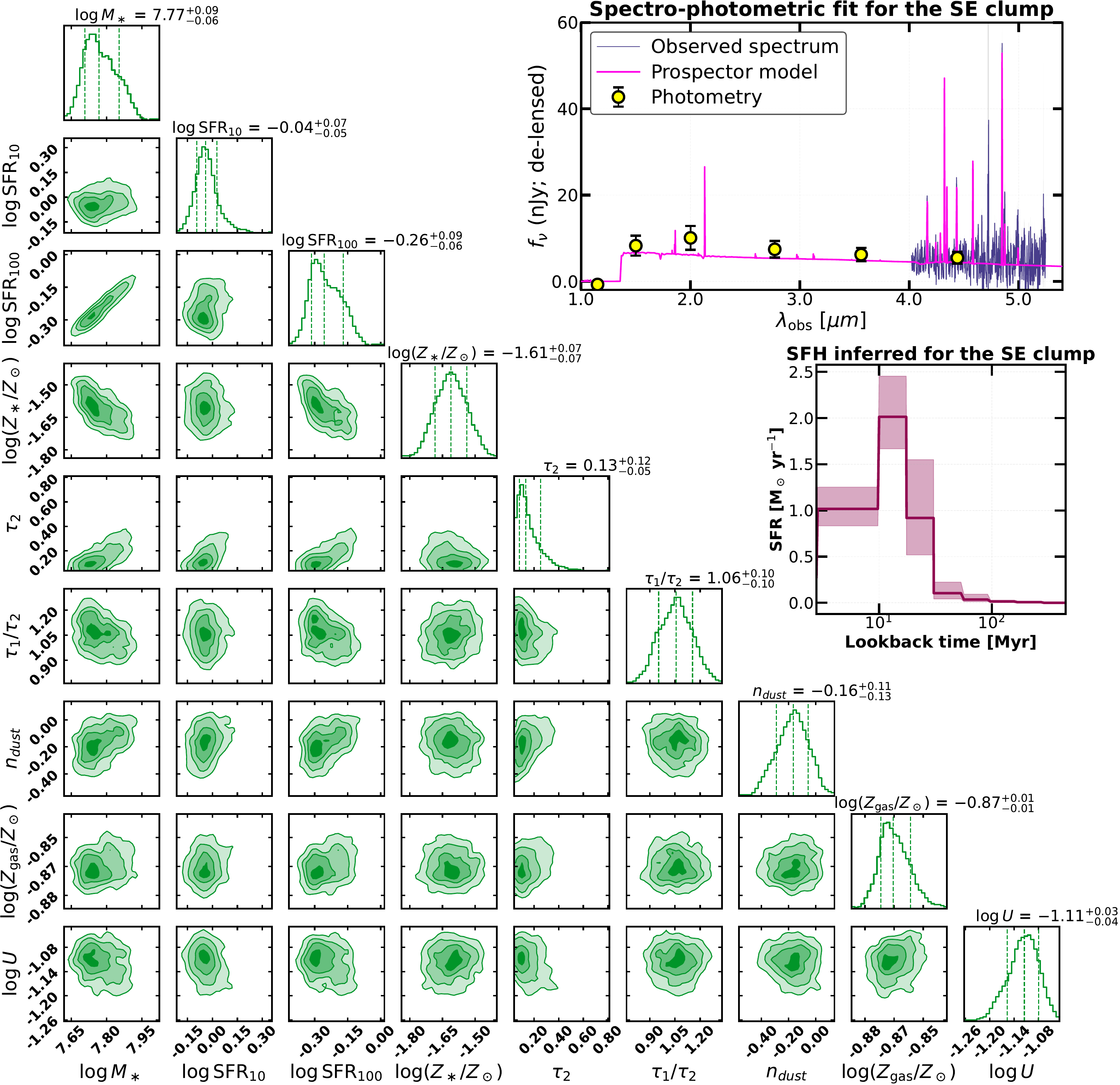}

  \caption{In this figure, we present the \textsc{Prospector} fitting results for the more massive stellar component (the south-east clump). The corner plot displays the marginalised one-dimensional posterior distribution and the two-dimensional correlations involving the following parameters: the logarithm of the stellar mass (in solar masses), the logarithms of SFR averaged over the 10 Myr or 100 Myr prior to observation (in $M_{\odot}\, \rm yr^{-1}$), stellar metallicity, the optical depth of the diffuse ISM ($\tau_{2}$), the extra-additional factor of attenuation towards the birth clouds ($\tau_{1}/\tau_{2}$), the dust attenuation curve slope, $n_{\rm dust}$, the gas-phase metallicity and the ionisation parameter. The top-right inset panel shows the resulting SED model for the SE clump, while the central-right inset panel displays the SFH inferred by \textsc{Prospector.}}
  \label{fig:prospector_full_big}
\end{figure*}

\begin{figure*}
\centering
  \includegraphics[width=\linewidth]{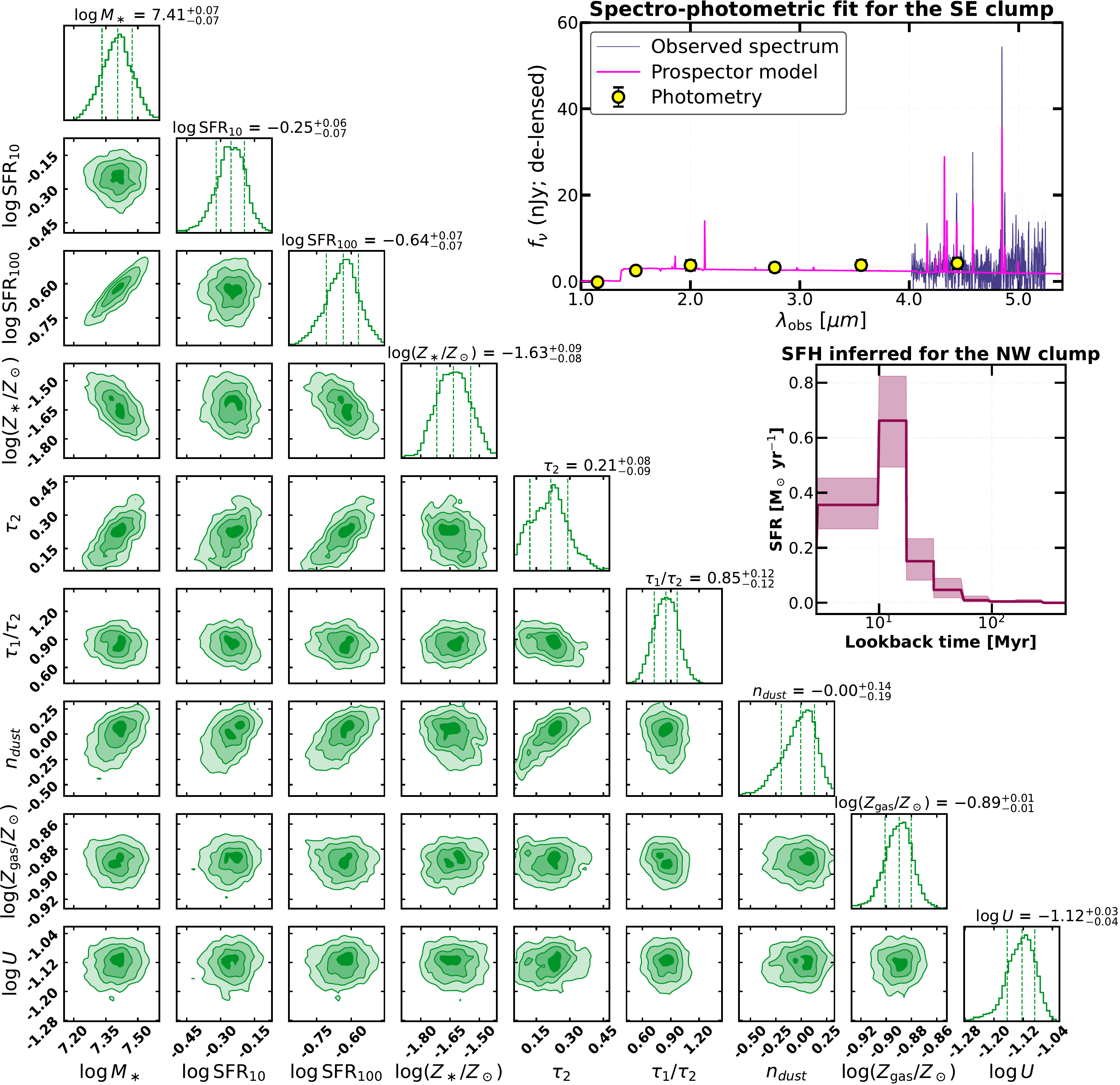}

  \caption{This figure is equivalent to \autoref{fig:prospector_full_big}, but it shows the \textsc{Prospector} best-fit parameters in the case of the less massive stellar component (the north-west clump). The top-right inset panel shows the SED model for the NW clump, while the central-right inset panel displays the SFH inferred by \textsc{Prospector.}}
  \label{fig:prospector_full_small}
\end{figure*}

In \autoref{fig:prospector_full_big} and \autoref{fig:prospector_full_small}, we show our results from the \textsc{Prospector} fits based on the spectra and photometric fluxes corresponding to the apertures of the two stellar components (SE-C and NW-C in \autoref{fig:astrometry_b}). Our fitting method is described in Section \ref{sec:sed}.

\bsp	
\label{lastpage}
\end{document}